\documentclass[preprint,12pt]{elsarticle}

\usepackage{amssymb}
\usepackage{graphicx}
\usepackage{amsfonts}
\usepackage{amsmath}
\usepackage{amssymb}
\usepackage{bbold} 
\usepackage[margin=.7in]{geometry}
\usepackage{booktabs}
\usepackage{float}
\usepackage{mathrsfs}
\usepackage{hyperref}

\journal{Nuclear Physics B}

\newcommand{\de}{\mathrm{d}}
\newcommand{\Tr}{ \operatorname{Tr}}

\newcommand{\diag}{\mathrm{diag}}
\newcommand{\ord}{\mathcal{O}}

\newcommand{\ket}{\rangle}
\newcommand{\bra}{\langle}

\newcommand{\op}{\mathscr{O}}

\newcommand{\Gr}{\mathcal{G}}

\newcommand{\nextline}{\notag \displaybreak[0]\\&}
\newcommand{\nline}{ \displaybreak[0]\\}
\newcommand{\mbf}{\mathbf}
\newcommand{\mbfb}[1]{\overline{ \mathbf{ #1}} }

\newcommand{\Br}{\mathcal{B}\mathrm{r}}

\newcommand{\SM}{\mathrm{SM}}
\newcommand{\LQ}{\mathrm{LQ}}
\newcommand{\FT}{\mathrm{FT}}
\newcommand{\GUT}{\mathrm{GUT}}

\renewcommand{\Re}{\operatorname{Re}}
\renewcommand{\Im}{\operatorname{Im}}

\begin{document}

\begin{frontmatter}



\title{Phenomenology of $\rm SU(5)$ low-energy realizations: the diphoton excess and Higgs flavor violation}


\author[label1]{Andrea Di Iura}
\author[label2]{Juan Herrero-Garcia}
\author[label1]{Davide Meloni}

\address[label1]{Dipartimento di Matematica e Fisica, Universit\`a  di Roma Tre;\\
INFN, Sezione di Roma Tre,\\
Via della Vasca Navale 84, 00146 Rome, Italy}
\address[label2]{Department of Theoretical physics, School of Engineering Sciences, \\KTH Royal Institute
of Technology,\\ AlbaNova University Center, 106 91 Stockholm, Sweden}

\begin{abstract}
We discuss different $\rm SU(5)$ low-energy realizations and illustrate their use with the 
diphoton excess and Higgs  flavor violation, which require new physics at the TeV scale. In particular, we study two scenarios for a $750$ GeV resonance: 
in the first one the resonance belongs to the adjoint of $\rm SU(5)$, being either an $\rm SU(2)_L$ singlet or a triplet, 
while in the second case the signal is due to the CP-even and CP-odd states of a new $\rm SU(2)_L$ Higgs doublet belonging to a ${\bf 45}_H$ or a ${\bf 70}_H$ representations, 
giving rise to a two-Higgs doublet model at low energies. We study the fine-tuning 
needed for the desired members of the multiplets to be light enough, while having the rest at the GUT scale. In these scenarios, the production 
and decay into photons of the new resonance are mediated by the leptoquarks (LQ) present in these large $\rm SU(5)$ representations. 
We analyse the phenomenology of such scenarios, focusing on the most relevant predictions that can help to disentangle the different models, 
like decays into gauge bosons, Standard Model (SM) fermions and LQs pair production. In the case of the ${\bf 45}_H$ (the Georgi-Jarlskog model), 
we also study the possibility to have Higgs flavor violation. We find that $B_s$ mixing limits (in addition to $\tau\rightarrow \mu \gamma$) always imply 
that $\Br(h\rightarrow \tau\mu,\,bs)\lesssim 10^{-5}$.
\end{abstract}

\begin{keyword}



\end{keyword}

\end{frontmatter}

\section{Introduction}

Most hints of new physics, either coming from LHC searches or flavor anomalies in the quark and lepton sectors, require new degrees of freedom at the TeV scale. 
In this paper we want to study low-energy realizations of $\rm SU(5)$, which can have TeV-scale members of the multiplets that can address such signals. 
Although our analysis of the different $\rm SU(5)$ frameworks will be completely general, and can be of use whenever TeV-scale particles are required, 
we are specially motivated by the diphoton excess observed at the LHC by ATLAS~\cite{Aad:2015mna, ATLASconf2, ATLASconf, Aaboud:2016tru} and 
CMS~\cite{CMSconf,CMSconf2}. Moreover, we will also study Higgs lepton and quark flavor violating decays, motivated by recent slight hints 
of a large $h\rightarrow \tau \mu$ signal~\cite{Aad:2016blu, Khachatryan:2015kon}, although, as we will discuss, we find that this cannot be accommodated, at least in our minimal scenario.

Regarding the diphoton signal, the local (global) significances of the excess of events in their combined $8+13$ TeV data are in the $\sim 3-4\,\sigma$ ($\sim 2\sigma$) range, and both experiments observe 
the excess at the same invariant mass of $750$ GeV. If interpreted as a new resonance, CMS prefers a narrow decay width 
$\Gamma_S\sim \mathcal{O}(10)$ MeV, while the ATLAS data seems to favor a large width, $\Gamma_S\sim 45$ GeV. 

Although more data are needed before claiming the discovery of a new particle or particles, several explanations for the diphoton excess have 
been proposed, introducing new TeV scale particles beyond the SM spectrum. In this paper we are mainly interested in scalar (pseudo-scalar) 
particles as the new resonance. The simplest models involve either $\rm SU(2)_L$ singlets, doublets or higher representations, 
that couple to TeV-scale degrees of freedom with large electric charges, and/or large 
multiplicities/couplings~\cite{Franceschini:2015kwy, Kamenik:2016tuv, Ellis:2015oso, Low:2015qho, Buttazzo:2015txu, Gupta:2015zzs, Dev:2015vjd}. 
Many models introduce, in addition to the resonance, new vector-like fermions, mostly in the context of two-Higgs doublet models 
(2HDM)~\cite{Angelescu:2015uiz, Djouadi:2016eyy} and/or leptoquarks (LQ), see for 
instance refs.~\cite{Bauer:2015boy, Murphy:2015kag, Hati:2016thk, Dey:2016sht}. In all these models, assuming that the new scalar resonance is produced 
through gluon-gluon fusion, the excess can be accommodated with $\Gamma_{\gamma\gamma}/m_S \approx 10^{-6}\,(10^{-4})$ in the case 
of a narrow (large) width. 

The cross section in the invariant mass of two photons can be estimated in 
accordance with ref.~\cite{Strumia:2016wys} at $\sqrt{s} = 13\ \mathrm{TeV}$ for a narrow (broad) resonance as:
\begin{equation}
\label{eq:results}
 \sigma_{\rm obs}(pp \to S \to \gamma \gamma) = \begin{cases}
                                       4.8 \pm 2.1\ (7.7 \pm 4.8)\ \mathrm{fb} & \text{CMS}\\
                                       5.5 \pm 1.5\ (7.6 \pm 1.9)\ \mathrm{fb} & \text{ATLAS}\\
                                      \end{cases}\, 
\end{equation}
while at  $\sqrt{s} = 8\ \mathrm{TeV}$ 
\begin{align}
 \sigma_{\rm obs}(pp \to S \to \gamma \gamma) = \begin{cases}
                                        0.63 \pm 0.31\ (0.99 \pm 1.05)\ \mathrm{fb} & \text{CMS}\\
                                       0.21 \pm 0.22\ (0.88 \pm 0.46)\ \mathrm{fb} & \text{ATLAS}\\
                                      \end{cases}\, .
\end{align}

In this paper we want to go one step further in the interpretation of the excess and study well-motivated and minimal scenarios 
in the context of Grand Unified Theories (GUT). The main motivation is that the needed new degrees of freedom are already 
part of GUT multiplets and, it turns out, this implies interesting correlations among processes involving the same multiplets. 
As a prototype of a GUT group, we will focus on non-supersymmetric $\rm SU(5)$~\cite{Georgi:1974sy}, with some well-known nice features like charge quantization 
or the prediction of the weak mixing angle 
$\sin\theta_W$, although
we are well aware of the fact that limits on proton decays, compatibility with charged fermion masses and mixings, unification of interactions and naturality considerations 
call for a  beyond the SM theory presumably  more complex than the one analyzed in this paper. 

Although in some cases unification 
of couplings can be improved with respect to the SM case
(with a light $({\bf 3},{\bf 3},1/3)\subset \mbf{45}_H$~\cite{Dorsner:2016ypw,Giveon:1991zm}, but we also found 
that the $({\bf \bar 3},{\bf 3},-4/3)\subset \mbf{70}_H$ is a good option), our set-up is to be seen as a minimal one, involving the less number of 
ingredients, while not attempting to solve all the known drawbacks of the SM. In this view, we devote particular attentions to 
the conditions that are necessary  for some useful fragments of  $\rm SU(5)$ representations 
to be at the TeV scale, and with those we will work in an effective field theory framework. Even though the main motivation of the paper is to accommodate 
the diphoton excess, the $\rm SU(5)$ analysis presented here 
with TeV-scale fields is completely general, and can be used to study other scenarios that require low energy $\rm SU(5)$ fields.

There have been some studies of the diphoton excess in $\rm SU(5)$ scenarios~\cite{Dorsner:2016ypw,Patel:2015ulo,Dutta:2016jqn}. 
In this paper, we analyze what we believe are the best-motivated cases. In the first one, we study the $\mbf{24}_H$ 
representation, which is the minimal addition needed to directly break $\rm SU(5)$ into the SM group. This representation has 
singlet and triplet fields that, if light enough, can be considered as the new observed resonance. 
Beside the singlet case (already studied in some detail in ref.~\cite{Dorsner:2016ypw}), we also analyse the triplet case and the possibility that color octets are at the TeV scale 
(see also refs.~\cite{Dorsner:2012pp, Bai:2010dj}). Furthermore we devote some attention on the fine-tuning (FT) needed to 
realize the needed mass splitting among the $\rm SU(5)$ multiplets. In our numerical results, we have taken into account 
all relevant phenomenological constraints from the LHC data.

In the second framework, we study the possibility that the resonance $S$ belongs to the Higgs doublets of the $\mbf{5}_H$ and $\mbf{45}_H$ or $\mbf{70}_H$ representations;  
thus, we are effectively left with a 2HDM \cite{Badziak:2015zez}, when one of the doublets is taken from the $\mbf{45}_H$, well motivated by charged-lepton and down-quark masses (Georgi-Jarlskog model~\cite{Georgi:1979df}), or from the $\mbf{70}_H$ (type I 2HDM). In these cases the resonance can be the heavy combination of the two neutral CP-even members of the Higgs doublets, or 
the CP-odd state $A$, or a combination of both. In addition to accommodating the observed diphoton signal,
we also look for distinctive collider signatures that can help to disentangle the different scenarios.

Regarding Higgs flavor violation, the second scenario with a $\mbf{5}_H$ and a $\mbf{45}_H$ (Georgi-Jarlskog model) immediately leads to violation of lepton and down-quark flavors. Moreover, as we will show, decays of the light Higgs to leptons and quarks are related and completely fixed in terms of charged lepton and quark masses, and the CKM mixings. We will study the most promising ones, $h\rightarrow \tau \mu$ and $h\rightarrow b s$, discussing the possibility of addressing the hint of 
a $1\,\%$ $\Br(h\rightarrow \tau \mu)$~\cite{Khachatryan:2015kon}. Furthermore, extensions of the minimal model with the light fields belonging to the 
$\mbf{24}_H$ will be discussed.

The paper is structured as follows. In sec.~\ref{sec:singlet} we study the $\mbf{24}_H$ and the different possibilities offered by it
for the new resonance, being either an $\rm SU(2)_L$ singlet or a triplet, also taking into account the possibility of light color octets,
 the needed fine-tuning in these set-ups and the relevant collider signatures.
In sec.~\ref{sec:2HDM_model} we focus on 2HDM in which one of the Higgses comes from the $\mbf{45}_H$ or the $\mbf{70}_H$ representations. We study Higgs lepton and quark flavor violation in sec.~\ref{sec:hlfv}. In sec.~\ref{sec:LQpheno} we study the related LQ phenomenology and how these new states affect GUT unification. 
Finally, we summarize the different predictions and draw our 
conclusions in sec.~\ref{sec:discussion}.

\section{A new resonance from the singlet and/or triplet of the $\mbf{24}_H$}
\label{sec:singlet}

\subsection{The model}
We assume an $\rm SU(5)$ framework with scalar fields in the $\mbf{5}_H$ and $\mbf{45}_H/\mbf{70}_H$ representations. 
All these fields contain an $\rm SU(2)_L$ doublet. In order to have a GUT breaking 
$\rm SU(5) \to \Gr^{\SM} \equiv SU(3)_c \otimes SU(2)_L \otimes U(1)_Y$ the minimal way is to introduce a scalar field $
\Sigma$ that transforms as the adjoint representation of $\rm SU(5)$, the $\mbf{24}_H$. The decomposition of $\Sigma$ under the SM gauge group $\Gr^{\SM}$ is 
the following
\begin{align}
 \Sigma \sim \Sigma_0 \oplus \Sigma_3 \oplus \Sigma_8 \oplus \Sigma_{3, 2} \oplus \Sigma_{\overline{3},2}\,,
\end{align}
where $\Sigma_0 \sim (\mbf{1}, \mbf{1}, 0),   \Sigma_3 \sim (\mbf{1}, \mbf{3}, 0),  \Sigma_8 \sim (\mbf{8}, \mbf{1}, 0), 
\Sigma_{3, 2}\sim (\mbf{3}, \mbf{2}, 5/6), \Sigma_{\overline{3}, 2}\sim (\mbfb{3}, \mbf{2}, -5/6)$. 
The $\Sigma$ field can be cast in the form $\Sigma = \sum_A \Sigma^A L^A$, where $L^A$ are the generators of $\rm SU(5)$~\cite{Ellis:2015oso}. 
The scalar potential can then be written as:
\begin{align}
\label{full_potential_SU5}
 V = V_{\mbf{5}} + V_{\mbf{24}} + V_{\mbf{45}} +V_{\mbf{70}}  + V_I\,,
\end{align}
where $V_I$ refers to the interaction potential among the scalar fields and $V_{\mbf{5},\mbf{24},\mbf{45},\mbf{70}}$ correspond to the self-interacting terms. In particular, for the adjoint representation $\Sigma$ the explicit form of the potential is
\begin{align}
\label{scalar_potential_24}
 V_{\mbf{24}} = - \frac{\mu_{\mbf{24}}^2}{2} \Tr(\Sigma^2) + \frac{a}{4} \Tr^2(\Sigma^2)+ \frac{b}{2} \Tr(\Sigma^4) + \frac{c}{3} \Tr(\Sigma^3) \,.
\end{align}
The minimum is proportional to the diagonal $\rm SU(5)$ generator $L^{12}$ and it is given by the well-known expression 
$\bra \Sigma \ket = v_{\mbf{24}}\ \diag \{2,2,2,-3,-3 \}/\sqrt{30}$~\cite{Guth:1981uk, Guth:1979bh} with
\begin{align}
\label{vev_adjoint_field}
 v_{\mbf{24}} = \frac{c}{b} \sqrt{\frac{\beta}{\gamma}}h(\beta \gamma)\,, \qquad h(x)\equiv  \sqrt{1 + \frac{1}{120x}}+ \frac{1}{\sqrt{120x}}\,.
\end{align}
The dimensionless parameters $\beta$ and $\gamma$ are useful to investigate the condition for the minimum. These are
\begin{align} \label{dless}
 \beta \equiv b\frac{\mu^2_{\mbf{24}}}{c^2}\,,\qquad \gamma \equiv \frac{a}{b}+\frac{7}{15}\,.
\end{align}
A non-zero VEV requires $\gamma >0$, while a definite positive ground state needs $b >0$, see ref.~\cite{Guth:1981uk} for further details. 
To have a local minimum and a positive mass spectrum the parameter $\beta$ is such that
\begin{align}
\label{beta_bound_SU5_SB}
 \beta > \left\{\begin{array}{l l}
                 \dfrac{15}{32}\left(\gamma - \dfrac{4}{15}\right) & \gamma > \dfrac{2}{15}\\
                 -\dfrac{1}{120\gamma} & 0 < \gamma < \dfrac{2}{15}
                \end{array}
 \right. .
\end{align}
After the $\rm SU(5)$ symmetry breaking to $\Gr^{\SM}$ the mass spectrum of the $\Sigma$ particles is the following: 
$\Sigma_{3, 2}$ and $\Sigma_{ \overline{3},2}$ are eaten by $X^\mu$ and $Y^\mu$, the twelve gauge bosons of $\rm SU(5)$. 
These are degenerate in mass with $m_X^2 = m_Y^2 = 5g_5^2 v_{\mbf{24}}^2/12 \sim m_{\GUT}^2$, where $ m_{\GUT}$ is the typical 
GUT mass scale of $\ord(10^{16})\ \mathrm{GeV}$ and $g_5$ is the $\rm SU(5)$ coupling constant. 
The other particles, that is $\Sigma_0,   \Sigma_3$ and $  \Sigma_8 $, have the following spectrum:
\begin{subequations}
\label{m_square_spectrum}
 \begin{align}
 \label{m0_square}
  m_0^2 &= 2 \gamma \left[ 1 - \frac{1}{1+ \sqrt{1+120 \beta \gamma}}\right]bv_{\mbf{24}}^2\,, \nline
  \label{m3_square}
  m_3^2 &= \left[ \frac{4}{3} - \frac{5}{h(\beta \gamma)}\sqrt{\frac{\gamma}{30\beta}}\right]bv_{\mbf{24}}^2\,, \nline
  \label{m8_square}
  m_8^2 &= \left[ \frac{1}{3} + \frac{5}{h(\beta \gamma)}\sqrt{\frac{\gamma}{30\beta}}\right]bv_{\mbf{24}}^2\,, 
 \end{align}
\end{subequations}
where $m_0, m_3$ and $m_8$ are the masses of $\Sigma_0,   \Sigma_3$ and $  \Sigma_8 $ respectively. 
The interaction potential $V_I$ introduces corrections of order $\ord(v_{\mbf{5}}/v_{\mbf{24}}) = \ord(10^{-14})$ where $v_{\mbf{5}}$ is 
the VEV of the doublet contained in $\mbf{5}_H$ scalar field~\cite{Buras:1977yy}. 
Corrections of similar size exist also for $\mbf{45}_H$ and $\mbf{70}_H$ and can be safely neglected.\\
In the following we define $S$ as the resonance observed at LHC in the diphoton channel, thus we can have 
$S = \Sigma_0$ or $S = \Sigma_3^0$ in the case $S$ is the neutral component of the $\Sigma_3$ field; to obtain a signal in the channel $S \to \gamma \gamma$ compatible with the diphoton 
excess we need to consider the effect of light particles contained in $\Sigma$ or decay modes mediated by 
leptoquarks, as discussed in \cite{Dorsner:2016ypw}. Instead of introducing ad-hoc fragments of $\rm SU(5)$ multiplets, we take into account that the needed particles are already contained in the higher dimensional representation of $\rm SU(5)$, 
such as $\mbf{45}_H$ and $\mbf{70}_H$. Under $\Gr^{\SM}$ we have:
\begin{align}
 \mbf{45} &\sim (\mbf{1}, \mbf{2}, -1/2) \oplus (\mbf{3}, \mbf{1}, 1/3) \oplus (\mbfb{3}, \mbf{1}, -4/3) \oplus (\mbfb{3}, \mbf{2}, 7/6) \oplus (\mbf{3}, \mbf{3}, 1/3) \oplus (\mbfb{6}, \mbf{1}, 1/3) \oplus (\mbf{8}, \mbf{2}, -1/2)\,, \nline
 \mbf{70} &\sim (\mbf{1}, \mbf{2},-1/2)  \oplus  (\mbf{3}, \mbf{1},1/3)  \oplus  (\mbf{1}, \mbf{4}, -1/2)  \oplus  (\mbf{3}, \mbf{3}, 1/3)  \oplus  (\mbfb{3}, \mbf{3}, -4/3)  \oplus  (\mbf{6}, \mbf{2}, 7/6)  \oplus  (\mbf{8}, \mbf{2}, -1/2) \nextline\oplus  (\mbf{15}, \mbf{1}, 1/3)\,,
\end{align}
hence in general several scenarios are possible for production and decay. We will analyze in the following some of the most interesting ones.

\subsection{TeV-scale representations from the $\mbf{24}_H$}
\label{spectra}
In the following we will distinguish among several orderings of the mass eigenstates singlet/triplet (and octet) of the ${\bf 24}_H$.
 
\subsubsection{Singlet case: $m_0 = 750\ \mathrm{GeV} \ll m_3 \sim m_8$}
Assuming that the singlet is the particle observed at LHC, 
a cancellation must occur in the coefficient in front of 
$v_{\mbf{24}}^2 \sim m_{\GUT}^2$ in eq.~\eqref{m0_square}. This is possible when
\begin{align}
 m_0 = 750\ \mathrm{GeV}: \quad 1 - \frac{1}{1+ \sqrt{1+120 \beta \gamma}} \simeq 0 \Longrightarrow \beta = 
 -\frac{1}{120\gamma} + \epsilon, \quad 0 < \epsilon \ll 1\,, \label{eps}
\end{align}
which requires $0 < \gamma < 2/15$ and thus, in general, $\beta$ is negative. 
In this case the square root of $\gamma/\beta$ is an imaginary number, 
also $h(\beta \gamma)$ is imaginary and the mass spectrum is necessarily positive. 
Although on the basis of eqs.~\eqref{dless} and \eqref{eps}
one would expect  $m_{3,8}$ at the GUT scale, we numerically 
verified that this is only possible at the prize of a large  fine-tuning  among the potential parameters; 
on the other hand,  our numerical scan seems to favor a less fine-tuned solution with particle  masses of order 
$m_3 \sim m_8 = \ord(10^6)\ \mathrm{GeV}$.

\subsubsection{Triplet case: $m_3 = 750\ \mathrm{GeV} \ll m_0 \sim m_8$}
Another possibility is that the neutral component of the triplet is the particle responsible for the excess in the diphoton channel. 
In this case the whole triplet is degenerate in mass at tree level and we have
\begin{align}
 m_3 = 750\ \mathrm{GeV}: \quad \frac{4}{3} - \frac{5}{h(\beta \gamma)}\sqrt{\frac{\gamma}{30\beta}} \simeq 0 \Longrightarrow \beta = \frac{15}{32}\left(\gamma - \frac{4}{15}\right)  + \epsilon, \quad 0 < \epsilon \ll 1\,, \label{eps_triplet}
\end{align}
so $\gamma > 2/15$, as can be seen from eq.~\eqref{beta_bound_SU5_SB}. 
In the region $2/15 < \gamma < 4/15$, the parameter $\beta <0$, while in the case $\gamma > 4/15$ the sign of $\beta$ changes. 
In order to have $b >0$, and thus a positive mass spectrum, the parameter $\mu_{\mbf{24}}^2$ changes sign as $\beta$. 
However the situation  $\beta < 0$ must be ignored since otherwise the octet does not have a real mass. 
As in the case of a light singlet a large fine-tuning is necessary to reproduce the
correct order of magnitude of $v_{\mbf{24}} \sim m_{\GUT}$ and having at the same time  $m_0, m_8 > 0$. 
The typical values for the masses obtained from our numerical scan are $m_0 \sim m_8 \sim \ord(10^7)\ \mathrm{GeV}$, although
with more and more fine-tuning the situation $m_0 \sim m_8 \sim m_{\GUT}$ can also be achieved. Notice that quantum corrections break the triplet mass degeneracy and we expect a decay $\Sigma_3^{\pm} \to W^{\pm} Z(\gamma)$ mediated by a loop of light LQs with an invariant mass $\gtrsim 750 \ \mathrm{GeV}$.

\subsubsection{Quasi-degenerate case: $m_0 = 750\ \mathrm{GeV} \lesssim m_3 \sim m_8 \sim 1\ \mathrm{TeV}$}
\label{sect223}
\begin{figure}[h!]
\centering
 \includegraphics[scale=.5]{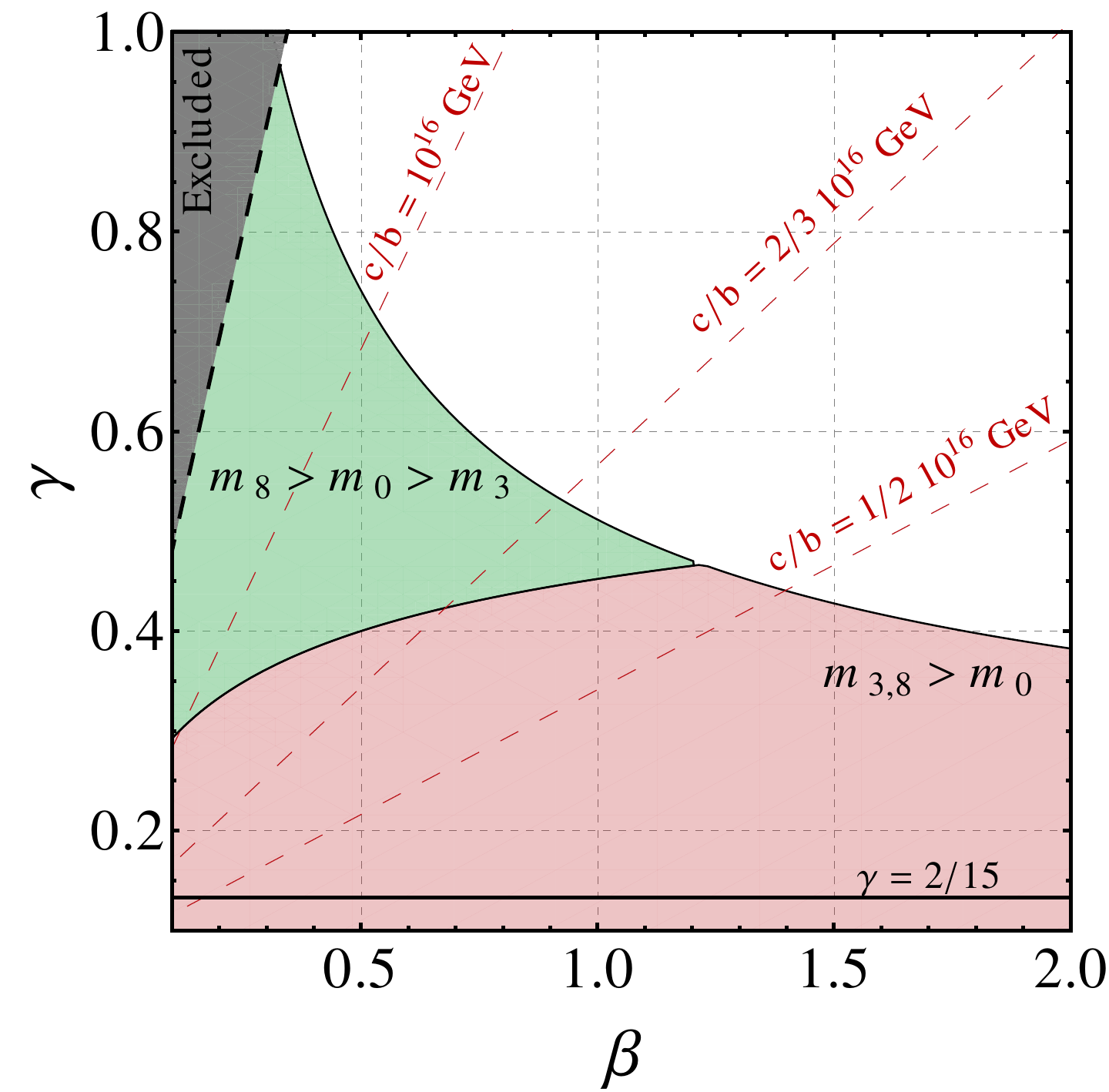}
  \caption{\it Region plot for $m_8 > m_0 > m_3$ (green) and $m_{8,3} > m_0$ (pink) in the $(\beta, \gamma)$ plane. 
  The darkest gray region is excluded by the constraints discussed in eq.~\eqref{beta_bound_SU5_SB}, 
  the red dashed lines are the region of fixed $v_{\mbf{24}} = 10^{16}\ \mathrm{GeV}$, obtained from eq.~\eqref{vev_adjoint_field} 
  for different values of $c/b$.}
 \label{fig:region_beta_gamma_mass_spectrum}
\end{figure}

The last interesting possibility considered here is a quasi-degenerate case, where all the masses are at the $\mathrm{TeV}$ scale.  
Requiring $m_0 \lesssim m_3 \sim m_8 \sim 1 \ \mathrm{TeV}$ we found that the 
region $ 2/15 \lesssim \gamma \lesssim 3/5$ is a good choice for $\beta >0$; however it is quite difficult to obtain $m_3 \gtrsim 1.7 \ \mathrm{TeV}$ for fixed $m_0 = 750\ \mathrm{GeV}$ and 
$v_{\mbf{24}} \sim m_{\GUT}$ (we need a fine-tuning of order $10^{-15}$ between $c$ and $b \times v_{\mbf{24}}$, 
as it can be seen in fig.~\ref{fig:region_beta_gamma_mass_spectrum}, where, for the sake of illustration,  
we take $m_0 = 750\ \mathrm{GeV}$). In order to understand the relations among the potential parameters we can study what happens in the simple case $m_0 \lesssim m_{3} \simeq  m_8$, as done for the singlet case in eq.~\eqref{eps}. From eq.~\eqref{m_square_spectrum} we get
\begin{align}
  m_3 = m_8: \frac{\sqrt{120 \beta  \gamma +1}+30 \gamma +1}{4 \sqrt{120 \beta  \gamma +1}-30 \gamma +4}\simeq 1 \Longrightarrow \beta =  \frac{10\gamma - 1}{3} + \epsilon, \quad 0 < \epsilon \ll 1\,. \label{eps_qd}
\end{align}
We obtain $m_0^2 = c^2(20\gamma-1)/3b + \ord(\epsilon)$ and $m_{8}^2 = m_3^2 = 25c^2/9b+ \ord(\epsilon)$. Since we require $m_0$ to be the lightest particle we get $\gamma \subset [2/15, 7/15]$.

As in the case of a light triplet we expect events with an invariant mass of order $1 \ \mathrm{TeV}$ from the decay of $\Sigma_3^{\pm}$, 
which can be pair-produced, into $W^{\pm}Z(\gamma)$. Notice that there is no mixing between $\Sigma_0$ and $\Sigma_3^0$, 
which is easy to check by expanding the potential $V_{\mbf{24}}$ defined in eq.~\eqref{scalar_potential_24} around the minimum. 
The signal of light scalar octets can also observed at LHC pair production of scalar 
octets $pp \to \Sigma_8\Sigma_8$ and their subsequent decays through processes like $\Sigma_8 \to gg$. 

From fig.~\ref{fig:region_beta_gamma_mass_spectrum} we also notice that a region exists where $m_3 \lesssim m_0\lesssim m_8$; we thus expect the direct decay 
$\Sigma_0 \to \Sigma_3^{+} \Sigma_3^{-}/\Sigma_3^{0} \Sigma_3^{0}$. However, from the scalar potential $V_{\mbf{24}}$ in eq.~\eqref{scalar_potential_24}, 
the trilinear interactions between $\Sigma_0$ and $\Sigma_3(\Sigma_8)$ are proportional to $bv_{\mbf{24}}\sim \ord(10^{-24})v_{\mbf{24}}$. 
Thus $g_{\Sigma_0 \Sigma_3\Sigma_3} \simeq g_{\Sigma_0 \Sigma_8\Sigma_8}= \ord(10^{-9}) \ \mathrm{GeV}$ and  these 
contributions are negligible.

These same particles can mediate the light ($125$ GeV) Higgs process $pp \to h \to \gamma\gamma$ observed so far at LHC Run 1. 
The explicit couplings $g_{h \Sigma_3\Sigma_3} $ and $g_{h \Sigma_8\Sigma_8}$ depend on the $V_I$ parameters
and, in principle, can cause an overproduction of final di-gamma pairs.
Assuming a quasi degenerate mass spectrum for the $\Sigma$ field and, for the sake of illustration,  only the $\mbf{5}_H$ Higgs, the interaction potential  
is given by:
 \begin{align} \label{coupl_24_5}
  V_I = \mu_1 \mbfb{5}_H \Sigma \mbf{5}_H + \lambda_1 \mbfb{5}_H \Sigma^2 \mbf{5}_H +  \lambda_2 \mbfb{5}_H \mbf{5}_H \Tr(\Sigma^2)\,.
 \end{align}
Neglecting the small mixing between $\Sigma_0$ and the neutral component in $\mbf{5}_H$ we have the 
following interaction between the physical Higgs boson $h$ and the triplet/octet:
 \begin{align}
  g_{h\Sigma_3\Sigma_3} = \sqrt{2}v (\lambda_1 + 2 \lambda_2)\,, \qquad g_{h\Sigma_8\Sigma_8} = 2\sqrt{2}v  \lambda_2\,,
 \end{align}
where $v = v_{\mbf{5}}[1 + \ord(v_{\mbf{5}}/v_{\mbf{24}})] = 246\ \mathrm{GeV}$. Therefore it is always possible to find a region in 
the parameter space $(\lambda_1, \lambda_2)$ compatible with the current LHC data on $h$ decays. 
Notice that an interaction potential between $\Sigma$  and $\mbf{45}_H$ or $\mbf{70}_H$ certainly contains more $\rm SU(5)$ singlets, 
so the potentially dangerous $g_{h\Sigma_3\Sigma_3}$ and $g_{h\Sigma_8\Sigma_8}$ couplings can be easily made vanishingly small.

\subsection{Fine-tuning}
\label{sec:fine}

In this section we explore in some detail the naturalness issues related to our models. In particular, as we have already
discussed in the previous section, we need to invoke some cancellation in the $b/c$ ratio in order to correctly reproduce both 
$v_{\mbf{24}} \sim m_{\GUT}$ and $m_{S} = 750\ \mathrm{GeV}$. To quantify the amount of required fine-tuning for $S = \Sigma_0, \Sigma_3^0$, we use the following dimensionless 
quantity:
\begin{align}
\label{DeltaFT_definition}
 \Delta^{\rm tree}_{\FT} \equiv -\log_{10} b \frac{\bra m\ket}{c }\,,
\end{align}
where the mean mass $\bra m\ket = \frac{1}{3}\sum_j m_j$.

\begin{figure}[h!]
\centering
 \includegraphics[scale=.5]{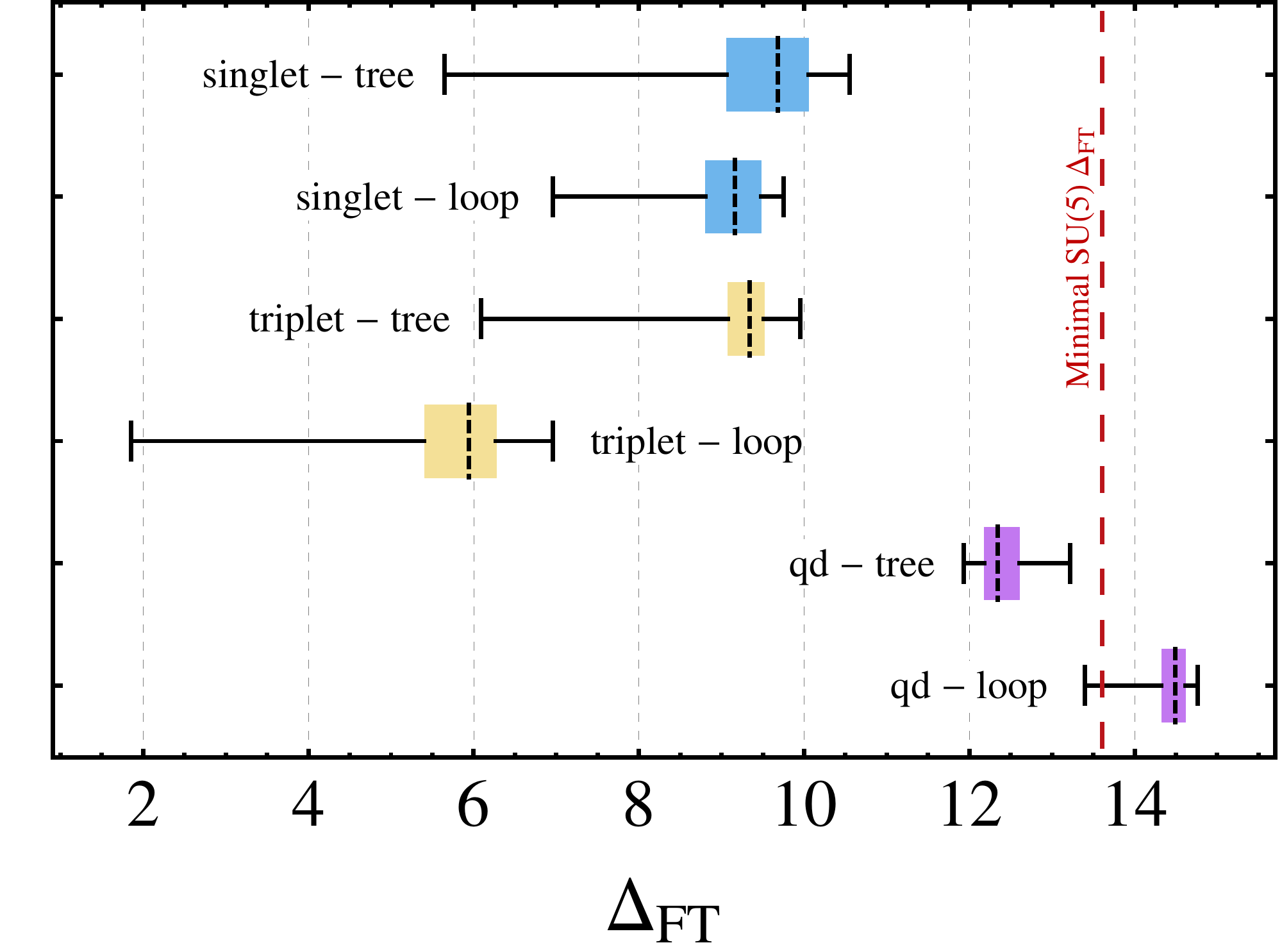}
  \caption{\it Summary of the values of the $\Delta_{\FT}$ parameter at tree level (upper boxcharts) and at one-loop (lower boxcharts) 
  obtained in our numerical scan over the potential parameters that reproduce the different spectra discussed in sec.~\ref{spectra}. The black lines are the allowed region for $\Delta_{\FT}$, the color bands are the $25-75\%$ percentiles and the solid dashed lines are the median. The vertical red dashed line is the typical value of $\Delta_{\FT}$ for the minimal $\rm SU(5)$ model.}
 \label{fig:bar_plot_FT}
\end{figure}

In addition, we also have to take into account naturality limits coming from the fact that loops 
(self-energies of the scalars) involving the $c$ parameter contribute to the scalar masses. 
Thus in the absence of cancellations, we should require:
\[
c \lesssim 4\pi\, m_{\text{lightest}}\,,
\]
that is
\[
c \lesssim 4\pi\, m_{0\,(3)}\lesssim 10\,\text{TeV}\,.
\]
If these latter naturality constraints are not fulfilled, there is some level of fine-tuning to deal with, that we 
can quantify using the following definition:
\begin{align}
\label{DeltaFT_definition2}
 \Delta^{\rm loop}_{\FT} \equiv \log_{10} \frac{ 4 \pi\,m_{\text{lightest}}}{c}\,.
\end{align}
For instance in the case of a light singlet and triplet this reads:
\begin{align}
 \Delta_{\FT}^{\rm loop\,\,singlet} \simeq \frac{1}{2} \log_{10} \sqrt{\frac{2}{15\gamma }}\frac{8  \pi^2 \epsilon^{1/2} }{b}\,, \qquad \Delta_{\FT}^{\rm loop\,\,triplet} \simeq \frac{1}{2} \log_{10} \frac{160 \pi^2 \epsilon}{15 b \gamma -2b}\,,
\end{align}
where the small parameter $\epsilon$ has been introduced in eq.~\eqref{eps} and eq.~\eqref{eps_triplet}. In the case of quasi-degenerate mass spectrum, assuming $m_0 \lesssim m_3 \sim m_8$ and the expansion discussed in eq.~\eqref{eps_qd} we get
\begin{align}
 \Delta_ {\FT}^{\rm loop\, \, qd}\simeq\frac{1}{2}\log_{10}\frac{16\pi^2}{3 b} (20\gamma- 1)\,.
\end{align}
Notice that, given the dependence on $\epsilon$, we expect the following scaling:
\begin{equation}
\Delta_ {\FT}^{\rm loop\, \, qd} > \Delta_ {\FT}^{\rm loop\, \, singlet} > \Delta_ {\FT}^{\rm loop\, \, triplet}\,.
\end{equation}
The numerical results obtained from eqs.~\eqref{DeltaFT_definition} and \eqref{DeltaFT_definition2} are shown in 
fig.~\ref{fig:bar_plot_FT}. In the plot, the range of the $\Delta_{\FT}$ values is built from the values of potential parameters
that realize each one of the mass spectra discussed in sec.~\ref{spectra}, with the additional constraint 
$v_{\GUT} =   10^{16} \ \mathrm{GeV}$.

We first observe that  for $\Delta^{\rm tree}_{\FT}$ we have $\Delta^{\rm tree\,\,qd}_{\FT} \sim 12$, while in the case of a light singlet we a have a wide 
range $\Delta^{\rm tree\,\,singlet}_{\FT} \sim 5.5 \div 10.5$ 
where the bulk of the distribution is around $\Delta^{\rm tree\,\,singlet}_{\FT}\sim 9.5$. This is similar
to the case of a light triplet,  where $\Delta^{\rm tree\,\,triplet}_{\FT} \sim 9$, although
a  smaller region is contemplated, $\sim 6 \div 10$.
Then, we also have 
$\Delta_{\FT}^{\rm loop\,\,triplet} \sim 2 \div 7$ with a  typical value of $6$  while for the singlet 
 we have 
$\Delta_{\FT}^{\rm loop\,\,singlet} \sim 7\div 10$ with a typical value of $9$. We also observe a large fine-tuning for the 
quasi-degenerate case ({\it qd}) where $\Delta_{\FT}^{\rm loop\,\,qd} \sim 14.5$.\\

These values can be compared to the $\Delta_{\FT}$ of the usual minimal $\rm SU(5)$, that we can estimate from the ratio between the 
EW vacuum $v$ and the GUT vacuum, $-\log_{10} v/ v_{\GUT} \gtrsim 13.6$. 
This value is the typical fine-tuning that one finds in order to have doublet-triplet splitting (it is also of the same order of magnitude in the 2HDM that we will discuss in the following). 
We see that the scenarios taken into account in this paper imply a degree of fine-tuning which is typically smaller than the 
usual minimal $\rm SU(5)$ one.

Note that there exist also upper bounds on the trilinear $c$ in order to avoid spontaneous symmetry breaking
of electric/color charges. 
The key point is that, if $c$ is very large, for large values of the field $\Sigma$ for any direction rather than the singlet 
$(\mbf{1},\mbf{1},0)$ there could be a minimum that violates a conserved charge. The constraints depend on the spectrum and the different quartic couplings, and are typically of the form:
\[
\mu \lesssim \mathcal{O}(10)\, m_{\text{heavy}}\,,
\]
where $m_{\text{heavy}}$ is the largest mass involved in the potential. Similar bounds have been obtained for the trilinear couplings in supersymmetric theories, see for instance \cite{Casas:1996de}.

\subsection{Collider phenomenology}

In the case $S = \Sigma_0$ or $\Sigma_3^0$ the decay in two photons/gluons can be mediated through a loop of LQs so, in order 
to evaluate the relevant decay rates, we need the couplings of $S$ with LQs and the couplings of LQs with gauge bosons.
The effective operator describing the former can be obtained from the potential in eq.~\eqref{full_potential_SU5} after 
the $\rm SU(5)$ symmetry breaking and it reads:  
\begin{align}
\label{effective_operators_LQs}
 \op_{eff} = m_S\sum_{\LQ \in \mbf{r}} c^{\mbf{r}}_{\LQ} \, \overline{\LQ}\, \LQ\, S\,,
\end{align}
where $\mbf{r} = \{\mbf{45}_H, \mbf{70}_H\}$. We consider all the LQs involved in the process, and we will discuss the limits on their masses in sec.~\ref{sec:LQpheno}. 
The explicit form of the dimensionless coefficients is obtained after the matching with the relevant terms in $V_I$; 
the existence of non-vanishing such coefficients can be tested using 
for example, a simple cubic and quartic interactions between $\mbf{70}_H$ and $\mbf{24}_H$ in $V_I$ 
(at this level we can safely ignore all possible $\rm SU(5)$ contractions), 
\begin{align}
 V_{\mbf{70}_H, \mbf{24}} = c_1[\mbfb{70}_H \mbf{70}_H \Sigma]_{\mbf{1}} + c_2 [\mbfb{70}_H \mbf{70}_H \Sigma\Sigma]_{\mbf{1}}\,.
\end{align}
Within this convention, the coupling between $S$ and a LQ is given by:
\begin{align}
 g_{S \LQ\LQ} = \frac{v}{2} m_S c^{\mbf{r}}_{\LQ}\,,
\end{align}
where $c^{\mbf{r}}_{\LQ}=c^{\mbf{r}}_{\LQ} (c_1,\,c_2)$.

On the other hand, the coupling of LQs to gluons  (summing over the $\rm SU(2)_L$ states) is
\begin{align}
 g_{gg} = T_R(\mbf{r})(2T +1)\,,
\end{align}
where $T_R(\mbf{3}) = 1/2$, $T_R(\mbf{6}) = 5/2$, $T_R(\mbf{8}) = 3$ and $T_R(\mbf{15}) = 10$
for the different $\rm SU(3)$ representations~\cite{Sierra:2015zma} and $T$ is the weak isospin. 
The coupling between a LQ pair and photons can be estimated as
\begin{align}
 g_{\gamma\gamma} =  d_c\sum_{ T_3 = -T}^{+T}\left(Y + T_3 \right)^2 = d_c (2T +1) \left[Y^2 + \frac{T(T+1)}{3}\right]\,,
\end{align}
where $Y$ is the hypercharge  and $d_c$ is the color multiplicity of the LQ. 
In the same spirit we can estimate the coupling involved in the $Z\gamma$ decay:
\begin{align}
 g_{Z\gamma} = {d_c} \sum_{ T_3 = -T}^{+T} \frac{(Y + T_3 )\left(T_3 - s^2_W (Y + T_3 )\right)}{s_{W} c_W} =  
 d_c(2T +1)\left[-t_WY^2 + \cot_W\frac{T(T+1)}{3}\right]\,,
\end{align}
where we used the short-hand notation $(s_W, c_W,t_W, \cot_W)= (\sin \theta_W,\cos \theta_W,\tan \theta_W,\cot \theta_W)$,
where
$\theta_W$ is the weak mixing angle. 
The presence of the Weinberg angle makes this coupling quite large (compared to $g_{\gamma\gamma}$) 
 for higher dimensional representation of $\rm SU(2)_L$. 
Finally, the couplings to two vector bosons ($W/Z$) mediated by a LQ loop are given by
\begin{align}
g_{WW} &= {d_c} \sum_{ T_3 = -T}^{+T} \frac{T_3^2}{s^2_W} =\frac{d_c(2T +1)}{s_W^2}  \frac{T(T+1)}{3}\,, \nline
 g_{ZZ} &= {d_c} \sum_{ T_3 = -T}^{+T} \frac{\left(T_3 - s^2_W (Y + T_3 )\right)^2}{s_W^2 c_W^2} = 
 {d_c(2T +1)}\left[t_W^2 Y^2 +  \cot^2_W\frac{T(T+1)}{3}\right]\,.
\end{align}
Hence, barring accidental cancellations that can appear in the $g_{Z\gamma}$ coupling, 
we expect the following hierarchy in the limit of large hypercharge:
\begin{align}
 g_{WW} \lesssim g_{ZZ}\lesssim g_{Z\gamma} \lesssim g_{\gamma\gamma}\Longleftrightarrow \Gamma(S \to W^+W^- ) 
 \lesssim \Gamma(S \to ZZ ) \lesssim \Gamma(S \to Z\gamma ) \lesssim  \Gamma(S \to \gamma\gamma ) \,,
\end{align}
that is:
\begin{align}
 \Gamma(S \to \gamma \gamma ) : \Gamma(S \to Z\gamma  ) : \Gamma(S \to ZZ ) = 1 : 2 \,t_W^2 : t_W^4 \simeq 1 : 0.6 : 0.09 \,,
\end{align}
where the factor two in the second estimate is a consequence of having identical particles in the final state for $\gamma\gamma$ and $ZZ$ 
decays.\\
Notice that the above couplings are not independent due to the presence of the following sum rules:
\begin{align}
\label{couplings_sum_rules}
 g_{ZZ} + t_W g_{Z\gamma} = g_{WW}\,, \qquad g_{ZZ}- \cot_W g_{Z\gamma} = \frac{g_{\gamma\gamma}}{c_W^2}-t_W^2 g_{WW}\,.
\end{align}
Thus if $g_{ZZ} = g_{WW} = 0$ we get $g_{\gamma \gamma } = g_{Z \gamma} = 0$. To give some numerical estimates, consider the case 
where $g_{WW} = 0$, 
which is the most favorable scenario: we then obtain $g_{ZZ}  =  t_W^2 g_{\gamma \gamma}\simeq 0.30g_{\gamma \gamma}$ and 
$g_{Z\gamma}  =  -t_W g_{\gamma \gamma}\simeq -0.55g_{\gamma \gamma}$, for $s^2_W = 0.23$;
on the other hand, assuming $g_{ZZ} = 0$, we get  
$g_{WW}  =  -g_{\gamma \gamma}s_W^2/c_{2W} \simeq -0.43g_{\gamma \gamma}$ and 
$g_{Z\gamma}  = -g_{\gamma \gamma} t_{2W}/2 \simeq -0.78 g_{\gamma \gamma}$.\\

We now specialize the previous considerations to our models.
In the case of the $\mbf{45}_H \in SU(5)$, the only leptoquark to be taken into account is $(\bar{\mbf{3}}, \mbf{2},7/6)$, 
for which we get the same results as those of ref.~\cite{Dorsner:2016ypw} and reported in the first line of 
tab.~\ref{tab:coupling_representation_45_70}. 
On the other hand, for the $\mbf{70}_H$ we have a richer phenomenology since many representations 
can have non-vanishing couplings to the gauge bosons; they are also summarized 
in tab.~\ref{tab:coupling_representation_45_70}. According to our naive estimates, we expect that the 
representations $(\mbf{\bar 3}, \mbf{3},-4/3)$ and $(\mbf{6}, \mbf{2}, 7/6)$ are the best candidates to enhance the 
singlet decays to $\gamma\gamma$.\\

\begin{table}[h!]
\small
\begin{center}
\resizebox{\columnwidth}{!}{%
\begin{tabular}{c  c c c c c }
\toprule
\toprule
$\LQ$  			  & $g_{gg}$	 & $g_{\gamma\gamma}$ 	& $g_{Z\gamma}$          	   	& $g_{WW}$ 				& $g_{ZZ}$\\
\midrule
$(\mbfb{3}, \mbf{2},7/6)_{\mbf{45}_H}$  & $1$	 	& $29/3 \simeq 9.7$ 	& $(9- 58s^2_W)/6s_Wc_W \simeq -1.7$  	& $3s^{-2}_W/2 \simeq 6.5$		& $(9 - 18 s_W^2 + 58 s_W^4)/6 s^2_Wc^2_W \simeq 7.5$	\\ 
\midrule
$(\mbf{3}, \mbf{1},1/3)_{\mbf{70}_H}$  & $1/2$	 & $1/3$		& $- s_W/3c_W \simeq -0.18$         	& $0$					& $t^2_W/3 \simeq 0.1$	\\ 
$(\mbf{3}, \mbf{3}, 1/3)_{\mbf{70}_H}$ & $3/2$	 & $7$			& $(6 -7s^2_W)/s_Wc_W \simeq 10$   	& $6 s^{-2}_W \simeq 26$		& $(6 -12 s_W^2 +7s_W^4)/s^2_W c^2_W \simeq 20$\\
$(\mbfb{3}, \mbf{3}, -4/3)_{\mbf{70}_H}$& $3/2$	 & $22$			& $(6 - 22s^2_W)/s_Wc_W \simeq 2.2$ 	& $6 s^{-2}_W \simeq 26$		& $2(3 -6 s_W^2 +11s_W^4)/s^2_W c^2_W \simeq 25$\\
$(\mbf{6}, \mbf{2}, 7/6)_{\mbf{70}_H}$ & $5$		 & $58/3 \simeq 19$     & $(9 - 58s^2_W)/3s_Wc_W \simeq -3.4$   & $3s^{-2}_W \simeq 13$ 		& $(9 - 18 s_W^2 + 58s^4_W)/3s^2_W c^2_W \simeq 15$\\  
\bottomrule
\bottomrule
\end{tabular}
}
\caption{\it Effective couplings for LQs candidates in the representations $\mbf{45}_H$ (first line) and $\mbf{70}_H \in SU(5)$ (lower lines). In the numerical evaluation we used $s_W^2 = 0.23$.}
\label{tab:coupling_representation_45_70}
\end{center}
\end{table}

Let us now consider the decay processes of our resonance candidates.\\
For loop mediated processes we use the same conventions  for the loop functions as those of ref.~\cite{Carena:2012xa}; 
in particular, we report in \ref{sec:LoopFunctionsAppendix} the decay width for $S\to \gamma \gamma$, $S\to Z\gamma$ 
and $S\to gg$ and the relevant loop functions.

Notice that, due to the sum rules presented in eq.~\eqref{couplings_sum_rules}, it is quite
difficult to have both $\Gamma(S \to W^+W^-)$ and $\Gamma(S \to ZZ)$ small. Using the data reported in 
ref.~\cite{Strumia:2016wys}, we have checked  that even in the case of large $\rm SU(2)_L$ quantum numbers the current bounds on dibosons in the 
final state are fulfilled.\\
The relevant decay rates in our model can be estimated from tab.~I in ref.~\cite{Dorsner:2016ypw}:
\begin{align}
\frac{\Gamma(S \to Z\gamma)}{\Gamma(S \to \gamma \gamma)} &\simeq 4.3\ \left(\frac{g_{Z\gamma}^{\LQ}}{g_{Z\gamma}^{(\mbf{3},\mbf{3},1/3)}}\right)^2 = 0.2\, 
 [0.5]\,\qquad \text{for}\qquad   (\mbfb{3}, \mbf{3}, -4/3)\,\,\, [(\mbf{6}, \mbf{2}, 7/6)]\,,\\
 \frac{\Gamma(S \to Z Z)}{\Gamma(S \to \gamma \gamma)} &\simeq 7.8\ \left(\frac{g_{ZZ}^{\LQ}}{g_{ZZ}^{(\mbf{3},\mbf{3},1/3)}}\right)^2= 12\, 
 [4.4]\,\qquad \text{for}\qquad   (\mbfb{3}, \mbf{3}, -4/3)\,\,\, [(\mbf{6}, \mbf{2}, 7/6)]\,,\\
  \frac{\Gamma(S \to W^+ W^-)}{\Gamma(S \to \gamma \gamma)} &\simeq 26\ \left(\frac{g_{WW}^{\LQ}}{g_{WW}^{(\mbf{3},\mbf{3},1/3)}}\right)^2= 
  26\, [6.5]\,\qquad \text{for}\qquad   (\mbfb{3}, \mbf{3}, -4/3)\,\,\, [(\mbf{6}, \mbf{2}, 7/6)]\,,
  \\
  \frac{\Gamma(S \to g g )}{\Gamma(S \to \gamma \gamma)} &\simeq 54\ \left(\frac{g_{gg}^{\LQ}}{g_{gg}^{(\mbf{3},\mbf{3},1/3)}}\right)^2= 
  54\, [540]\,\qquad \text{for}\qquad   (\mbfb{3}, \mbf{3}, -4/3)\,\,\, [(\mbf{6}, \mbf{2}, 7/6)]\,.
 \end{align}
We clearly see that the decay to gluon final states is the most relevant one. However, the use of large $\rm SU(3)$ representations is not really an issue in this context (they can 
lead to an overproduction of the SM Higgs through gluon-gluon fusion and a too fast Higgs decay rates)
because the scalar potential involving $\mbf{5}_H$ and $\mbf{70}_H$ fields contain
three $\rm SU(5)$ singlets whose couplings can be rearranged as to satisfy the Run-I results. Anyway, in all our numerical analysis below we include all partial widths discussed so far.

If the decays of $S$ into LQ are kinematically closed, the diphoton branching ratio can be estimated under the assumption that $\Gamma(S \to g g ) $ dominates the 
total width and  that only one LQ dominates in the loop, thus obtaining:
\begin{align}
 \Br(S \to \gamma\gamma)\simeq \frac{\Gamma(S\to \gamma\gamma)}{\Gamma(S \to  gg)} \simeq  \frac{32}{9}\left( \frac{\alpha_{em}}{\alpha_S}\right)^2
 \left|\frac{g_{\gamma\gamma}^{\LQ}}{g_{gg}^{\LQ}}\right|^2\approx\ord(10^{-3}) \left|
 \frac{g_{\gamma\gamma}^{\LQ}}{g_{gg}^{\LQ}}\right|^2\,.
\end{align}
Hence the decay in two photons is more suppressed as the LQ $\rm SU(3)$ representation gets larger.\\

On the other hand, if kinematically open, the decay into a LQ pair is a tree-level process and can be easily computed from the effective operator defined in 
eq.~\eqref{effective_operators_LQs}. Notice that the lower bounds on LQ masses from LHC data are typically $\ord(1)\ \mathrm{TeV}$, but these are model dependent. We will discuss in sec.~\ref{sec:LQpheno} how to relax such bounds. For a LQ in representation $\mbf{r} \in SU(5)$ we expect:
\begin{equation}
\Gamma(S \to  \overline{\LQ} \LQ ) \sim \sum_{\LQ}d_c^{\LQ} \,|c_{\LQ}^{\mbf{r}}|^2\times\ord(10)\ \mathrm{GeV}\,,
\end{equation}
where $d_c^{\LQ}$ explicitly takes into account the color factor of the final LQ and $10$ GeV is a good estimate for the 
phase-space contribution (see eq.~\eqref{S_LQLQ_Width} in \ref{sec:LoopFunctionsAppendix}). For  
$d_c^{\LQ} \,|c_{\LQ}^{\mbf{r}}|^2\sim{\cal O}(1)$ this estimate is 
consistent with the ATLAS observation on a large $\Gamma$. A naive estimate assuming only one type of LQ and neglecting the gluon contribution in the total width is given by
\begin{align}
\label{BR_S_LQLQlimit}
 \Br(S\to \gamma\gamma) \simeq \frac{\Gamma(S\to \gamma\gamma)}{\Gamma(S \to  \overline{\LQ} \LQ )} = 
 \left(\frac{\alpha_{em}^2}{8\pi^2}\right)\frac{ m_S^4}{ M_{\LQ}^4} \times \ord(1)\,,
\end{align}
where the $\ord(1)$ is the loop factor that also takes into account the color factors. \\

Notice that the LQs couplings to $h$ come from the quartic scalar potential
\begin{equation}
V_{\mbf{70}_H, \mbf{5}_H} = [\mbf{5}_H \mbfb{5}_H \mbf{70}_H \mbfb{70}_H]_{\mbf{1}}\,,
\end{equation}
which contains six invariants. A similar relation also occurs if the LQ belongs to the $\mbf{45}_H$. 
After GUT and EW symmetry breaking we obtain an effective operator of the form: 
\begin{align}
\op_{eff}^h = m_h \sum_{\LQ \in \mbf{r}} c^{\mbf{r}}_{h\overline{\LQ}\LQ}  \overline{\LQ}\, \LQ\, h\,,   
\end{align}
 that in general will modify the Higgs properties. However, the effective coupling $c^{\mbf{r}}_{h\overline{\LQ}\LQ}$ contains different $\rm SU(5)$ 
 Lagrangian parameters with respect to those involved in the coupling of the heavy resonance, $c^{\mbf{r}}_{\LQ}$, see eq.~\eqref{effective_operators_LQs}. Thus, it is possible to have a hierarchy
  $|c^{\mbf{r}}_{h\overline{\LQ}\LQ}| \ll |c^{\mbf{r}}_{\LQ}| \sim \ord(1)$. A detailed analysis involving the coupling of the Higgs with the scalar LQs was performed in ref.~\cite{Dorsner:2012pp}.

 \subsubsection{Singlet decay }
 \label{sec:singlet_decay}

We can now use the estimates on the branching ratios given in the previous section 
to put some bounds on the relevant parameters of our models, that is 
the mass of the leptoquarks $M_{\LQ}$ and the effective couplings of $S$ to LQs, $c^{\mbf{r}}_{\LQ}$, using 
the combined  CMS and ATLAS data at $\sqrt{s} =8\ \mathrm{TeV}$ reported in ref.~\cite{Strumia:2016wys}.

\begin{figure}[h!]
\centering
 \includegraphics[scale=.33]{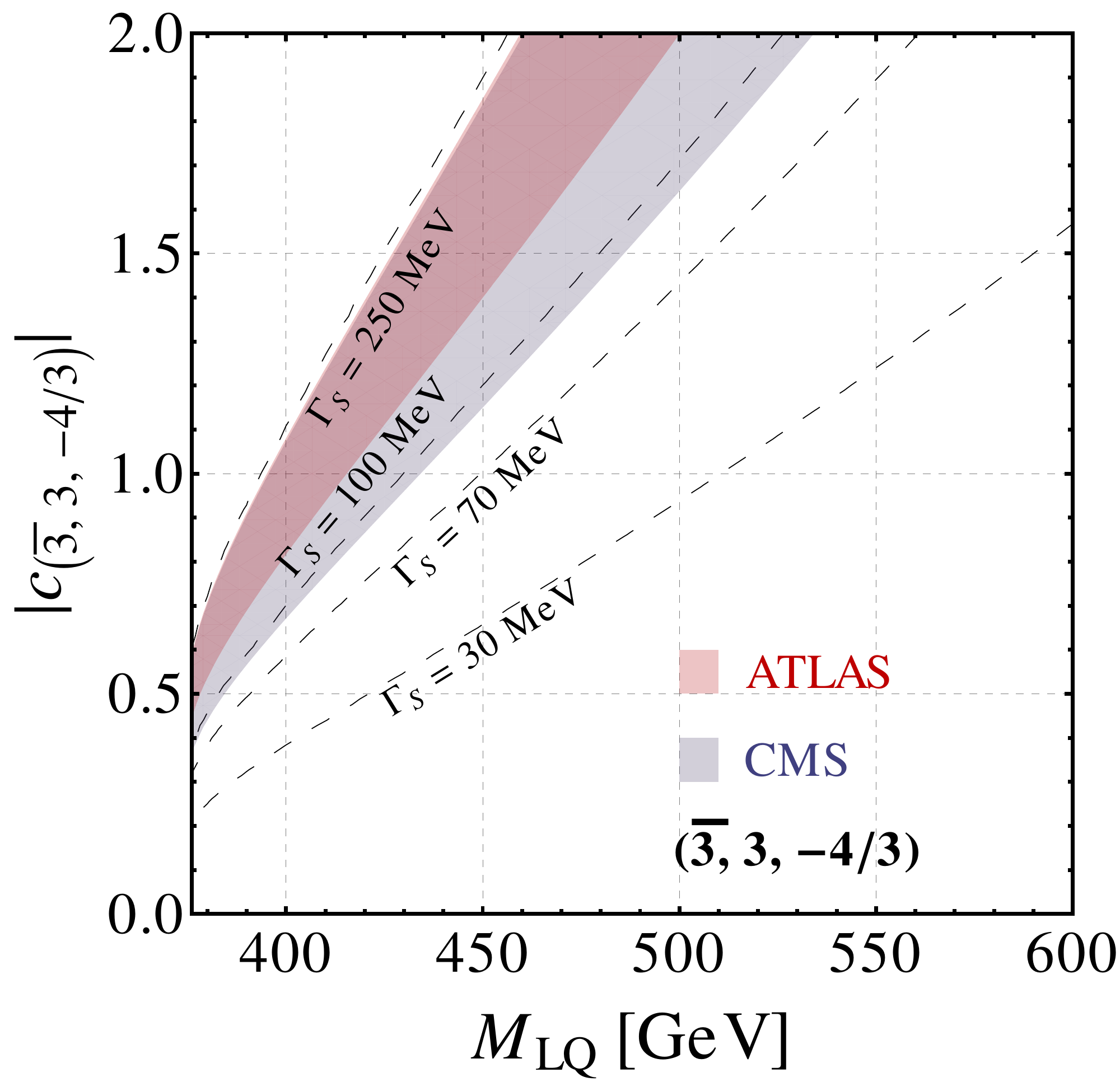}
 \includegraphics[scale=.33]{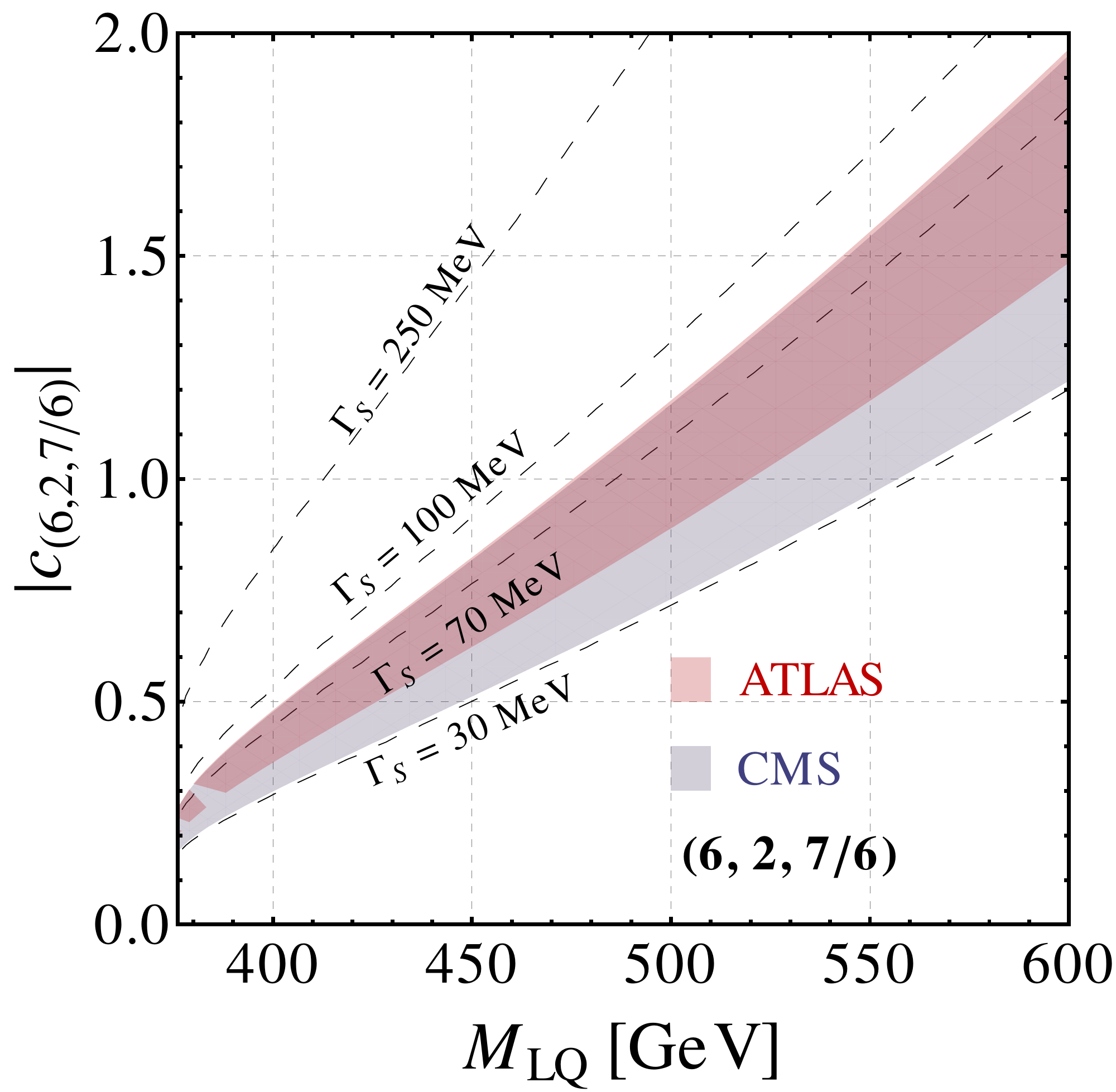}
  \caption{\it Region plot of $\sigma(pp\to S \to \gamma\gamma)$ in the plane $(M_{\LQ}, c_{\LQ})$ assuming 
  that only one heavy LQ mediates the $S$ decay: $(\mbfb{3}, \mbf{3}, -4/3)$ on the left panel and $(\mbf{6}, \mbf{2}, 7/6)$ on the right. 
  In the blue (red) region the values of $M_{\LQ}$ and $c_{\LQ}$ is in accordance with the excess observed at CMS (ATLAS). 
  The black dashed lines are the contour lines of constant $\Gamma_S$.}
 \label{fig:region_c70_MLQ}
\end{figure}

In order to do that, we use eq.~\eqref{cross_section_theo}, reported 
in \ref{sec:LHC_production}, where we also take into account 
that, for a large coupling with the photon and a large width, also 
the photo-production mechanism can be important, see  ref.~\cite{Csaki:2015vek}.
At LHC with $\sqrt{s} = 8/13\ \mathrm{TeV}$ the gluon PDF constitutes  the main contribution, while only a few percent of the 
event is related to vector-boson fusion, thus we can ignore this channel. These very same estimates can also be used to check the compatibility with the diphoton excess in the 
$(M_{\LQ},c^{\mbf{r}}_{\LQ})$ plane, fig.~\ref{fig:region_c70_MLQ}.

The first case we analyze is $S = \Sigma_0$, with large $m_{3, 8}$ under the simplifying assumptions that 
only one LQ is responsible for the decays (if we consider more than one LQ with large mass we expect similar results 
since the new contributions, for a given $g_{S \LQ \LQ}$, will always interfere positively, except in the $Z\gamma$ channel, see tab.~\ref{tab:coupling_representation_45_70}). 
Since the results for the LQ in the $\mbf{45}_H$ have been already discussed in 
ref.~\cite{Dorsner:2016ypw}, we focus on the LQs contained in the $\mbf{70}_H$, in particular on the $(\mbfb{3}, \mbf{3}, -4/3)$
 and $(\mbf{6}, \mbf{2}, 7/6)$ representations\footnote{Notice that none of these LQs mediate proton decay at tree level. 
 However, some of them can mediate it a loop level, 
like the $(\mbf{\bar 3},\mbf{3}, -4/3)$, thus their mass can not be too light or the relevant couplings have to be suppressed \cite{Dorsner:2012nq}.}. Our results are presented in fig.~\ref{fig:region_c70_MLQ}, 
where we assume that $M_{\LQ} \geq m_S/2$. 
We  consider the strong coupling constant $\alpha_S$ evaluated at the scale $\mu_R = m_S$ including the effect of LQs in the running 
whereas the electromagnetic coupling in $S \to \gamma \gamma$ is evaluated at $\mu_R = 0$ since the photons are real \cite{Djouadi:2005gi}.

We clearly see that for both states a narrow width $\Gamma_S = \ord(10^2) \ \mathrm{MeV}$ is compatible with the diphoton excess
 in the region of relatively low LQ masses and ${\cal O}(1)$ couplings, blue (CMS) and red (ATLAS) regions.

In the case of $M_{\LQ} \lesssim m_S/2$ the situation is quite different because also the contribution from photoproduction becomes important.  If we assume that the $\sigma_{\gamma\gamma}$ function given in eq.~\eqref{cross_section_photon} is the main contribution 
 and we ignore the color factor $d_c$, the experimental data can be fit only for 
 $M_{\LQ} \simeq 60\ \mathrm{GeV}$, see eq.~\eqref{BR_S_LQLQlimit}, a value excluded from direct searches; we interpret this result as that 
 a large part of the signal should be a consequence of the gluon fusion mechanism. 

\begin{figure}[h!]
\centering
 \includegraphics[scale=.33]{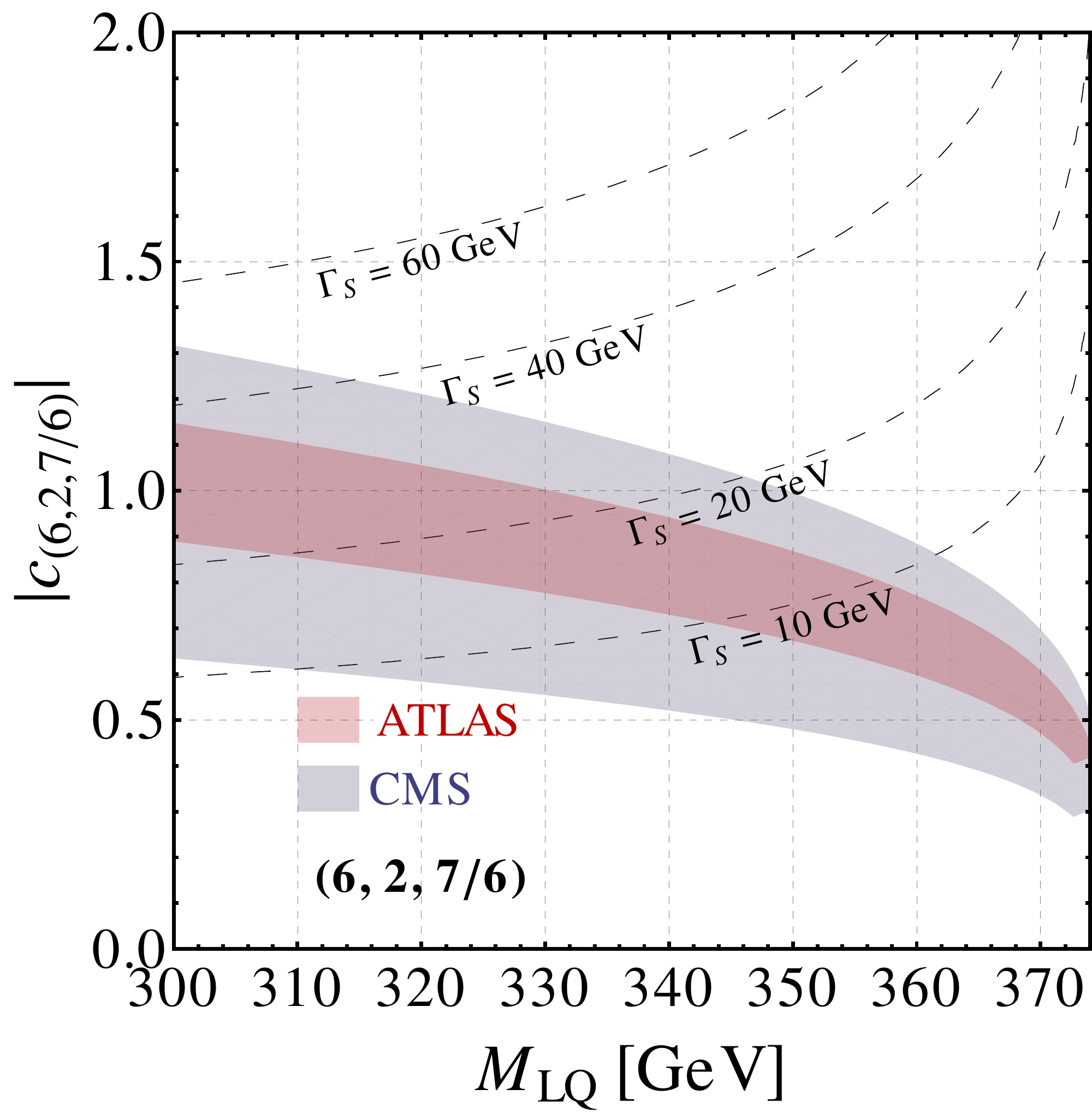}~~ \includegraphics[scale=.33]{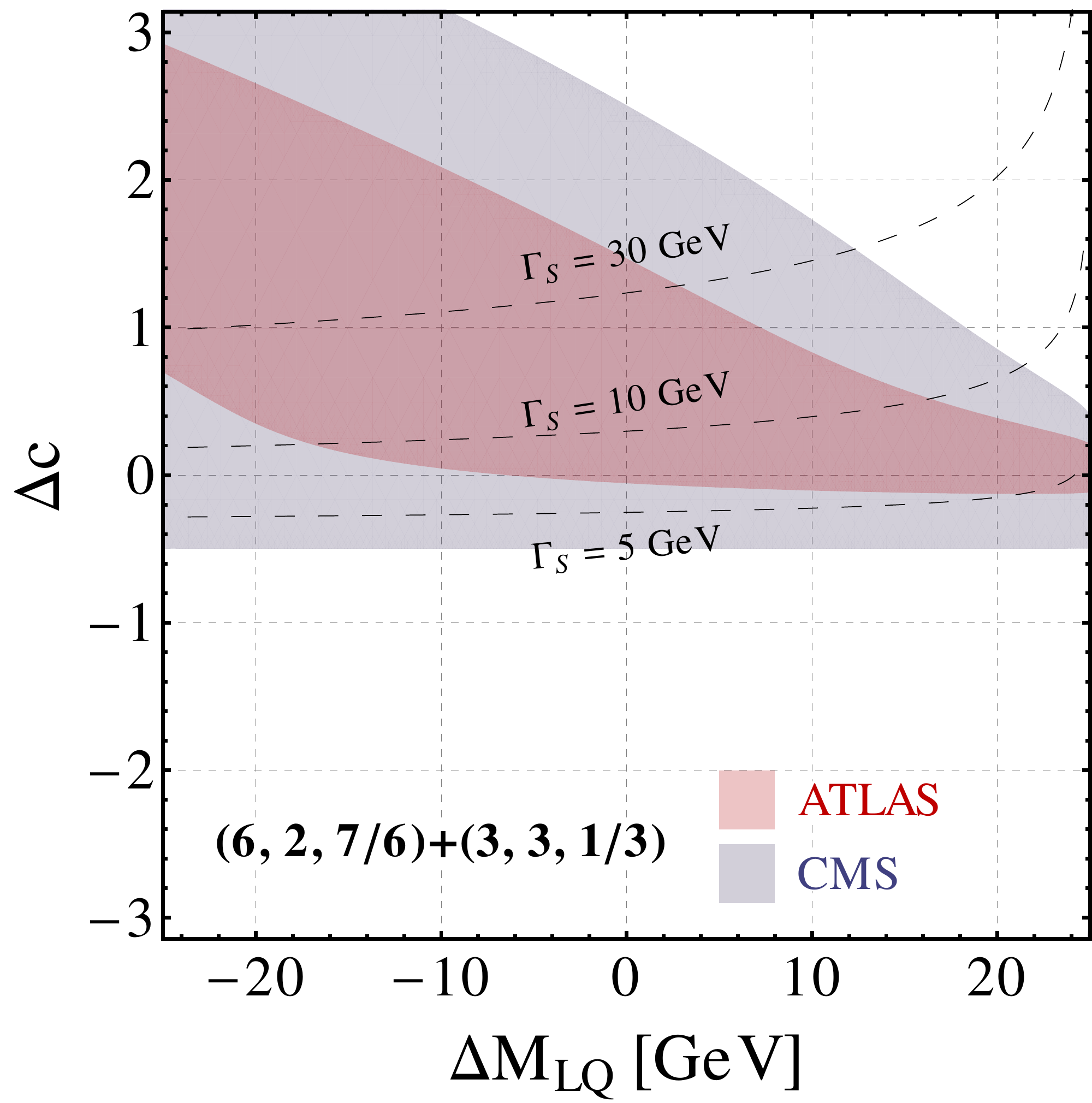}
  \caption{\it Left) Region plot of $\sigma(pp\to S \to \gamma\gamma)$ in the plane $(M_{\LQ}, c_{\LQ})$ assuming one 
  light LQ $(\mbf{6}, \mbf{2}, 7/6)$ and $M_{\LQ} \lesssim m_S/2$. The blue (red) region is in accordance with the excess observed at CMS (ATLAS). 
  The black dashed lines are the contour lines of constant $\Gamma_S$.
  Right) Same, but in the plane $(\Delta M_{\LQ}, \Delta c)$ assuming two light LQs: $(\mbf{6}, \mbf{2}, 7/6)$ and $(\mbf{3}, \mbf{3}, 1/3)$.}
 \label{fig:region_c70_MLQ_LowMass}
\end{figure}

For one light LQ we report the excluded and allowed regions in the left panel of fig.~\ref{fig:region_c70_MLQ_LowMass}. \\

We can now open the possibility  that two LQs $(\mbf{3}, \mbf{3}, 1/3)$ and $(\mbf{6}, \mbf{2}, 7/6)$ are simultaneously light. 
In our numerical analysis we assume that 
the lightest LQ is $(\mbf{6}, \mbf{2}, 7/6)$ with a  mass $M_{\LQ} = 350\ \mathrm{GeV}$.
We indicate with $\Delta M_{\LQ} \equiv M_{\LQ} - M_{\LQ'}$ the mass difference and 
we focus on $|\Delta M_{\LQ}| \leq 25 \ \mathrm{GeV}$ to assure that both LQ masses satisfy 
$M_{\LQ} \lesssim m_S/2$. We also fix the coupling of the lightest LQ to be $c_{\LQ} = 1/2$ 
while the other coupling is $ c_{\LQ} + \Delta c$, where no assumptions on $\Delta c$ are done; 
in particular, with the help of the \texttt{Susyno} package \cite{Fonseca:2011sy}, we have checked that the couplings 
for these two different representations have different Clebsh-Gordan coefficients at least in one $\rm SU(5)$ invariant built with the 
$\mbf{70}_H$, so a priori $\Delta c$ could be different from zero. 
Notice that the decays 
to $ZZ$ and $W^+W^-$ can be safely neglected because the width to a pair of LQs ($\sim {\cal O}(10)$ GeV) is much larger than that to 
$Z$ and $W$'s (around 100 MeV). In the right panel of fig.~\ref{fig:region_c70_MLQ_LowMass} we show the allowed regions in the plane 
$(\Delta M_{\LQ}, \Delta c)$, with $\sigma(pp \to S \to \gamma\gamma)$ compatible with the excess observed at LHC.

 \subsubsection{Triplet decay}
In this section we assume that the neutral component of the triplet $\Sigma_3$ is the resonance behind the diphoton excess. 
The bounds on $M_{\LQ}$ and $c_{\LQ}$ are the same of the previous case because the coupling with the LQs can be expressed using 
an effective operator of the same structure as in eq.~\eqref{effective_operators_LQs}, although with different coefficients. 
We should notice that it is not possible to generate the operator $\bra \Sigma_3^{0} \ket \Sigma_3^{0} \overline{\LQ} \LQ$ because 
the neutral component of the triplet cannot acquire a VEV if we want to break $SU(5) \to \Gr^{\SM}$.\\
A new interesting observable in this case is the decay of the charged components $\Sigma_3^{\pm}$ in $W^{\pm} \gamma$ or $W^{\pm} Z$, governed by the effective couplings $g_{\Sigma_3W\gamma}$ and $g_{\Sigma_3WZ}$, respectively. 

Following ref.~\cite{Asakawa:2006gm} they can be estimated as:
\begin{align}
 g_{\Sigma_3W\gamma\, (\Sigma_3WZ)} \sim e\frac{g_{W\gamma\,(WZ)}}{(4\pi)^2M_{\LQ}} \log\frac{M_{\LQ}}{m_{\Sigma_3^{\pm}} }\,,
 \end{align}
where the logarithmic dependence comes from the loop function. 
The LQs couplings for $W\gamma$ and $WZ$ are instead:
 \begin{align}
 g_{W\gamma} &= {d_c}\sum_{ T_3 = -T}^{+T}\frac{T_3\left(Y + T_3 \right)}{s_W} = \frac{d_c (2T +1)}{s_W} \frac{T(T+1)}{3} = s_W g_{WW}\,,\nline
  g_{WZ} &= {d_c} \sum_{ T_3 = -T}^{+T}\frac{T_3\left(T_3 - s_W^2 (Y + T_3 )\right)}{s_W^2 c_W} = \cot_W\frac{d_c (2T +1)}{s_W} \frac{T(T+1)}{3} = \cot_W g_{W\gamma} \,.
 \end{align}
These couplings are summarized in tab.~\ref{tab:coupling_representation_45_70_triplet} for the LQs in representation $\mbf{70}_H$ 
(since there is no dependence of the couplings of the triplets on the hypercharge we left it indicated with a generic index $n$).

\begin{table}[h!]
\begin{center}
\begin{tabular}{c  c c }
\toprule
\toprule
$\LQ$  			   & $g_{W\gamma}$ 		& $g_{WZ}$          	 \\
\midrule
$(\mbf{3}, \mbf{3}, n)$		& $6/s_W \simeq12.5$		& $6/t_Ws_W \simeq 23$ 	\\
$(\mbf{6}, \mbf{2}, 7/6)$	& $3/s_W \simeq 6.3$     	& $3/t_Ws_W\simeq 11.4$   \\  
\bottomrule
\bottomrule
\end{tabular}
\caption{\it Effective couplings for LQs candidates in representation $\mbf{70} \in SU(5)$.} 
\label{tab:coupling_representation_45_70_triplet}
\end{center}
\end{table}
A rough estimate of the decay widths gives
$\Gamma(\Sigma_3^{\pm} \to W^{\pm} Z) \sim {\cal O}(1)\, \mathrm{GeV}$ for the LQ in representation $(\mbfb{3}, \mbf{3}, -4/3)$ and 
$\Gamma(\Sigma_3^{\pm} \to W^{\pm} Z) \sim {\cal O}(0.1 \div 1)\, \mathrm{GeV}$ for $(\mbf{6}, \mbf{2}, 7/6)$ 
for $M_{\LQ} \gtrsim m_S/2$ and $\left(m_{3^{\pm}} - m_3\right) \in [0, 100]\ \mathrm{GeV}$.

\subsubsection{Degenerate mass spectrum}
\label{sec:qd_pheno}
In this case we assume that the singlet has a mass of $750\ \mathrm{GeV}$. Since no mixing between $\Sigma_3^0$ and $\Sigma_0$ 
is allowed, the singlet phenomenology is the same as the one described in sec.~\ref{sec:singlet_decay}. 
In principle, in this scenario the decay of a $\Sigma_8$ in a pair of gluons could be detected at LHC. 
However, the triple coupling of the scalar octet is $g_{\Sigma_8 \Sigma_8 \Sigma_8} \sim \ord(10^{-9})\ \mathrm{GeV}$ in this region of the parameter space (see sec.~\ref{sect223}), and in addition there is also a suppression due to the kinematic factor $\pi^2/9 -1$~\cite{Bai:2010dj}.
Thus the main contribution to $\Sigma_8$ decays is through a loop with LQs. The effective operator at the scale $m_8 \gtrsim m_0$ 
can be extracted from the interaction potential
\begin{align}
\label{effective_operators_LQs_Sigma8}
 \op_{eff} = m_8\sum_{\LQ \in \mbf{r}}   \overline{c}^{\mbf{r}}_{\LQ}\overline{\LQ} \LQ \Sigma_8\,,
\end{align}
where notice that we use $\overline{c}^{\mbf{r}}_{\LQ}$ for the octets, keeping $c^{\mbf{r}}_{\LQ}$ for the singlet/triplet. 

For every color representation $\mbf{r}$ several $\rm SU(3)$ contractions can be worked out. For example:
\begin{gather}
 \LQ \sim \mbf{3} \in SU(3) : \lambda^a_{ij} \overline{\LQ}^i \LQ^j \Sigma_8^a\,,  \nline
 \LQ \sim \mbf{6} \in SU(3) : \lambda^a_{li} \overline{\LQ}^{lj} \LQ^{ij} \Sigma_8^a + \lambda^a_{lj} \overline{\LQ}^{il} \LQ^{ij} \Sigma_8^a \,,
\end{gather}
where $i, j, l = 1,2, 3$ and the $\lambda^a$ matrices are the usual Gell-Mann matrices, with $a = 1, \dots 8$.\\ 

The effective coupling between LQs and the scalar octet in our notation, neglecting the $\rm SU(3)$ contraction, is
\begin{align}
 g_{\Sigma_8 \LQ \LQ} = \frac{v}{2} m_8  \overline{c}^{\mbf{r}}_{\LQ} \,.
\end{align}
The decay widths into a gluon pair or $Z(\gamma)g$ are summarized in \ref{sec:LoopFunctionsAppendix}, eqs.~\eqref{eq:S8ggamma} and \eqref{eq:S8gZ}.
The effective couplings for the decay into $\gamma/Z$ and a gluon mediated by LQs are
\begin{align}
 g_{\gamma g} &= d_c T_R(\mbf{r})\sum_{ T_3 = -T}^{+T} (T_3 +Y) = d_c T_R(\mbf{r}) (2T +1)Y\,,  \nline
 \label{coupling_S8_Zg}
 g_{Zg} &= {d_c} T_R(\mbf{r})\sum_{ T_3 = -T}^{+T} \frac{T_3 - s_W^2 (Y + T_3 )}{s_W c_W} = -t_Wd_c T_R(\mbf{r}) (2T +1)Y = -t_W g_{\gamma g}\,,
 \end{align}
where we have also performed the sum over the colour states (see also tab.~\ref{tab:coupling_representation_45_70_degenerate_mass_spectrum} 
for a numerical estimate of such couplings).

\begin{table}[h!]
\begin{center}
\begin{tabular}{c  c c }
\toprule
\toprule
$\LQ$  			  	 & $g_{\gamma g}$ 	& $g_{Zg}$          	 \\
\midrule
$(\mbfb{3}, \mbf{2},7/6)$  	 & $7/2$ 		& $-t_W7/2 \simeq -1.91$  	\\ 
\midrule
$(\mbf{3}, \mbf{1},1/3)$  	 & $1/2$		& $- t_W/2 \simeq -0.27$         	\\ 
$(\mbf{3}, \mbf{3}, 1/3)$ 	 & $3/2$		& $- t_W3/2 \simeq -0.82$   	\\
$(\mbfb{3}, \mbf{3}, -4/3)$	 & $-6$			& $t_W 6 \simeq 3.28$ 	\\
$(\mbf{6}, \mbf{2}, 7/6)$ 	 & $35$     		& $-t_W35\simeq -19$   \\  
\bottomrule
\bottomrule
\end{tabular}
\caption{\it Effective couplings for LQs candidates in the representations $\mbf{45}_H$ (upper) and $\mbf{70}_H \in SU(5)$ (lower).}
\label{tab:coupling_representation_45_70_degenerate_mass_spectrum}
\end{center}
\end{table}
The factor $T_R(\mbf{r})$ is given by the contraction of two $\rm SU(3)$ adjoint external lines. 
We expect that the decay into $Z\,g$ is always suppressed with respect to the $\gamma\,g$ one. 
Notice also that the coupling to $\gamma$ and $g$ is a factor $35$ for the LQ $(\mbf{6}, \mbf{2}, 7/6)$ due to the large 
$T_R(\mbf{6})$ coefficient.  

In the following we give a naive comparison between some of the most interesting singlet to octet decay rate ratios:
\begin{align}
\label{ratio_S_gammagamma_S8_gammag}
\frac{\Gamma(\Sigma_0 \to gg)}{\Gamma(\Sigma_8 \to gg)} &=\frac{8}{\rho_{\LQ}}\left(\frac{m_0}{m_8}\right)^5 \left| 
 \frac{\sum_{\LQ}c^{\mbf{r}}_{\LQ} }{\sum_{\LQ}\overline{c}^{\mbf{r}}_{\LQ}} \right|^2 \times \ord(1)\,,
 \nline
  \frac{\Gamma(\Sigma_0 \to \gamma\gamma)}{\Gamma(\Sigma_8 \to \gamma g)} 
  &\simeq t_W^4\frac{\Gamma(\Sigma_0 \to Z\gamma)}{\Gamma(\Sigma_8 \to Zg)} = 
  8\frac{\alpha_{em}}{\alpha_S}\left(\frac{m_0}{m_8}\right)^5 
  \left| \frac{\sum_{\LQ}g_{\gamma \gamma}^{\LQ}c^{\mbf{r}}_{\LQ} }{\sum_{\LQ}g_{\gamma g}^{\LQ}\overline{c}^{\mbf{r}}_{\LQ} } \right|^2 
  \times \ord(1)\,,  \nline
   \frac{\Gamma(\Sigma_0 \to \gamma\gamma)}{\Gamma(\Sigma_8 \to gg)} &=\frac{1}{\rho_{\LQ}}\left(\frac{\alpha_{em}}{\alpha_S}\right)^2
   \left(\frac{m_0}{m_8}\right)^5 \left|  \frac{\sum_{\LQ}c^{\mbf{r}}_{\LQ} 
   g_{\gamma\gamma}^{\LQ}}{\sum_{\LQ}\overline{c}^{\mbf{r}}_{\LQ}g_{gg}^{\LQ}} \right|^2 \times \ord(1)\,,
 \end{align}
where in $\ord(1)$ we consider the loop factors and $\rho_{\LQ}$ is an order one coefficient that takes into account the color structure in the decay $\Sigma_8 \to gg$, see eq.~\eqref{rho_LQ}. We first observe that a larger decay width into two gluons is expected for the singlet, assuming the same order of magnitude for the $\ord(1)$ coefficients and $m_0\simeq m_8$. For the other decay channels, the presence of the strong couplings 
always favors the decays of $\Sigma_8$ into final states containing at least one gluon jet.

\begin{figure}[h!]
\centering
 \includegraphics[scale=.33]{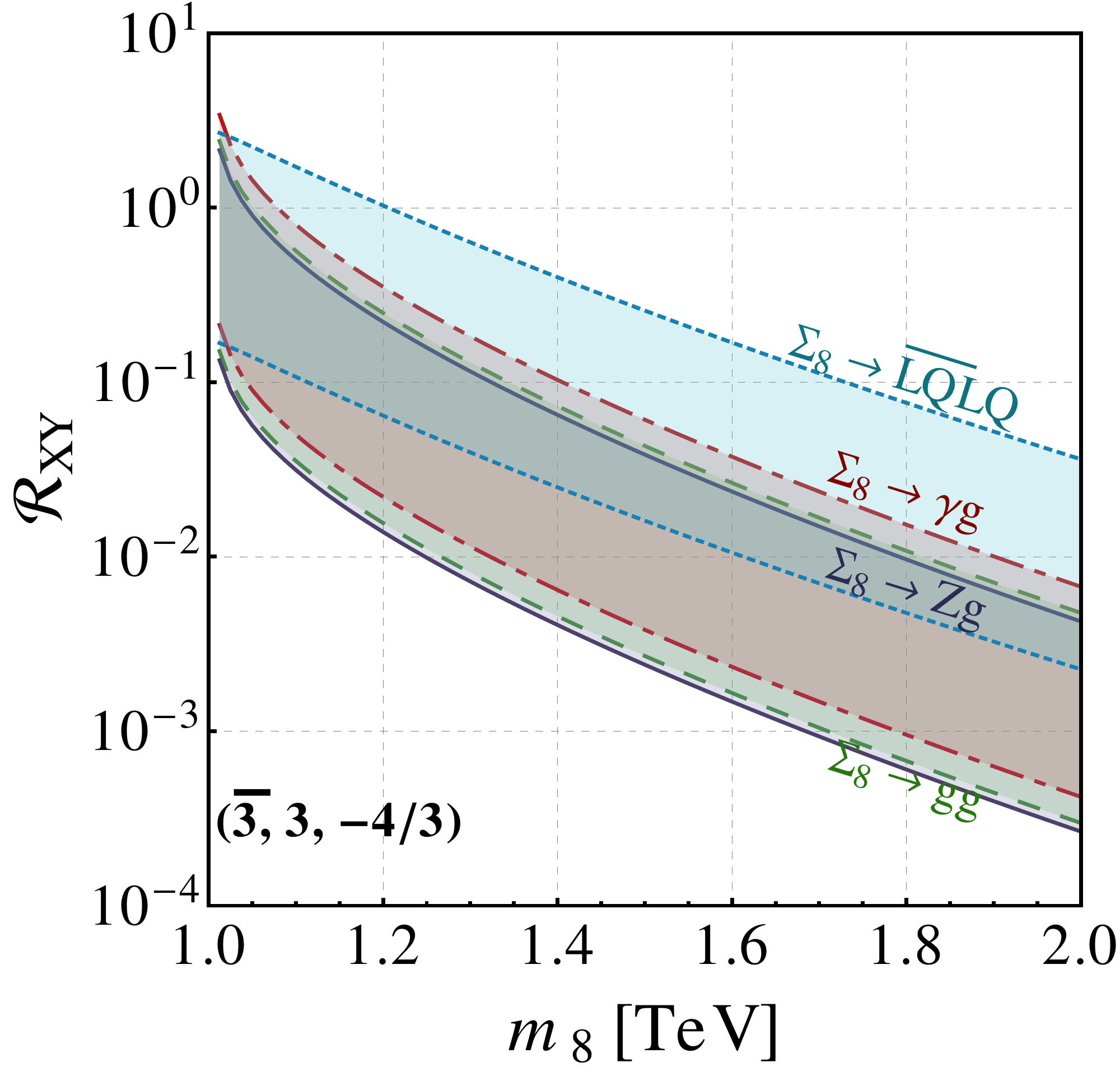}
 \includegraphics[scale=.33]{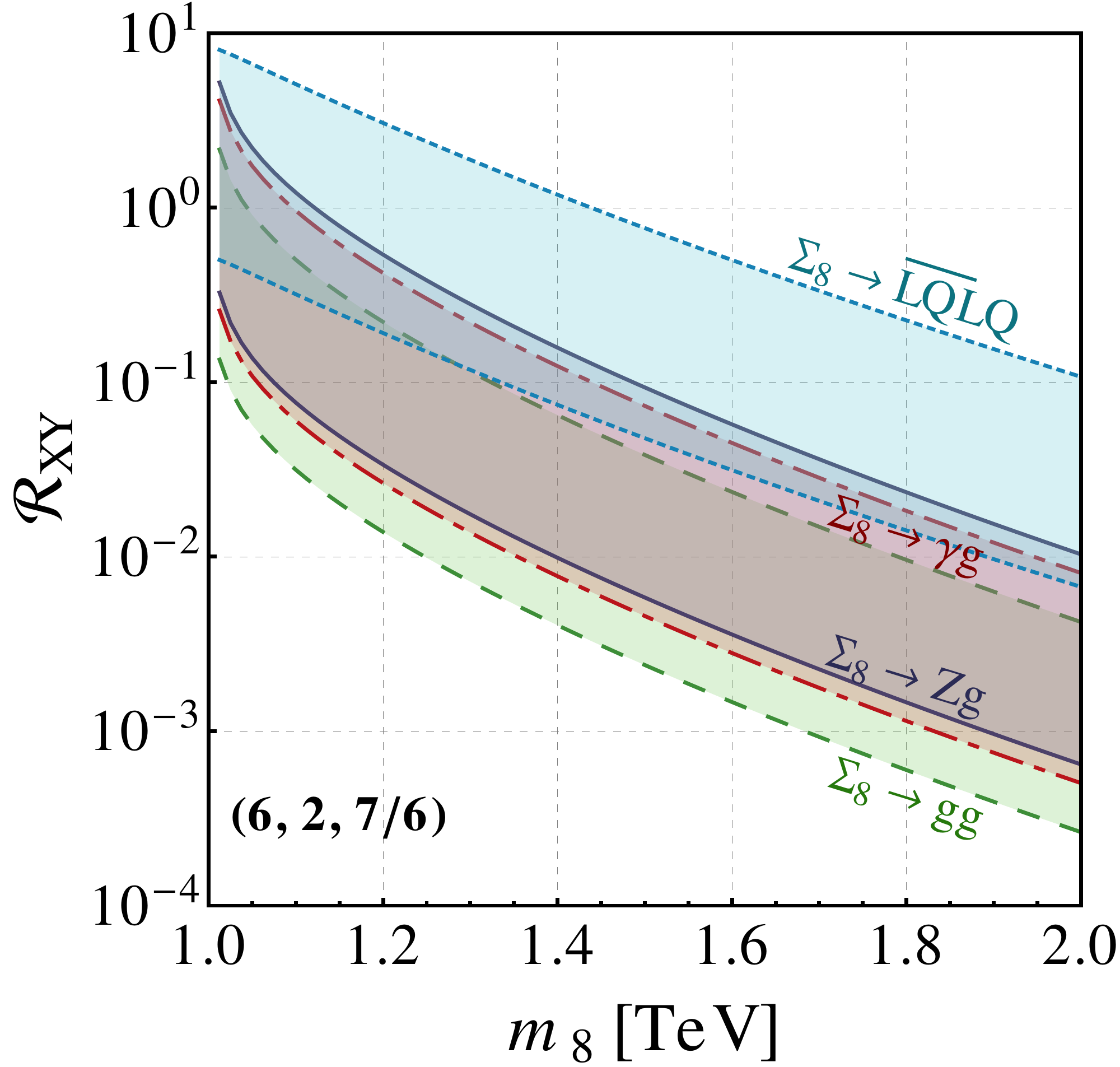}
  \caption{\it $\mathcal{R}_{XY}$  as a function of $m_8$ assuming one light LQ with $M_{\LQ} = 500\ \mathrm{GeV}$ and $c_{70} = 1$. 
  In the left panel we show the contribution of $(\mbfb{3}, \mbf{3}, -4/3)$ while in the right panel that of $(\mbf{6}, \mbf{2}, 7/6)$. 
  The green dashed, red dot-dashed and blue solid lines are the signal region in the channels $gg$, $\gamma g$ and $Zg$, respectively. 
  The azure dotted lines contain the region $\sigma(pp \to \Sigma_8 \to \overline{\LQ}\LQ)/\sigma(pp \to  \overline{\LQ}\LQ)$.}
 \label{fig:plotS8}
\end{figure}

In order to have a large enough $\sigma(pp \to S \to \gamma\gamma)$ we observed in sec.~\ref{sec:singlet_decay} that we need 
$M_{\LQ} \sim 400\ \mathrm{GeV}$ with an order one coefficient. 
For $m_8 \sim 1 \ \mathrm{TeV}$ the tree-level decay to a LQ pair, given in eq.~\eqref{width_S8_LQLQ},  dominates the total width, so that
$\Br(\Sigma_8 \to XY) \simeq \Gamma(\Sigma_8 \to XY)/\Gamma(\Sigma_8 \to  \overline{\LQ} \LQ )$
where $XY$ is any possible final state. 
The expected signal at LHC can be computed in a similar way to the diphoton channel for $\Sigma_0$, 
see eq.~\eqref{cross_section_theo_S8} in \ref{sec:LHC_production}.
To compare the cross section for a given final state $XY$ produced through $\Sigma_8$ with the cross section 
$\sigma(pp \to S \to \gamma\gamma)$, we can define the following dimensionless quantity:
\begin{align}
\label{definition_R_XY}
\mathcal{R}_{XY} \equiv \frac{\sigma(pp \to \Sigma_8 \to XY)}{\sigma(pp \to \Sigma_0 \to \gamma \gamma)} = \frac{C_{gg}(\mu_F = m_8)}{C_{gg}(\mu_F = m_0)} \times \frac{m_0}{m_8} \frac{\Gamma(\Sigma_8 \to gg)}{\Gamma(\Sigma_0 \to gg)} \frac{\Br(\Sigma_8 \to XY)}{\Br(\Sigma_0 \to \gamma \gamma)}\,,
\end{align}
where our estimates do not take into account the 
photoproduction mechanism $\sigma_{\gamma \gamma}$. 
Here $\mu_F = m_8$ is the factorization scale.
To produce valuable numerical results, we fix the LQ mass to be 
$M_{\LQ} = 500\ \mathrm{GeV}$ and $c_{\LQ} = 1$. We obtain $\sigma(pp \to \Sigma_0 \to \gamma \gamma) = 1.0\,[5.1]\ \mathrm{fb}$ for a $(\mbfb{3}, \mbf{3}, -4/3)$ [$(\mbf{6}, \mbf{2}, 7/6)$] LQ, that we can use together with $\mathcal{R}_{XY}$ to extract information 
on the various decay widths of the $\Sigma_8$ state. 
We report our results in fig.~\ref{fig:plotS8} as a function of the octet mass $m_8$. 
The green dashed, red dot-dashed and blue lines are the signal region in the channels $gg$, $\gamma g$ and $Zg$, respectively. 
For comparison, we also show  the ratio $\sigma(pp \to \Sigma_8 \to \overline{\LQ}\LQ)/\sigma(pp \to  \overline{\LQ}\LQ)$ with
the azure dotted lines (see sec.~\ref{sec:LQpheno2} for further details). The effective coupling 
$\overline{c}^{\mbf{r}}_{\LQ}$ are allowed to vary in the interval $[1/2, 2]$, so that the upper lines always correspond  
to $\overline{c}^{\mbf{r}}_{\LQ}=2$ and the lower ones to $\overline{c}^{\mbf{r}}_{\LQ}=1/2$.

We observe that in the case of $(\mbf{6}, \mbf{2}, 7/6)$ and for low octet masses there is an enhancement in the channel $
\Sigma_8 \to \gamma g$, while for $(\mbfb{3}, \mbf{3}, -4/3)$ the order of magnitude is the same as that of 
$\sigma(pp \to \Sigma_0 \to \gamma\gamma)$. The large suppression with $m_8$ 
is given by the small gluon contribution encoded in the $C_{gg}$ coefficient: at $\sqrt{s} = 13 \ \mathrm{TeV}$ we get $C_{gg}(m_8 = 1\ \mathrm{TeV}) \sim 448$ and $C_{gg}(m_8 = 2\ \mathrm{TeV}) \sim 7$ 
using the PDF set \texttt{mstw2008nlo}.

\section{A new doublet from the $\mbf{45}_H$ or the $\mbf{70}_H$ as the resonance}
\label{sec:2HDM_model}

\subsection{The model}
In this section we study the possibility that the resonance $S$ belongs to a second Higgs doublet. Assuming only two scalar fields $\mbf{5}_H$ and 
$\mbf{45}_H$ or $\mbf{70}_H$ we can construct the $\rm SU(5)$ invariant potential as discussed in Appendix B of ref.~\cite{Emmanuel-Costa:2013gia} for the $\mbf{45}_H$. 
In our study we assume some particular mass spectrum 
$m_h < m_S \ll m_A, m_{H^\pm}$, $m_h < m_A \ll m_S, m_{H^\pm}$ or $m_h < m_S \simeq  m_A \ll m_{H^\pm}$ and, for the sake of simplicity, 
we  neglect the effect of the charged particles in Higgs sector. This mass spectrum is allowed for 
$m_{H^{\pm}} \lesssim1\ \mathrm{TeV}$ by Electroweak Precision Tests, see ref.~\cite{Becirevic:2015fmu} for a recent analysis. 

Let us consider first that the second Higgs doublet is in the $\mbf{45}_H$. This setup corresponds to the well-known
Georgi-Jarlskog model~\cite{Georgi:1979df}. We will follow the notation of ref.~\cite{Khalil:2013ixa}. Neglecting the corrections of order 
$v_{\mbf{5,45}}/v_{\mbf{24}}$, we can normalize the relative weights in the electroweak vacuum as
\begin{align}
 v^2 = |v_{\mbf{5}}|^2 + 9 |v_{\mbf{45}}|^2 \Longrightarrow  t_\beta \equiv \frac{v_2}{v_1} \equiv  \frac{-3v_{\mbf{45}}}{v_{\mbf{5}}}\,.
\end{align}
Notice that this normalization is needed in order to reproduce the correct SM gauge boson masses and to be able to define the alignment limit, 
in which the lightest CP-even Higgs is SM-like, i.e., with SM-like couplings to both gauge bosons and fermions. 
And similarly for the $\mbf{70}_H$, with a different contribution to gauge boson masses from its VEV, and thus a different $t_\beta$.

From a low-energy perspective, we are dealing with an effective two-Higgs doublet model (2HDM). 
Therefore we have two CP-even states, namely $h$ and $S$ which are a linear 
superposition of the real neutral components in $H_{\mbf{5}} \sim (\mbf{1}, \mbf{2}, 1/2) \in \mbf{5}_H$ and 
$H_{\mbf{45}}\sim (\mbf{1}, \mbf{2}, 1/2) \in \mbf{45}_H$~\cite{Branco:2011iw}: $\Re\{H_{\mbf{5}}^0\} = 
-h s_\alpha + S c_\alpha$ and 
$\Re\{H_{\mbf{45}}^0\} = h c_\alpha + S s_\alpha$, where $\sin \alpha\equiv s_\alpha,\, \cos \alpha\equiv c_\alpha$. The mixing angle $\alpha$ is related to the parameters in the potential 
defined in eq.~\eqref{full_potential_SU5}. 
Another possibility is that the resonance observed at the LHC is the pseudoscalar $A$ of the 2HDM, so that
$\Im\{H_{\mbf{5}}^0\} = G^0 c_\beta - A s_\beta$ and $\Im\{H_{\mbf{45}}^0\} = G^0 s_\beta + A c_\beta$, 
where $G^0$ is the would-be Goldstone boson, eaten by the Z boson. Finally, it could be that the signal is really due to both the CP-even and CP-odd resonances, 
if CP is not a good symmetry and they are sufficiently close in mass.\\

We can compute the couplings of the neutral Higgses to SM gauge bosons, as derived
from the kinetics terms in the $\rm SU(5)$ Lagrangian. Notice that such a kind of couplings are zero in the case 
$S = \Sigma_0, \Sigma_3^0$ and $A$. We obtain
\begin{align}
\label{coupling_2HDM_bosons}
 c_{hVV} = \frac{m_V^2}{2 v^2} s_{\beta-\alpha}\,\qquad c_{SVV} = \frac{m_V^2}{2 v^2} c_{\beta-\alpha}\,,  \qquad c_{AVV} = 0\,.
\end{align}
If we want to reduce the decay width into a boson pair for the heavy scalar $S$ while at the same time have a SM-like Higgs we have to go to the alignment 
or decoupling limit:
$\beta-\alpha = \pi/2$. 

The presence of the $\mbf{45}_H$ gives an extra term to the masses of the SM fermions, which in particular produces the desired factor of 3
among down and charged lepton masses~\cite{Georgi:1979df}. We can closely follow ref.~\cite{Khalil:2013ixa} regarding the Yukawa Lagrangian and,
for the sake of simplicity, we  take the Yukawa $Y_4$ of the $\mbf{45}_H\,\mbf{10}_F\, \mbf{10}_F$ term equal to zero 
(its presence will only open the allowed parameter space). The masses read:
\begin{align}
 M_{E} &= Y_1^T v_5-6 Y_2^T v_{45}\,,  \qquad M_{D}= Y_1 v_5+2 Y_2 v_{45}\,,  \qquad M_{U}= 4(Y_3+Y_3^T)v_5\,. \label{masses}
\end{align}
In this way, starting in the basis where up quarks and charged leptons are diagonal (with diagonal mass matrices $m_E,\,m_U$), we can completely determine the Yukawa interactions of the scalars with the SM fermions in terms of their masses and the CKM mixing matrix $V_{\rm CKM}\equiv V^d_L$. 

Using $\tan\alpha\equiv t_\alpha$, and equivalently for $\beta$, the neutral interactions of $h/S/A$ which we are interested in read, for the quarks ($+{\rm H.c}$ for the opposite chiralities):
\begin{subequations}
\begin{align}
 c_{h\overline{u}_R u_L} &= -Y^\prime_{3}\,\frac{s_\alpha}{c_\beta}\to Y^\prime_{3}\,,   \qquad\qquad \qquad c_{S\overline{u}_R u_L} = Y^\prime_{3}\,\frac{c_\alpha}{c_\beta}\to Y^\prime_{3} t_\beta\,,   \qquad \qquad\qquad c_{A\overline{u}_R u_L} = -i Y^\prime_{3}\, t_\beta\,,  \label{coupling_S_ff_2HDM} \nline
 c_{h\overline{d}_R d_L} &= -Y^\prime_{1}\,V_{\rm CKM}\,\frac{s_\alpha}{c_\beta} \to Y^\prime_{1}\,V_{\rm CKM}\,,  \qquad c_{S\overline{d}_R d_L} =Y^\prime_{1}\,V_{\rm CKM}\,\frac{c_\alpha}{c_\beta} \to Y^\prime_{1}\,V_{\rm CKM}\, t_\beta\,,  \qquad c_{A\overline{d}_R d_L} = -iY^\prime_{1}\,V_{\rm CKM}\,t_\beta\,,  \label{coupling_S_ff_2HDMb}
\end{align}
\end{subequations}
and for the charged leptons:
\begin{align}
 c_{h\overline{e}_Re_L} &= Y^\prime_{2}\frac{c_\alpha}{s_\beta}-Y^\prime_{1}\frac{s_\alpha}{c_\beta} \to Y^\prime_{2}+Y^\prime_{1}= \frac{m_E}{v}\,,\nonumber \nline
 c_{S\overline{e}_Re_L} &= Y^\prime_{2}\frac{s_\alpha}{s_\beta}+Y^\prime_{1}\frac{c_\alpha}{c_\beta}  \to Y^\prime_{1}\,t_\beta-\frac{Y^\prime_{2}}{t_\beta}\,,\nonumber   \nline
  c_{A\overline{e}_Re_L} &= -i\, \left(Y^\prime_{1}\,t_\beta-\frac{Y^\prime_{2}}{t_\beta}\right)\,,\label{coupling_S_ff_2HDMc}
\end{align}
where the couplings after the arrows show how they simplify in the decoupling limit $\beta-\alpha=\pi/2$, and we have defined the effective Yukawas in the mass basis as:
\begin{align}
 Y^\prime_{1}\equiv Y_{1} c_\beta &= \frac{3 m_D V^\dagger_{\rm CKM}+ m_E }{4 v}\,,  \qquad Y^\prime_{2}\equiv 2 Y_{2} s_ \beta = -3 \frac{m_D V^\dagger_{\rm CKM}- m_E}{4 v}\,,  \qquad Y^\prime_{3}\equiv 4(Y_{3}+Y_{3}^T) c_\beta = \frac{m_U}{v}\,.
  \label{yukawas}
\end{align}
Constraints on the mixing angles $\alpha$  and $\beta$ can be obtained from the observed values on the light Higgs ($h$) channels, 
as measured by CMS and ATLAS~\cite{ATLAS_CMS_HIGGS_CONFERENCE_2015}. We define the signal strength for the channel $ii$ as:
\begin{align} \label{muhii}
 \mu^h_{ii} \equiv \frac{\sigma(pp \to h)}{\sigma_{\SM}(pp \to h)} \times \frac{\Br(h \to ii)}{\Br_{\SM}(h \to ii)}=\left(\frac{s_\alpha}{c_\beta}\right)^2 \times \frac{\Br(h \to ii)}{\Br_{\SM}(h \to ii)} \,,
\end{align}
where $ii=\tau^+ \tau^-,\,b\bar b, \,t\bar t,\, W^+W^-,\,ZZ,\,\gamma\gamma$ and we assume dominant gluon gluon fusion production, as is the case of the LHC.

For the heavy scalars S, A we impose the upper limits (and the diphoton signal) from LHC searches on the different channels~\cite{Strumia:2016wys}
\begin{align} \label{muHii}
 \mu^H_{ii}& \equiv \sigma(pp \to S)\times \Br(S \to ii) = \sigma^0_S\, \left(\frac{c_\alpha}{c_\beta}\right)^2\times \Br(S \to ii)\,, \\
\mu^A_{ii} & \equiv \sigma(pp \to A )\times \Br(A \to ii) = \sigma^0_A\, t^2_\beta \times \Br(A \to ii)\,,
\end{align}
where $\sigma^0_{S,\,A}$ are the cross sections of $S$, $A$ at 750 GeV with full couplings to tops, 
which have a value $\sigma^0_{S} = 0.736\ \mathrm{pb}$ and $\sigma^0_{A} \simeq |\tilde{A}_{1/2}(\tau_t)|^2/|A_{1/2}(\tau_t)|^2\sigma^0_{S} = 1.039\ \mathrm{pb}$ 
\footnote{
See \href{https://twiki.cern.ch/twiki/bin/view/LHCPhysics/CrossSections}
{\texttt{https://twiki.cern.ch/twiki/bin/view/LHCPhysics/CrossSections}}.} . 

\begin{figure}[h!]
\centering
\includegraphics[scale=.33]{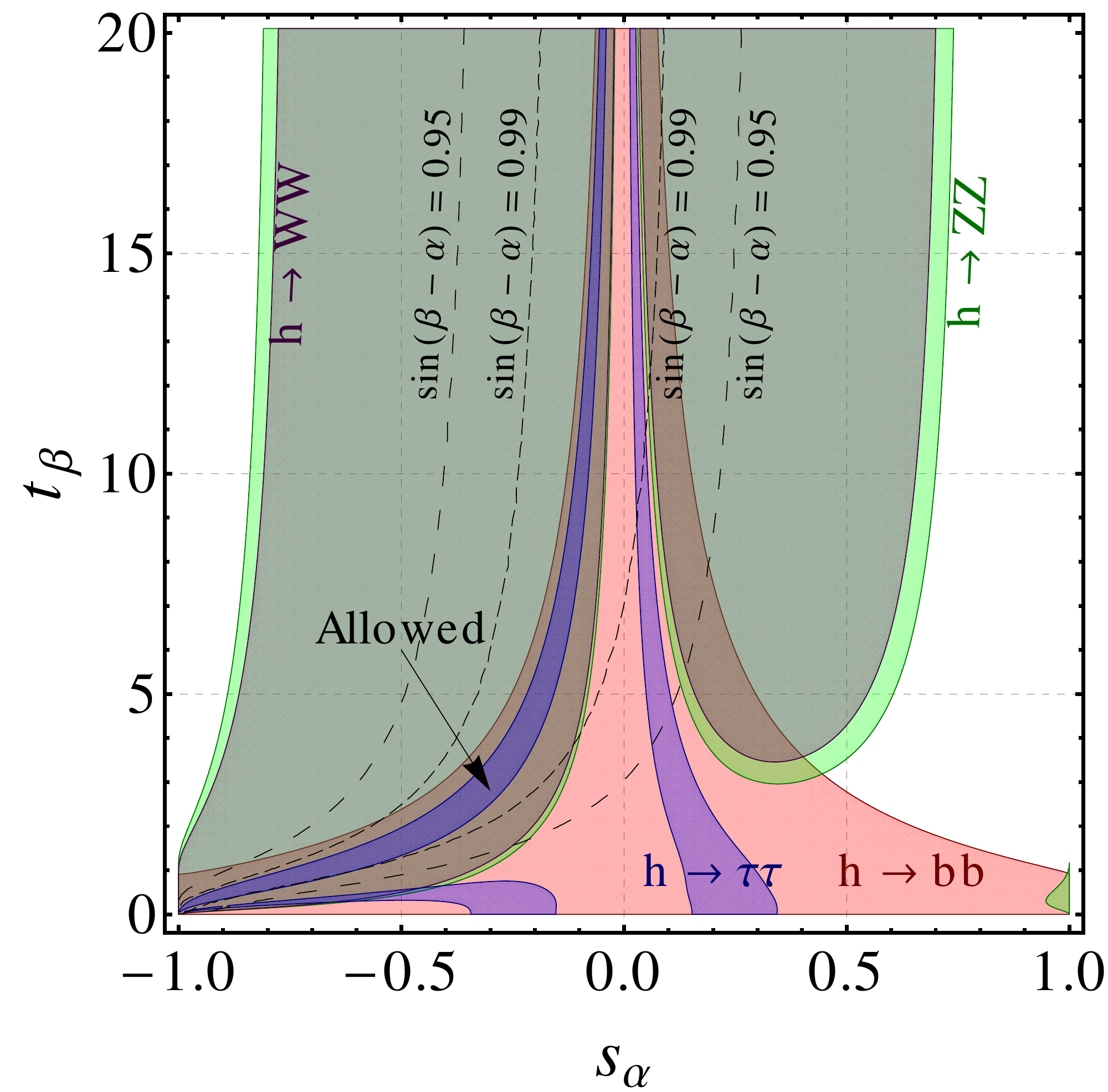}
  \caption{\it Allowed regions for Higgs tree-level decay channels at $3\sigma$ on the $t_\beta$ versus $s_\alpha$ plane. In blue $h\to\tau^+ \tau^-$, 
  in red $h \to b\overline{b}$, in green $h \to ZZ$ and in gray $h \to W^+W^-$. The black dashed lines are obtained for 
  constant values of $\sin(\beta - \alpha)$.
} \label{fig:channels}
\end{figure}

The couplings involved for the different channels can be straightforwardly obtained from 
eqs.~\eqref{coupling_S_ff_2HDM}, ~\eqref{coupling_S_ff_2HDMb} and \eqref{coupling_S_ff_2HDMc}, where $Y_{1,\,2}^\prime$ are defined in eq.~\eqref{yukawas}. 
The strongest constraints come from searches of heavy resonances decaying to SM quarks and leptons. In particular, we impose the limits on a $750$ GeV 
resonance decaying into gauge bosons, tops, bottoms, and tau-leptons~\cite{ATLAS_CMS_HIGGS_CONFERENCE_2015} (as well as 
the limits on the light Higgs decay channels  mentioned above). 
These limits are shown in fig.~\ref{fig:channels}, in the plane $t_\beta-s_\alpha$. Clearly, data impose to be close to the decoupling limit. 

The interaction between the heavy scalar $S$ or pseudoscalar $A$ and the $\LQ \sim (\mbfb{3}, \mbf{2}, 7/6) \in \mbf{45}_H$ 
can be obtained from the scalar potential involving $\mbfb{5}_H$ and $\mbf{45}_H$. We can 
construct the following two different quartic contributions: 
\begin{align}
\label{eq:pot45}
 V \supset [\mbf{45}_H\mbfb{45}_H\mbf{5}_H\mbfb{5}_H]_{\mbf{1}} + [\mbf{45}_H\mbfb{45}_H\mbf{45}_H\mbfb{45}_H]_{\mbf{1}}\,,
\end{align}
where we ignore all the possible $\rm SU(5)$ contractions. 
From the first type of terms $[\mbf{45}_H\mbfb{45}_H\mbf{5}_H\mbfb{5}_H]_{\mbf{1}}$ the coupling 
is proportional to $v  c_\beta c_\alpha $, while from the second term  we obtain $v s_\beta s_\alpha $. 
Instead, in the case of $A$  
we get $v s_\beta c_\beta $ from both terms. Therefore  
\begin{gather}
\label{coupling_S_LQLQ_2HDM}
 c_{S\overline{\LQ}\LQ} =  v \left( g_1 c_\beta c_\alpha + g_2 s_\beta s_\alpha\right) \to v \left(g_1- g_2\right) s_\beta c_\beta\,,  \nline
 \label{coupling_A_LQLQ_2HDM}
 c_{A\overline{\LQ}\LQ} =   -iv \left(g_1- g_2\right) s_\beta c_\beta \,,
\end{gather}
where $g_{1,2}$ are linear combination of the couplings involved in the potential in eq.~\eqref{eq:pot45}; the last equality 
in eq.~\eqref{coupling_S_LQLQ_2HDM} is computed in the decoupling limit. We see that an accidental suppression for 
the decay $S/A \to \overline{\LQ}\LQ$ is
possible for $g_1 - g_2 \simeq 0$. 
A similar coupling to eq.~\eqref{coupling_S_LQLQ_2HDM} is possible for the SM-like Higgs $h$ 
\begin{equation}
\label{hlqlq}
c_{h\overline{\LQ}\LQ} =v\left(-g_1 c_\beta s_\alpha +g_2 c_\alpha s_\beta \right) \to v \left( g_1c_\beta^2  +g_2 s_\beta^2 \right)\,.
\end{equation}
Thus for $t_\beta\sim1$ and $g_1\sim-g_2$, we  can suppress the light Higgs couplings to LQs while obtaining the desired enhancement. 
The alignment limit can be easily investigated in the basis where only the neutral component of one of the two Higgs doublets gets a vacuum expectation value, see for instance ref.~\cite{Gunion:2002zf}. In that case, our condition $g_1 \sim -g_2$ translates into a hierarchy among the quartic couplings involved in the interaction terms $h \overline{\LQ}\LQ$ and $S \overline{\LQ}\LQ$, the latter being much smaller.
The relation $g_1 \sim -g_2$ can be affected by loop corrections which are however suppressed by the loop factor $1/(16 \pi^2)$ and are expected to be under control.

Since $m_S \geq 2m_h$ we can have the tree-level decay mode $S \to hh$. The coupling is a non-trivial function of 
$\beta$ and $\alpha$. However, in the decoupling limit $c_{Shh}, c_{AZh}  \to 0$, like the decays into gauge bosons, 
given its proportionality  to $c_{\beta-\alpha}$. For the CP-odd state $c_{Ahh} = 0$.

A similar pattern occurs in the case of $\mbf{70}_H$; in fact, the coupling to vector bosons is exactly the same as 
eq.~\eqref{coupling_2HDM_bosons} and in this case we can work in the exact decoupling limit. Since the $\mbf{70}_H$ does not couple to fermions, 
this model is equivalent to a type-I 2HDM. Following ref.~\cite{Branco:2011iw}, in the decoupling limit we have:
\begin{align}
\label{coupling_S_ff_2HDM_typeI}
 c_{h\overline{f}_Lf_R} = - \frac{m_f}{v} \frac{s_\alpha}{c_\beta} \to \frac{m_f}{v}\,, \qquad c_{S\overline{f}_Lf_R} &= \frac{m_f}{v}\frac{c_\alpha}{c_\beta}\to \frac{m_f}{v} t_\beta\,,  
 \qquad c_{A\overline{f}_Lf_R} =i \frac{m_f}{v}t_\beta \,.
\end{align}
The couplings between LQs can be computed in the same 
way as in eq.~\eqref{coupling_S_LQLQ_2HDM} for $S$ or eq.~\eqref{coupling_A_LQLQ_2HDM} for $A$, with the obvious caveat that
they depend on different Clebsh-Gordan coefficients. 
The main difference with respect to the case of the $\mbf{45}_H$ is in the decay mode $S/A \to \gamma \gamma$ 
mediated by a loop of LQs because the $\mbf{70}_H$ contains larger $\rm U(1)_Y$ representations.

\subsection{Collider phenomenology}
In the case of 2HDM the phenomenology is quite similar to the singlet decay discussed in sec.~\ref{sec:singlet}. 
The effective operator describing the interactions of LQ and the heavy scalar is the same, 
but the coefficient is now given by eq.~\eqref{coupling_S_LQLQ_2HDM} 
and  with the right normalization, $g_{S\overline{\LQ}\LQ} = c_{S\overline{\LQ}\LQ}m_S/2$ and 
$g_{A\overline{\LQ}\LQ} = c_{A\overline{\LQ}\LQ}m_A/2$. 
In the loop processes we need to include the fermion contributions ($t$ and $b$ quarks or the $\tau$ lepton); the scalar/pseudoscalar decay widths into a fermion pair are given in eq.~\eqref{SA_decay_fermions}
of \ref{sec:LoopFunctionsAppendix}, with couplings as in eqs.~\eqref{coupling_S_ff_2HDM}, \eqref{coupling_S_ff_2HDMb} and \eqref{coupling_S_ff_2HDMc} for a type-III 2HDM and 
eq.~\eqref{coupling_S_ff_2HDM_typeI} for a type-I 2HDM. Working in the decoupling limit we can safely neglect the decay into light Higgses and vector bosons, 
while at one loop they contribute to the total decay width, see for example eq.~\eqref{sww} in \ref{sec:LoopFunctionsAppendix}.

The branching ratio for 2HDM is a function of the ratio $v_{\mbf{45}}/v_{\mbf{5}}$ or $v_{\mbf{70}}/v_{\mbf{5}}$
through the parameter $t_\beta$ and the effective couplings $g_{1,2}$. In order to obtain a large signal we expect that 
the coupling of $S$ and $A$ to LQs obey $g_1 \simeq - g_2$. Such a large coupling is also needed  in order 
to reduce the negative interference with the top quark contribution in the decay width. The region $t_\beta \geq 20$ is excluded from SUSY searches at the 
LHC~\cite{Aad:2014vgg,Khachatryan:2014wca}, while in the low $t_\beta$ region one should consider $t_\beta \gtrsim 1$, motivated by the ATLAS and CMS searches for spin one resonances decaying into a top 
pairs~\cite{Aad:2015fna, Khachatryan:2015sma}.

We expect that the largest part of the decay width, for $M_{\LQ} \gtrsim m_S/2$, is given by decay into a top pairs. The branching ratio can be estimated for $S$ as
\begin{align}
 \Br(S \to \gamma \gamma) \simeq \frac{\Gamma(S \to \gamma \gamma)}{\Gamma(S \to \overline{t}t)} = \frac{\alpha_{em}^2}{96\pi^2} \left(\frac{v}{m_t t_\beta}\right)^2 \times 
|c_{\LQ}g_{\gamma\gamma}|^2\dfrac{m_S^4}{M_{\LQ}^4} \times \ord(1)\,,
\end{align}
where in $\ord(1)$ we consider the loop contribution and $t_\beta \sim 1$ due to the signal strength $\mu_{\gamma\gamma}^h$. A similar result holds for the pseudoscalar $A$.

\subsubsection{Decays in 2HDM} 
\begin{figure}[h!]
\centering
 \includegraphics[scale=.33]{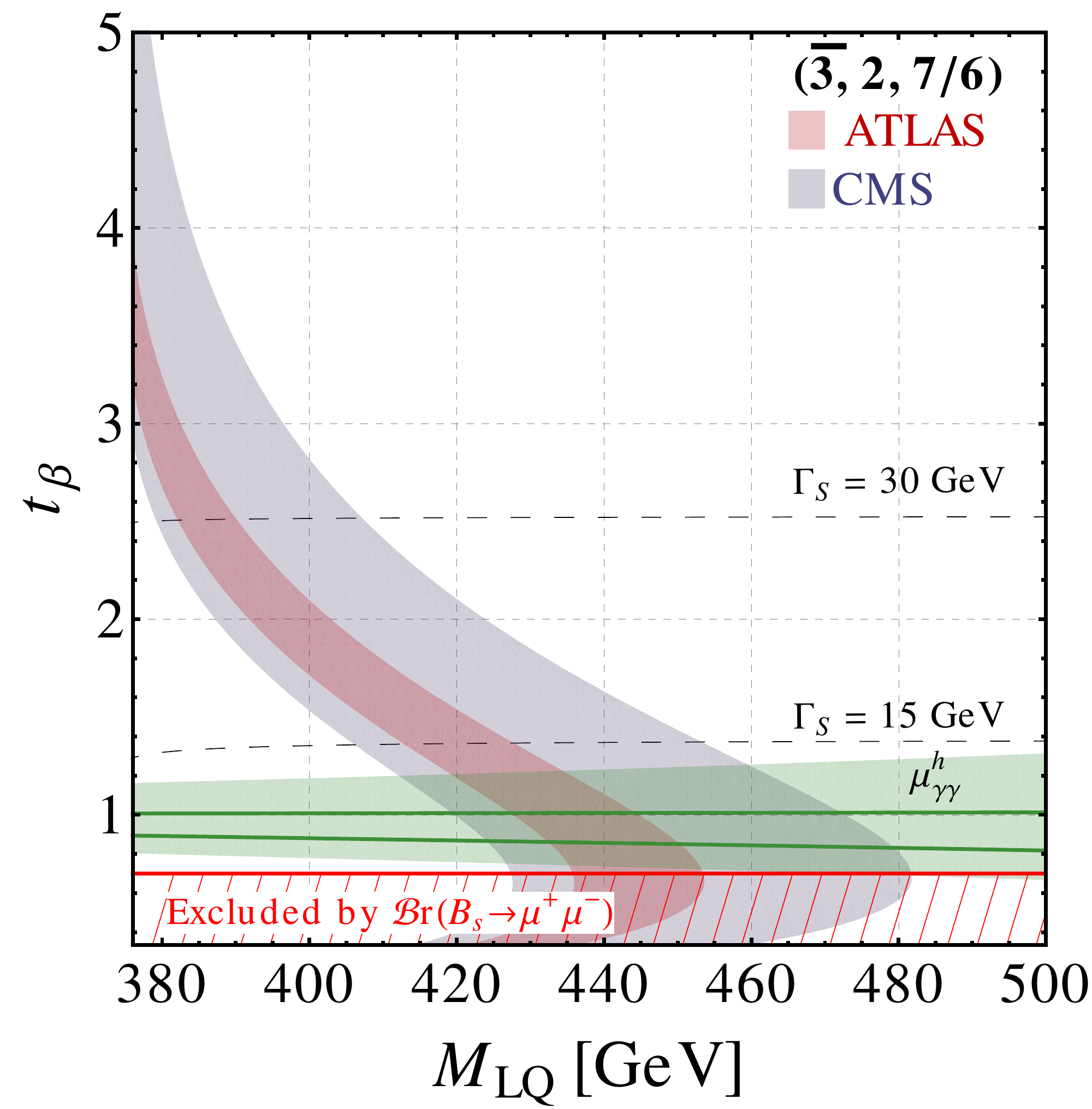}
 \includegraphics[scale=.33]{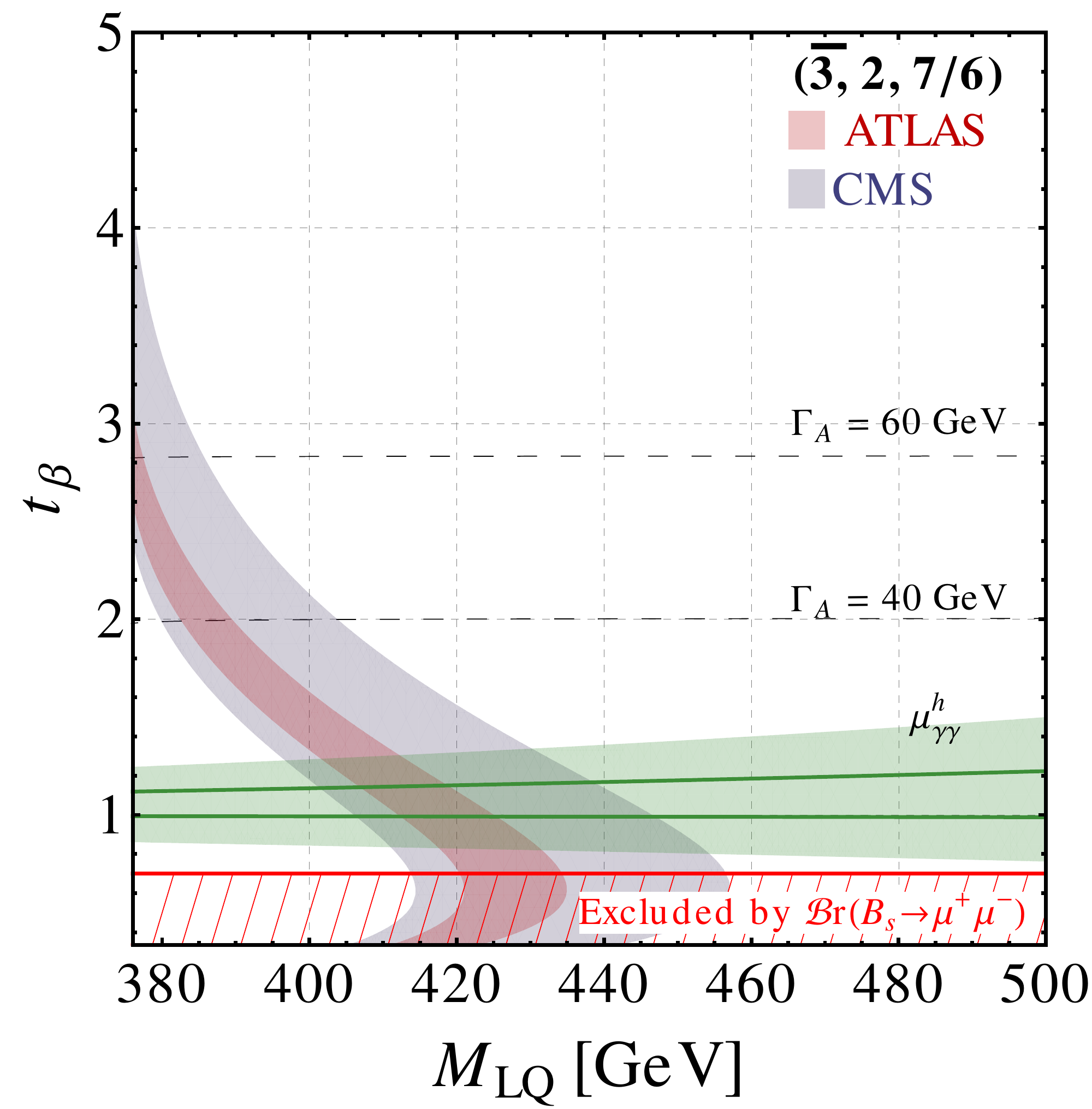}
  \caption{\it Signal region of $\sigma(pp \to S \to \gamma \gamma)$ (left) and $\sigma(pp \to A \to \gamma \gamma)$ (right) 
  in the plane $(M_{\LQ}, t_\beta)$ for $g_1 = -g_2 = \pm2\pi$ ($+$ for $S$ and $-$ for $A$) with $S,\,A,\,(\mbfb{3}, \mbf{2}, 7/6) \in \mbf{45}_H$. 
  The red (blue) region is allowed at ATLAS (CMS). The red hatched stripes correspond to low $t_\beta \leq 0.7$, excluded by $\Br(B_s \to \mu^+ \mu^-)$ data. 
  The black dashed lines are the isocontours of total width. The green lines indicate the allowed region of the parameter space $@\ 1\sigma$ CL for 
  $\mu^h_{\gamma \gamma}$, while the green area  $@\ 3\sigma$.  }
 \label{fig:scan_2HDM_45}
\end{figure}

\begin{figure}[h!]
\centering
 \includegraphics[scale=.33]{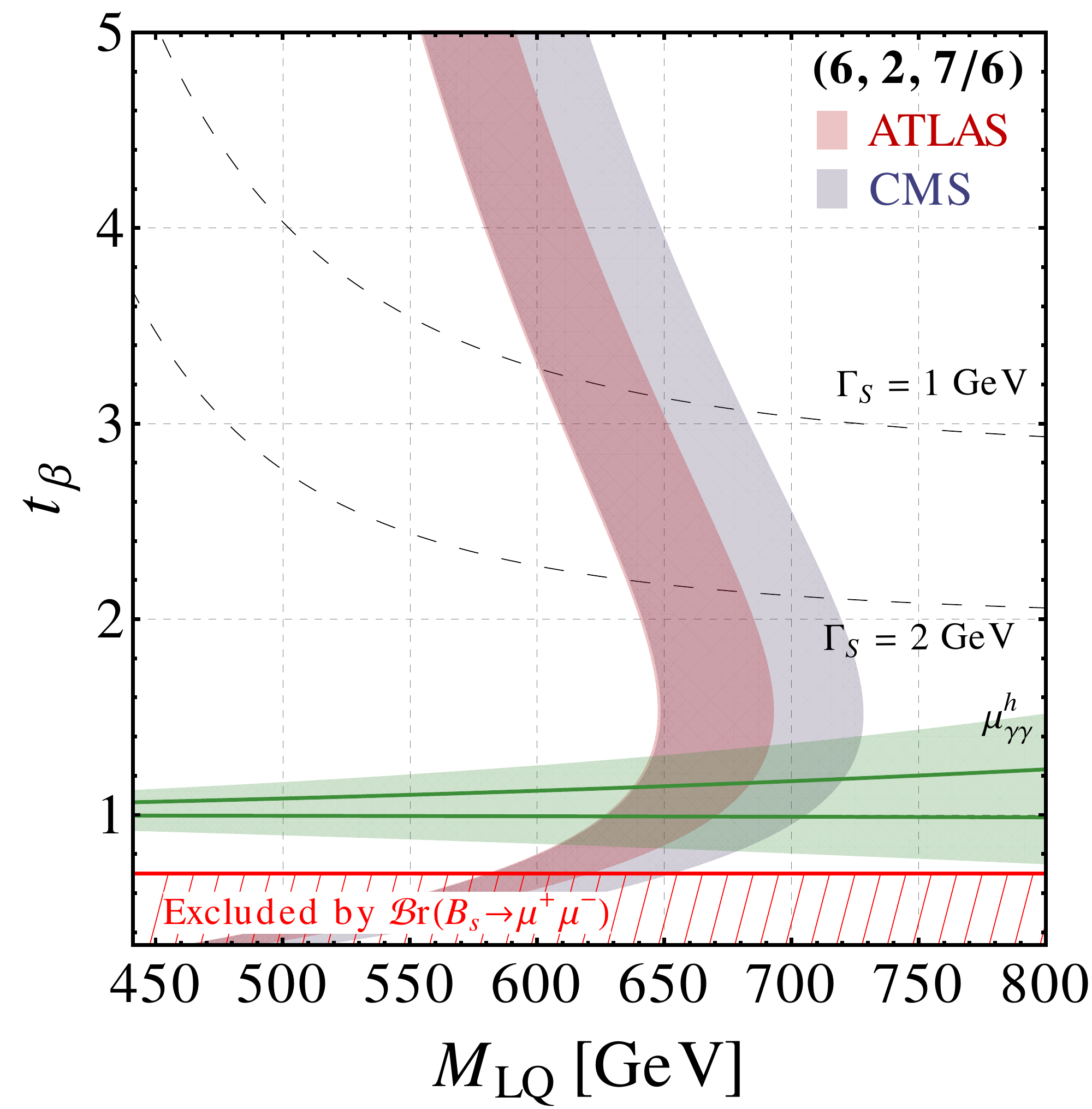}
 \includegraphics[scale=.33]{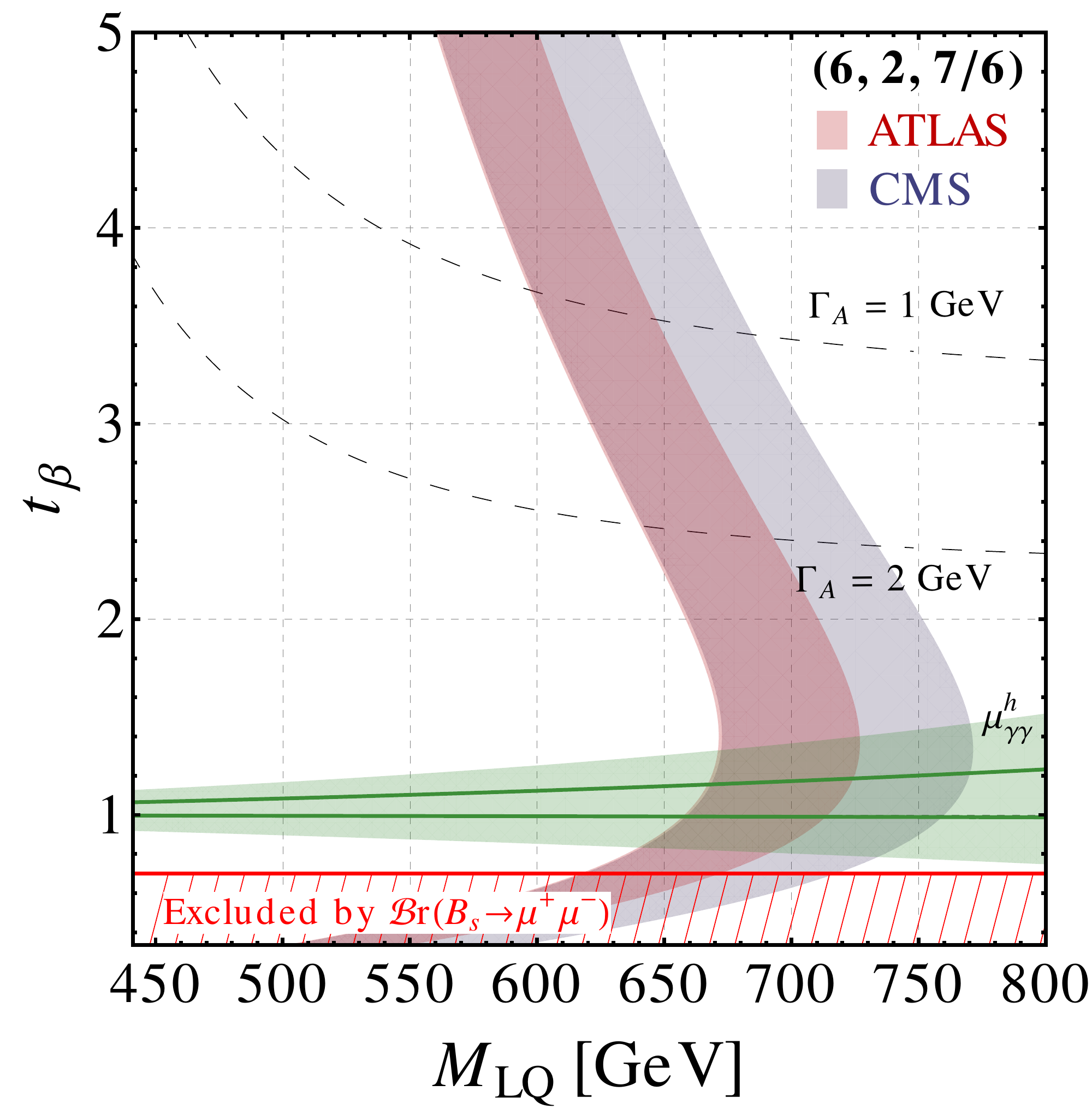}
  \caption{\it Same of fig.~\ref{fig:scan_2HDM_45} for $S,\,A,(\mbf{6}, \mbf{2}, 7/6) \in \mbf{70}_H$.}
 \label{fig:scan_2HDM_70}
\end{figure}

\begin{figure}[h!]
\centering
 \includegraphics[scale=.33]{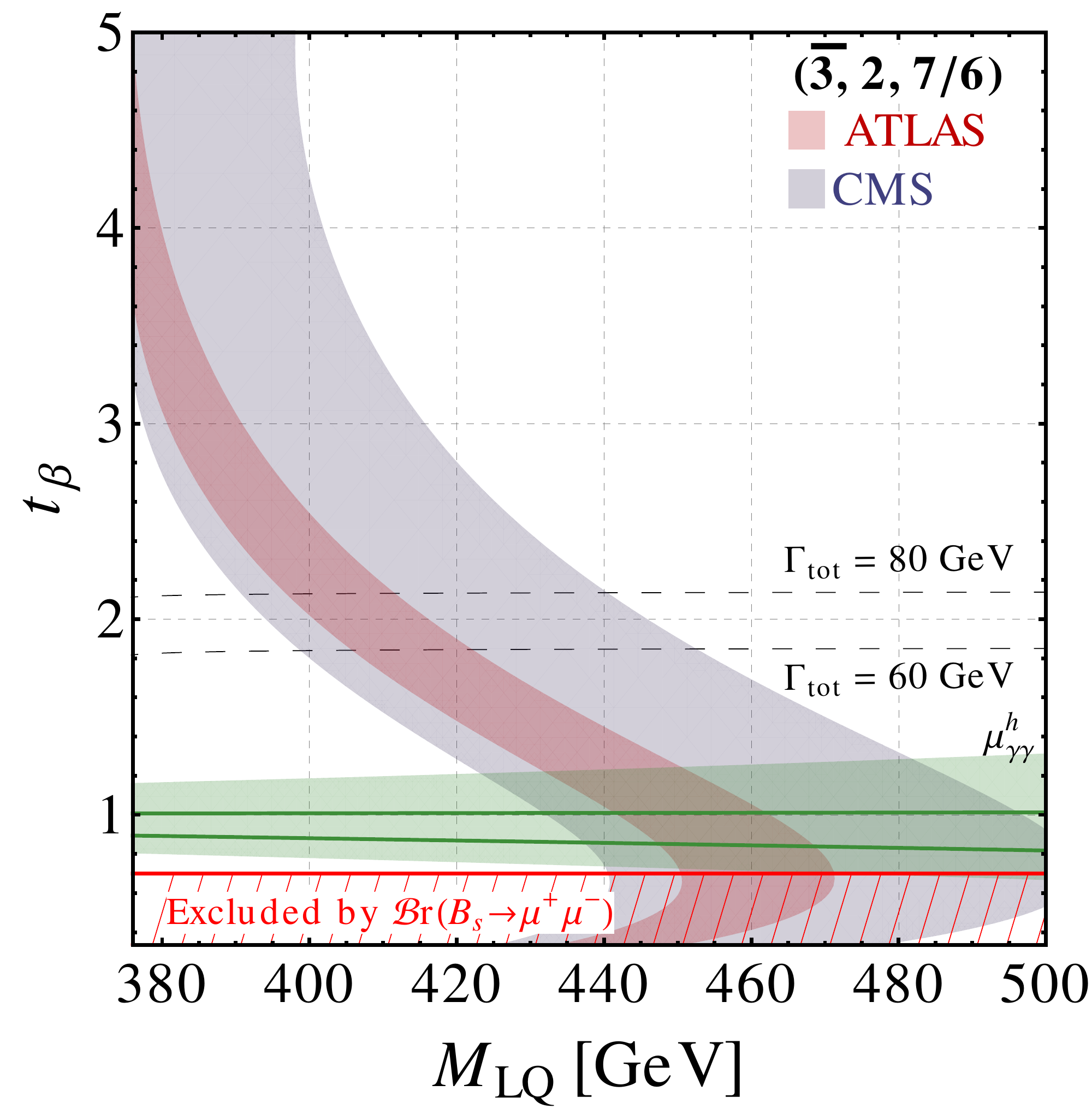}
  \includegraphics[scale=.33]{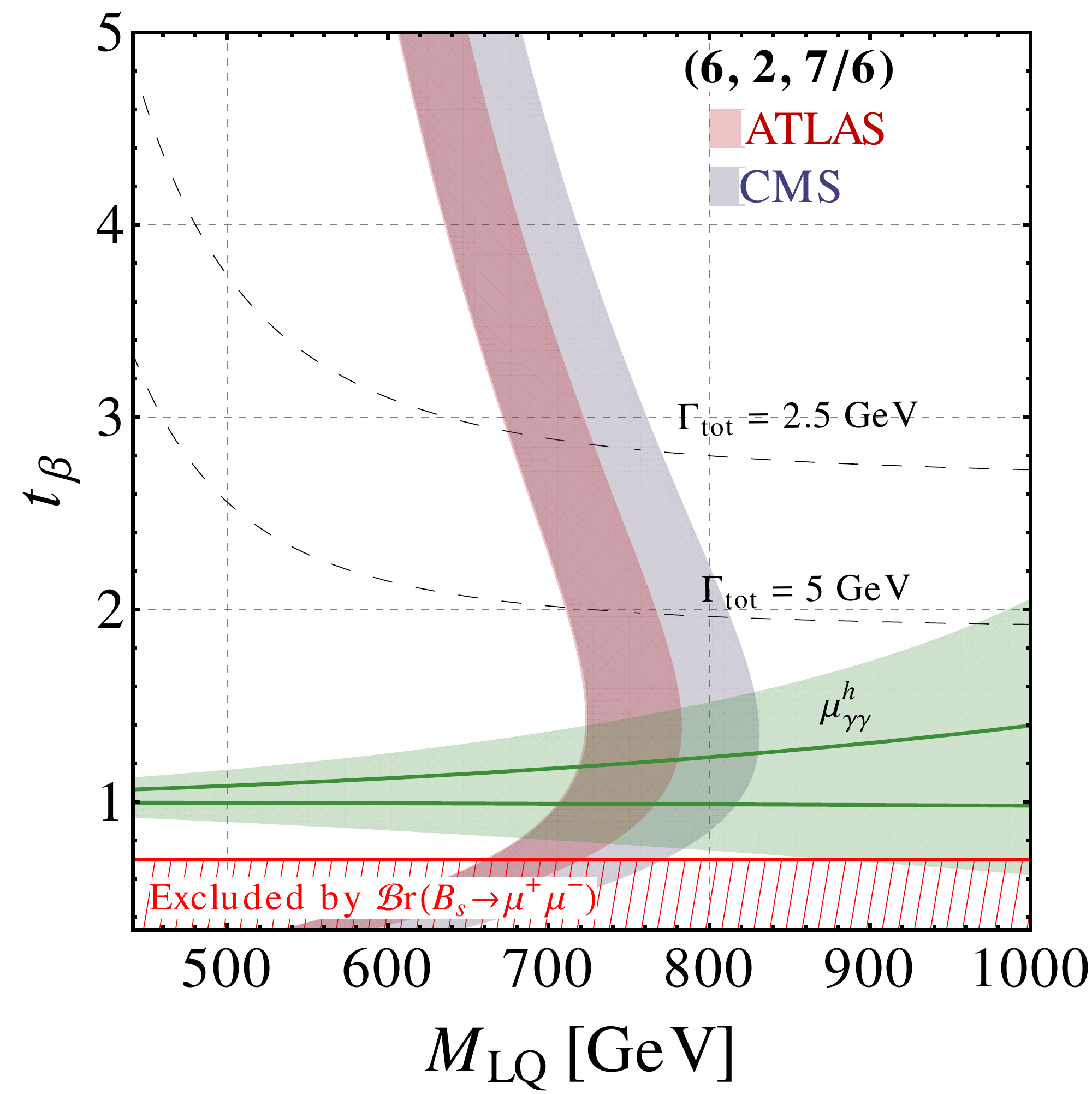}
  \caption{\it 
Similar to figs.~\ref{fig:scan_2HDM_45} and ~\ref{fig:scan_2HDM_70} for $\sigma(pp \to S \to \gamma \gamma) + \sigma(pp \to A \to \gamma \gamma)$, taking $g_1 = -g_2 = 2\pi$, assuming $\mbf{45}_H$ (left) or $\mbf{70}_H$ (right). We show in black dashed lines contours of constant total width, $\Gamma_{\mathrm{tot}} = \Gamma_A + \Gamma_S$.}
 \label{fig:scan_2HDM_AS}
\end{figure}

The decays of a heavy particle $S$ and/or $A$ can be mediated by a loop of LQs. 
In this exploratory study we assume the LQs in representations $(\mbfb{3}, \mbf{2}, 7/6) \in \mbf{45}_H$ and 
$(\mbf{6}, \mbf{2}, 7/6) \in \mbf{70}_H$. 
We fix the effective couplings $g_{1,2}$ to some representative values and leave the LQ mass $M_{\LQ}$ and $t_\beta$ 
as free parameters. 
We assume $t_\beta \gtrsim 1/3$ to keep the top Yukawa coupling perturbative. 
In fig.~\ref{fig:scan_2HDM_45} and \ref{fig:scan_2HDM_70} we show our results for the cross section 
$\sigma(pp \to S \to \gamma \gamma)$ (with very large $A$ mass) and $\sigma(pp \to A \to \gamma \gamma)$ (with very large $S$ mass) 
for the $(\mbfb{3}, \mbf{2}, 7/6)$  and $(\mbf{6}, \mbf{2}, 7/6)$ leptoquarks, respectively. 
In fig.~\ref{fig:scan_2HDM_AS} the allowed regions are instead obtained for the sum 
$\sigma(pp \to S \to \gamma \gamma) + \sigma(pp \to A \to \gamma \gamma)$ assuming a quasi-degenerate mass spectrum $m_A \simeq m_S$.
In the plots we also consider the effects of LQs on the production and decay of the SM $h$
through the signal strength $\mu^h_{\gamma \gamma} = 1.16^{+ 0.20}_{-0.18}$, 
reported in tab.~11 of~\cite{ATLAS_CMS_HIGGS_CONFERENCE_2015} and defined in eq.~\eqref{muhii}.

In our analysis we work in the decoupling limit, thus $\Gamma(h \to VV) = \Gamma_{\SM}(h \to VV)$ 
\footnote{For the decay into a vector boson pair we use the values quoted in LHC Cross Section Working Group: 
$\Gamma_{\SM}(h \to W^+W^-) = 8.815 \times 10^{-4}\ \mathrm{GeV}$ and $\Gamma_{\SM}(h \to ZZ) = 1.0824 \times 10^{-4}\ \mathrm{GeV}$. 
See \href{https://twiki.cern.ch/twiki/bin/view/LHCPhysics/CrossSections}
{\texttt{https://twiki.cern.ch/twiki/bin/view/LHCPhysics/CrossSections}}.} 
and we consider only production via gluon fusion, 
hence ${\sigma(pp \to h)}/{\sigma_{\SM}(pp \to h)} = \Gamma(h \to gg)/\Gamma_{SM}(h \to gg)$. 
The inclusion of the corrections given by vector-boson fusion are beyond the scope of this work and are neglected.

The modified couplings of $h$ are discussed in the previous subsections. 
The constraint from $\mu^h_{\gamma \gamma}$ implies that $t_\beta \sim 1$.
if $g_1 \sim - g_2$, see eq.~\eqref{hlqlq};
thus in this region we can neglect the $\overline{q}q$-channel in the $S$ 
production because the largest contribution comes from the $b$ PDF, which gives $C_{bb} \sim 15$ but it is sensibly smaller than the gluon 
contribution. 
If, on the other hand,  $t_\beta \gg 1$ then 
also the $b$-quark channel can be relevant; however this possibility is not generally realized in our models. 
Notice that the data $\Br(B_s \to \mu^+ \mu^-)$ tell us that $t_\beta \lesssim 0.7$ is excluded \cite{Becirevic:2015fmu}. 
This applies to the Georgi-Jarlskog model, where flavor is violated in the bs sector.

Assuming a type-III 2HDM our results give a large width, $\Gamma_{S,\,A} \simeq 20 \ \mathrm{GeV}$ dominated by tree level decays into SM fermions, 
however in the case of the pseudoscalar $A$ the allowed regions in the plane $(M_{\LQ}, t_\beta)$ are quite different 
because the coupling between LQs and $A$ is imaginary, 
thus the interference term in the $\Br(S \to \gamma\gamma)$ has a large impact on the observed signal, 
see right panel in  fig.~\ref{fig:scan_2HDM_45}.

If we consider the second Higgs doublet as a member of the $\mbf{70}_H$, 
the situation is qualitatively different to the previous case, since we have now  $\Gamma_S \sim \Gamma_A = \ord(1)\ \mathrm{GeV}$ and, 
due to different LQ representation, the allowed mass range is slightly different. 
For instance, in the case of $(\mbf{6}, \mbf{2}, 7/6)$, the 
$\sqrt{s} = 8\ \mathrm{TeV}$ data exclude a larger  region for  both for $S$ and $A$ if $M_{\LQ} \gtrsim m_{S, A}/2$;
thus we expect $M_{\LQ} \gtrsim 500\ \mathrm{GeV}$. 
The dependence on $t_\beta$ in $\mu^h_{\gamma\gamma}$ is similar to the scenario with the $\mbf{45}_H$ because 
the coupling among the SM-like Higgs $h$ and LQs is the same in the effective theory approach. The results for $S$ and $A$ are shown in fig.~\ref{fig:scan_2HDM_70}.

We can also consider the possibility that  $A$ and $S$ are quasi-degenerate particles, for instance with masses
$m_S = 750 \ \mathrm{GeV}$ and $m_A = 730 \ \mathrm{GeV}$. In this case, for $\mu_F = m_A = 730\ \mathrm{GeV}$, 
the gluon PDF slightly increases with respect to the values computed at $m_S = 750 \ \mathrm{GeV}$ (see \ref{sec:LHC_production})
and smaller effective couplings $|g_1| \sim |g_2| \lesssim 2\pi$, even 
for large LQ mass,  are needed to obtain a cross section compatible with the diphoton excess observed at LHC, fig.~\ref{fig:scan_2HDM_AS}.
This is quite different to the case of just S or A, as it can be seen in figs.~\ref{fig:scan_2HDM_45} and ~\ref{fig:scan_2HDM_70}.

\section{Higgs flavor violation from the Georgi-Jarlskog model} \label{sec:hlfv}

In this section we will discuss Higgs flavor violating (HFV) decays, both in the lepton and the down quark sectors. 
\footnote{In the up-quark sector, decays like $t\rightarrow hc,\,hu$ are absent, due to the fact that we take the minimal model with $Y_4=0$, see ref.~\cite{Khalil:2013ixa}.} In the Georgi-Jarlskog model, these decays are controlled by the CKM matrix. In the down-quark (charged-lepton) sector they are furthermore proportional to the charged-lepton masses (down-quark masses), see eqs.~\eqref{coupling_S_ff_2HDM} and~\eqref{coupling_S_ff_2HDMb}. Thus the most promising channels are $h\rightarrow bs$ and $h\rightarrow \tau \mu$. In the decoupling-limit, Higgs lepton flavor violating interactions are absent, as it should, while those in the down-quark sector are present, as in the SM, but controlled in this case by the charged-lepton masses. Thus, in order to have $h\rightarrow \tau \mu$, we will depart slightly from the decoupling limit, 
taking $\sin (\beta-\alpha)\gtrsim 0.9$, see fig.~\ref{fig:channels}. In this way we are able to open the parameter space, and furthermore study $h\rightarrow \tau\mu$, for which there is a hint of a signal. Indeed, CMS 8 TeV data show a $2.4 \sigma$ excess in the light Higgs channel $h\rightarrow \mu \tau$~\cite{Khachatryan:2015kon}, which is translated into a branching fraction: 
\begin{equation} \label{CMS_taumu}
\Br(h\rightarrow \mu \tau)=(0.84^{+0.39}_{-0.37})\,\%\,,
\end{equation} 
while ATLAS shows no significant deviation $\Br(h\rightarrow \mu \tau)=(0.53\pm 0.51)\,\%$~\cite{Aad:2016blu}.

If confirmed, this would be a clear signature of physics beyond the Standard Model, at the same level of the diphoton signal. 
There have been many works trying to explain this $\sim1\%$ signal, either using an EFT approach~\cite{Blankenburg:2012ex,Harnik:2012pb,Herrero-Garcia:2016uab} or 
focusing on a type III 2HDM~\cite{Davidson:2010xv,Sierra:2014nqa,Dorsner:2015mja,Omura:2015nja, Botella:2015hoa,Bizot:2015qqo} (also at loop level, see for instance an example in ref.~\cite{Arganda:2015naa}). Here we will study if the signal can be accommodated or not in the Georgi-Jarlskog model, with a $\mbf{5}_H$ and a $\mbf{45}_H$. Although the new LQ can provide a signal at one loop, only tree 
level topologies naturally allow for a $1\%\, \Br$, as shown in ref.~\cite{Herrero-Garcia:2016uab}. In particular, topologies with new scalars 
(a 2HDM, or a 2HDM plus new scalars) can explain the result~\cite{Herrero-Garcia:2016uab}. Recently, a 2HDM was employed to explain both the diphoton excess and the $h\rightarrow \tau \mu$ hint
simultaneously~\cite{Efrati:2016uuy} (see also ref.~\cite{Bizot:2015qqo,Han:2015qqj, Han:2016bus, Han:2016bvl}).

The $h\rightarrow \mu \tau$ branching ratio is given:
\begin{equation}  \label{BRH}
\Br(h\rightarrow \mu \tau) = \frac{m_h}{8\pi\Gamma_h} \,{\bar c}_{h\tau  \mu}^2\,, \qquad \bar c_{h\tau  \mu}=\sqrt{|c_{h\bar\tau  \mu}|^2+|c_{h\bar \mu \tau }|^2}\,,
\end{equation}
where the relevant couplings are given in eq.~\eqref{coupling_S_ff_2HDMc}, with $Y_{1,\,2}^\prime$ defined in eq.~\eqref{yukawas}.

Although the strongest constraints come in most models from $\tau \rightarrow \mu \gamma$,\footnote{We use the $\tau\rightarrow \mu \gamma$ expressions 
(including also the two-loops Barr-Zee diagrams.) given in refs.~\cite{Chang:1993kw, Goudelis:2011un, Blankenburg:2012ex,Harnik:2012pb}, summed over 
the different scalars $h,\,S,\,A$, and with their couplings as given in eqs.~\eqref{coupling_S_ff_2HDM}, ~\eqref{coupling_S_ff_2HDMb} and \eqref{coupling_S_ff_2HDMc}, 
where $Y_{1,\,2}^\prime$ are defined in eq.~\eqref{yukawas}. } in the Georgi-Jarlskog model \cite{Georgi:1979df} the Yukawas to down quarks and to charged leptons are completely related, c.f $c_{h \bar dd}$ in eq.~\eqref{coupling_S_ff_2HDMb} with $c_{h \bar ee}$ in eq.~\eqref{coupling_S_ff_2HDMc}. Furthermore, even though it is an effective low-energy 2HDM, both couplings involve $Y_{1,\,2}^\prime$, which are completely fixed by the down quark masses, the charged lepton masses and the CKM, 
as it  can be seen in eq.~\eqref{yukawas}. As we focus on the $\tau\mu$ sector, 
this means that there may be strong constraints from the $bs$ sector. Thus, we will furthermore impose, in addition to $\tau\rightarrow \mu \gamma$, 
the strongest constraints of the down-quarks sector, which come from $B_s$ meson mixing, in particular on the mass splitting $\Delta M_{B_s}$\cite{Gupta:2009wn}
 (see for instance tab. II of ref.~\cite{Harnik:2012pb} for the constraints on the $bs$ Yukawa):
\begin{align}
\label{MBs}
 \Delta M_{B_s} =\Delta M_{B_s}^{\rm SM}+ \frac{1}{M_{B_s}}\left[S_{B_s}\, \left(\frac{\bar c^2_{h b s}}{m_{h}^2}+ \,\frac{\bar c^2_{S b s}}{m_{S}^2}\right)+ P_{B_s}  
 \frac{\bar c^2_{A b s}}{m_A^2}\right]\,,
\end{align}
where we have defined $\bar c_{\Phi b s}=\sqrt{|c_{\Phi\bar b  s}|^2+|c_{\Phi\bar s b }|^2}$, for $\Phi=h,\,S,\,A$, and:
\begin{align}
 S_{B_s} = \frac{B_{B_s}f^2_{B_s}M_{B_s}^2}{6} \left[1+ \frac{M_{B_s}^2}{(m_b + m_s)^2} \right] \qquad P_{B_s} = \frac{B_{B_s}f^2_{B_s}M_{B_s}^2}{6} \left[1+ \frac{11M_{B_s}^2}{(m_b + m_s)^2} \right].
\end{align}
The SM value is assumed to be $\Delta M_{B_s}^{\rm SM} = (128.968 \pm 10.691) \times 10^{-13}\ \mathrm{GeV}$, \cite{Bazavov:2016nty}, 
and we consider the $3\sigma$ allowed region. The experimental value is $\Delta M_{B_s} = (116.834 \pm 0.138) \times 10^{-13}\ \mathrm{GeV}$ \cite{Agashe:2014kda}.

\begin{figure}[h!]
\centering
 \includegraphics[scale=.33]{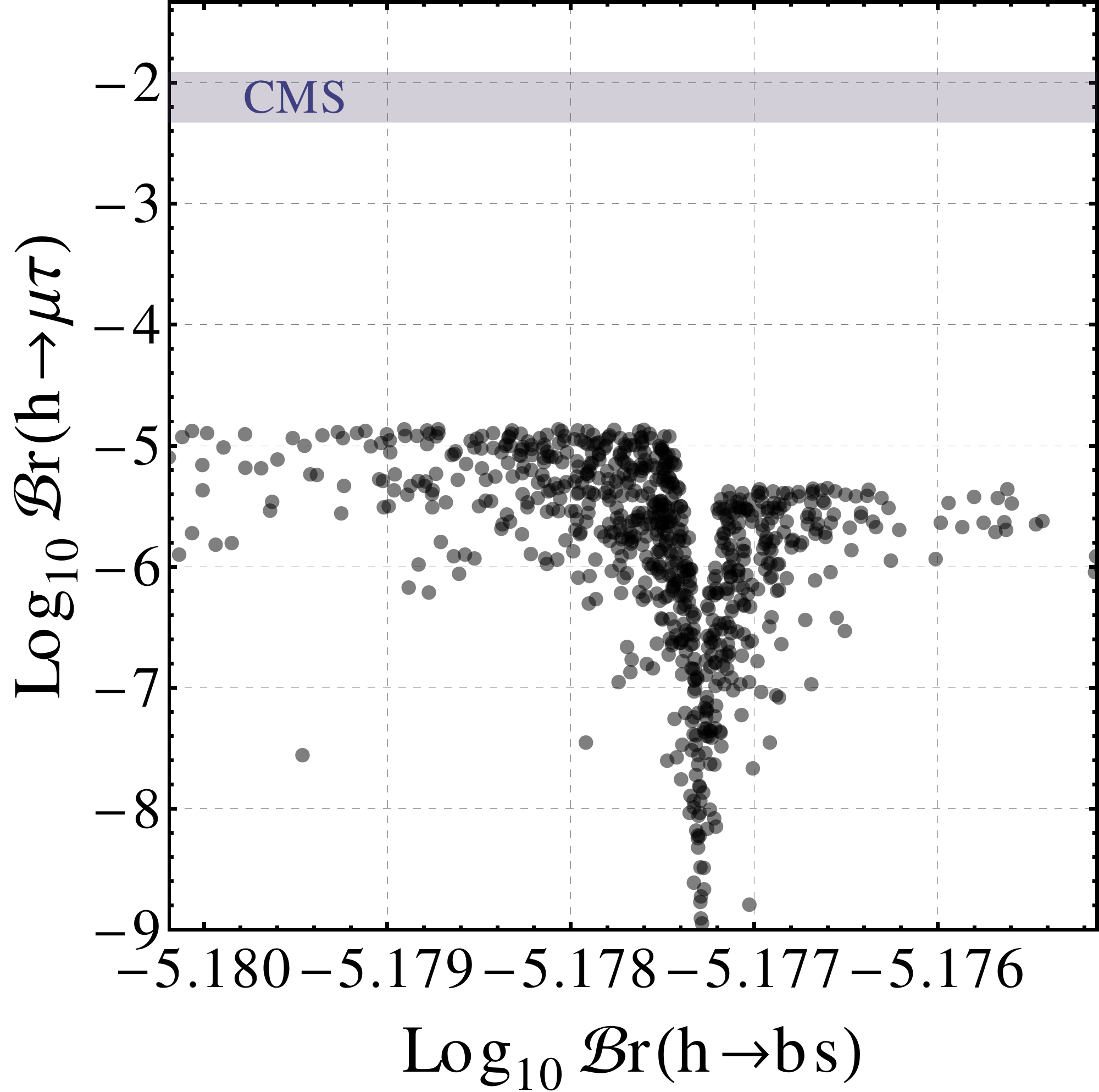}~~\includegraphics[scale=.35]{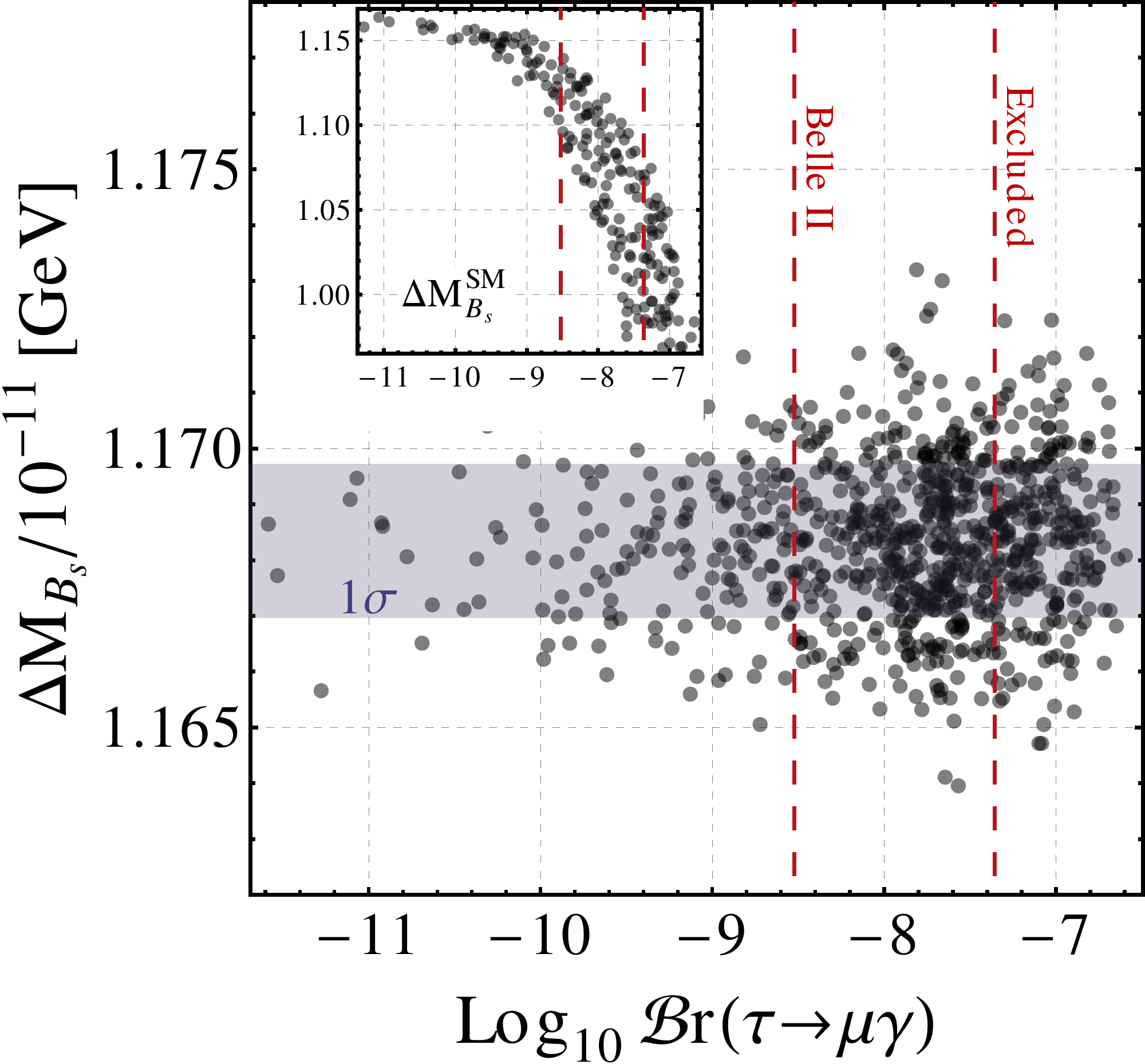}
  \caption{\it Left) $\Br(h\rightarrow \mu \tau)$ versus $\Br(h\rightarrow b s)$, for the points allowed by the 
  upper bound on  $\Br(\tau \rightarrow \mu \gamma)$. The blue band is the CMS hint region at $1\sigma$. 
  Right) $\Br(\tau \rightarrow \mu \gamma)$ versus $ \Delta M_{B_s}$. The red dashed lines are the upper bound at 90\% CL 
  $\Br(\tau \rightarrow \mu \gamma) \leq 4.4 \times 10^{-8}$ from BaBar \cite{Aubert:2009ag} and the projected sensitivity at Belle II 
  $\Br(\tau \rightarrow \mu \gamma) \leq 3 \times 10^{-9}$  \cite{Aushev:2010bq}. The blue region is the $1\sigma$ allowed region for $\Delta M_{B_s}$. 
  In the inset we report the related distribution of $ \Delta M_{B_s}^{\SM}$.
} \label{fig:hlfv}
\end{figure}

Clearly, in order to satisfy these last constraints and have large HFV the best case scenario is to have $m_S=750$ GeV and $m_A\gg750$ GeV.  
We scan over the relevant parameter space, $t_\beta \in [0, 50], s_\alpha \in [-1, 1]$, while $700\ \mathrm{GeV} \leq m_A \leq 5\ \mathrm{TeV}$ and $m_{H^{\pm}}$ 
is fixed using the best value of the $T$ parameter, $T = 0.01$ \cite{Agashe:2014kda} (in accordance with the relation discussed in 
ref.~\cite{Davidson:2010xv}, we observe $|m_A/m_{H^{\pm}} - 1| \lesssim 5\%$). For this scenario, in the left panel of fig.~\ref{fig:hlfv} we 
show $\Br(h\rightarrow \mu \tau)$ versus $\Br(h\rightarrow b s)$. We also plot $\tau \rightarrow \mu \gamma$ versus $ \Delta M_{B_s}$ in the right panel 
of fig.~\ref{fig:hlfv}. We find that $ \Delta M_{B_s}$ mixing always imposes stronger constraints than $\tau \rightarrow \mu\gamma$ in this model. 

The results show that $\Br(h\rightarrow \mu \tau),\,\Br(h\rightarrow b s)\lesssim 10^{-5}$ always, far below LHC and future expected sensitivity. 
Thus, the hint of an observed $\Br(h\rightarrow \mu \tau)$ reported in eq.~\eqref{CMS_taumu} cannot be explained in the minimal scenario and 
a confirmation of the $\Br(h\rightarrow \mu \tau)$ signal would rule-out the minimal model as the explanation. 
Notice that the dip in $\Br(h\rightarrow \mu \tau)$ comes from a cancellation in the relevant terms of the Higgs effective coupling to leptons, 
see first line of eq.~\eqref{coupling_S_ff_2HDMb}.

Let us conclude by mentioning that we have focused on topology A of ref.~\cite{Herrero-Garcia:2016uab} (a type III 2HDM) but a  topology B can also be realized 
in our set-up (see fig.~2 and tab.~3 of ref.~\cite{Herrero-Garcia:2016uab}), where the relevant scalars at low energy would be, in addition to the second Higgs 
doublet belonging to a $\mbf{5}_H$ or a $\mbf{45}_H$, a hypercharge-less singlet or triplet, fields which are precisely present in the $\mbf{24}_H$ as discussed 
extensively in sec.~\ref{sec:singlet}. In addition to the flavor-violating Yukawa of the second Higgs, 
the relevant term in the potential that can generate the
topology B (and thus $h\rightarrow \tau \mu$) is precisely $\mu_1$ in eq.~\eqref{coupl_24_5} for a second $\mbf{5}_H$ (and similarly for the case of a 
second $\mbf{45}_H$). This gives rise to mixing among the scalars, and will in general also give a contribution to the total rate. Indeed if these trilinear terms 
are larger than the scalar masses, topology B would be 
enhanced and could dominate. However, this would pose other problems, like naturality and or charge/breaking that we do not address here.

\section{Leptoquarks phenomenology}
\label{sec:LQpheno}
\subsection{Pair production and limits from direct searches}
\label{sec:LQpheno2}
Let us consider the pair production of LQs  $\subset \mbf{70}_H,\, \mbf{45}_H$ or $\Sigma_8\subset\mbf{24}_H$. 
As we have seen in the previous sections, 
LQs in representations $(\mbfb{3}, \mbf{3}, -4/3),\,(\mbf{6}, \mbf{2}, 7/6)\subset \mbf{70}_H$ are particularly 
interesting candidates to accommodate the diphoton excess. Thus, we briefly discuss here their main production and decays modes.

These LQs do not couple to fermions at renormalizable level. However, at the level of $D = 7$ for LQs in representations $(\mbfb{3}, \mbf{3}, -4/3),\,(\mbf{6}, \mbf{2}, 7/6)\subset \mbf{70}_H$ we can construct the following EFT operator
\begin{align}
  \op_{7} = \frac{C_7}{\Lambda^3} \mbfb{5}_F \mbfb{5}_F\mbfb{10}_F\mbfb{10}_F \mbf{70}_H =  \frac{C_7}{\Lambda^3}\times \left\{
  \begin{array}{l l}
   d^c_R d_L u_R Q_L^c\, \overline{\LQ},\, & \LQ \sim (\mbfb{6}, \mbf{2}, 7/6)\\
  d^c_R L_L e_R Q_L^c \, \overline{\LQ},\, & \LQ \sim (\mbfb{3}, \mbf{3}, -4/3)
  \end{array}\right. \, ,
\end{align}
where $\Lambda$ is the scale of the heavy degree of freedom that mediates the process. Assuming an order one coefficient and taking $M_{\LQ} = 500\ \mathrm{GeV}$, 
we obtain that short-lived LQs need a cutoff scales $\Lambda \lesssim  10^{5}\  \mathrm{GeV}$, while stability at collider scales is instead obtained for $10^{5}\  \mathrm{GeV} \lesssim \Lambda \lesssim  10^{7}\  \mathrm{GeV}$ (where the upper bound assures decay lifetimes below 100 s).

Both ATLAS and CMS have searched for single or pair production of the state $(\mbfb{3}, \mbf{2}, 7/6)$ 
(which could also be relevant for the 2HDM in the case of $\mbf{45}_H$).
Assuming that the LQ couples to the second generation of fermions with an $\ord(1)$ Yukawa, they found that $M_{\LQ} \gtrsim1 \ \mathrm{TeV}$, 
if the decay fraction into a charged lepton and a quark is $1$, see ref.~\cite{Dorsner:2016wpm} for a review on this topic. 
This bound can be avoided assuming a smaller decay fraction (while the Yukawa dependence is less relevant), otherwise 
we cannot achieve a signal for the diphoton excess in the case of 2HDM also for $|g_{1,2}| \leq 4\pi$, see 
fig.~\ref{fig:scan_2HDM_45}.

The largest production mechanism is through gluon fusion. We use the data quoted in tab.~9 of 
ref.~\cite{Dorsner:2016wpm} for $\LQ \sim \mbf{3} \in SU(3)$, where $\sigma(pp \to \overline{\LQ}\LQ) = 0.0461\ \mathrm{pb}$ at 
$\sqrt{s} = 14\ \mathrm{TeV}$ for $M_{\LQ} = 500\ \mathrm{GeV}$. We can estimate  the  $pp \to \mbfb{6}\mbf{6} $  cross section
using eqs.~\eqref{partonic_cross_section_pair} and \eqref{partonic_cross_section_pairb} (where the involved Casimir are
$C(\mbf{3}) = 4/3$ and $C(\mbf{6}) = 10/3$) as:
\begin{align}
 \frac{\sigma(pp \to \mbfb{6}\mbf{6} )}{\sigma(pp \to \mbfb{3}\mbf{3} )} = \frac{C(\mbf{6})^2 }{6 \times 2}\frac{3 \times 3}{C(\mbf{3})^2} = \frac{243}{64} \simeq 3.80\,,
\end{align}
hence we expect a larger production for the LQ in representation $(\mbf{6}, \mbf{2}, 7/6)$.

\begin{figure}[h!]
\centering
 \includegraphics[scale=.33]{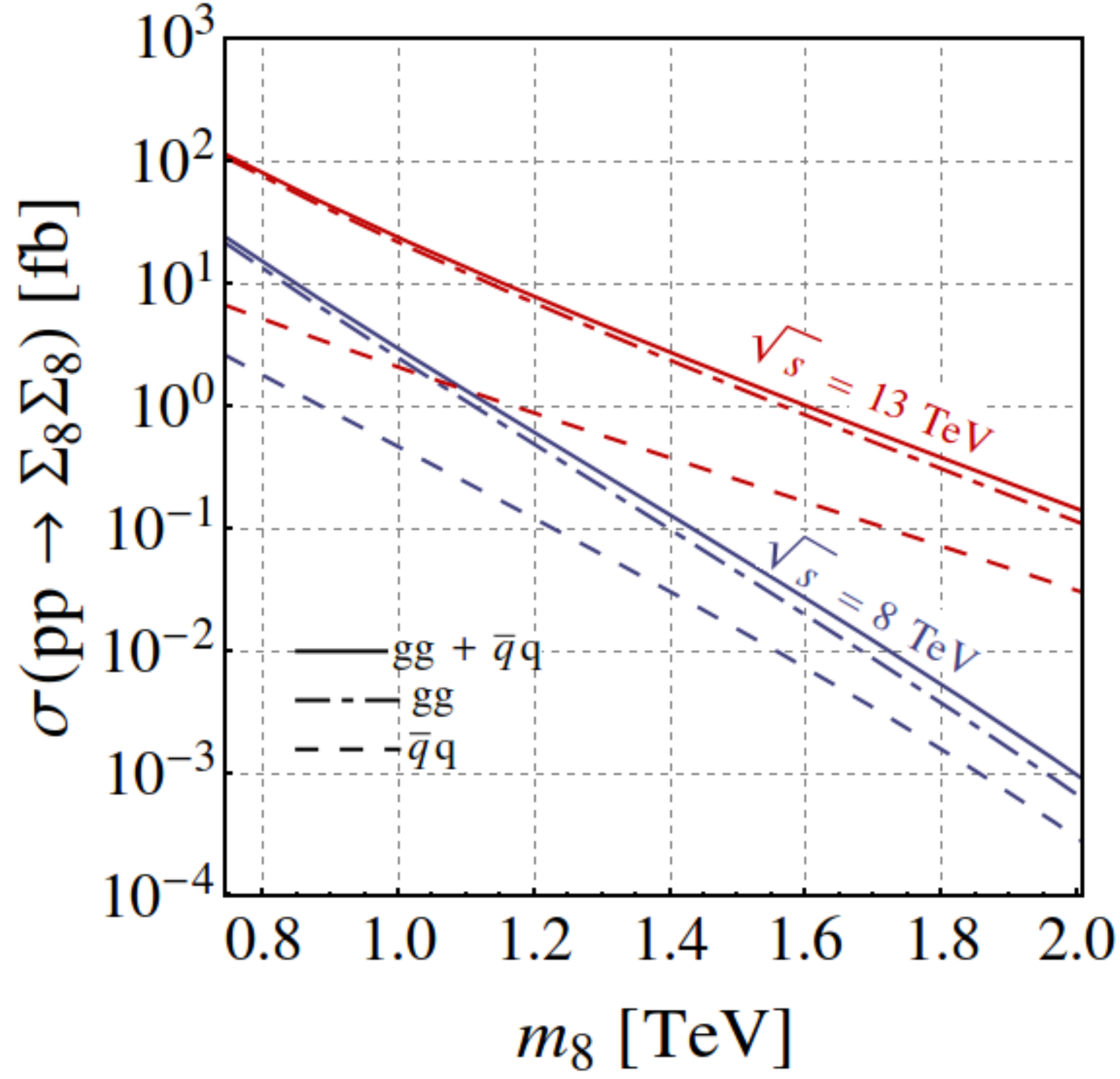}
  \caption{\it Cross section as function of $m_8$ at LHC for $\sqrt{s} = 8\ \mathrm{TeV}$, in blue, and $\sqrt{s} = 13\ \mathrm{TeV}$, in red. 
  The solid lines are the total cross section, the dashed the $\overline{q}q$-production and the dot-dashed the gluon production.
  We use the \texttt{mstw2008nlo} PDF set. }
 \label{fig:scan_m8_CrossSection}
\end{figure}

Another interesting signal is the pair production of two scalar octets through the kinetic term in the 
$\rm SU(5)$ Lagrangian. The partonic cross sections 
can be computed from eqs.~\eqref{partonic_cross_section_pair} and \eqref{partonic_cross_section_pairb} and are in agreement with refs.~\cite{Bai:2010dj, Chivukula:1991zk}, 
see further details in \ref{sec:LQAppendix}. In fig.~\ref{fig:scan_m8_CrossSection} we report the cross section for the pair 
production of $\Sigma_8$ as a function of the octet mass $m_8$, for $\mu_F = \mu_R = 2m_8$ and 
$\alpha_S(m_Z) = 0.1185$; the one loop correction to $\alpha_S$ do not take into account the contribution of the LQ, that we 
put at a mass larger than $2m_8$.
We also neglect QCD corrections. The results are in agreement with a similar analysis performed in ref.~\cite{Chivukula:2013hga}.

To conclude this section, we remind that the pair production of the octet $\Sigma_8$ 
with its subsequent decays to LQ, $\Sigma_8 \to\overline{\LQ}\LQ$ can compete with direct pair production of LQ, 
as shown in fig.~\ref{fig:plotS8}; from that, we observe that the LQ pair production has the same order of
magnitude of the decay mode $\Sigma_8 \to \overline{\LQ}\LQ$ in the region $m_8 = \ord(1)\ \mathrm{TeV}$ for order one $\overline{c}^{\mbf{r}}_{\LQ}$ couplings.

\subsection{Qualitative discussion on GUT unification}
The presence of massive LQs, with $M_{\LQ} \sim 1\ \mathrm{TeV}$, poses the question of how strong they affect the running
of the gauge couplings. In order to grasp the relevant effect, we consider all LQs already studied in this paper and, for each of 
them, we compute the shifts induced to the one-loop $\beta$ function coefficients $\delta b_i$, where $i=1,2,3$ refers to the $SU(3)$,
$\rm SU(2)_L$ and $\rm U(1)$ gauge groups \cite{Koh:1983ir,Jones:1981we}. For the sake of simplicity, we only take into account the 
contributions of LQs, no matter of whether we are dealing with a singlet or 2HDM scenarios. 
The {\it amount} of unification at one-loop is quantified by the parameter $A$, which is defined as the ratio between 
the area of the GUT triangle in one particular model (we take $M_{\LQ} = 1\ \mathrm{TeV}$), 
and that for the SM; for comparison, the latter is given by $A_{\SM} = 5.87 \times 10^{14}\ \mathrm{GeV}$. 
Notice that the lesser the $A$, the closer we are to having unification, being $A = 0$ the case of exact unification.
The results for the $\beta$ functions are summarized in tab.~\ref{tab:running_couplings}.

\begin{table}[h!]
\begin{center}
\begin{tabular}{c  c c c c c c c}
\toprule
\toprule
$\LQ$  			  & $D_{SU(3)}$  & $D_{SU(2)}$	 & $\delta b_1$	& $\delta b_2$ 	& $\delta b_3$ 	& $A$\\
\midrule
$(\mbf{3}, \mbf{2},7/6)$  & $1/2$	 & $1/2$	& $49/30$ 	& $1/2$ 	& $1/3$		& $0.5$\\ 
\midrule
$(\mbf{3}, \mbf{1},1/3)$  & $1/2$	 & $0$		& $1/15$	& $0$ 		& $1/6$		& $5.3$\\ 
$(\mbf{3}, \mbf{3}, 1/3)$ & $1/2$	 & $2$		& $1/5$ 	& $2$ 		& $1/2$		& $2.3 \times 10^{-2}$\\ 
$(\mbfb{3}, \mbf{3}, -4/3)$& $1/2$	 & $2$		& $16/5$ 	& $2$ 		& $1/2$ 	& $6.8 \times 10^{-5}$\\
$(\mbf{6}, \mbf{2}, 7/6)$ & $5/2$	 & $1/2$	& $49/15$	& $1$ 		& $5/3$		& $8.6 \times 10^{3}$\\   
\bottomrule
\bottomrule
\end{tabular}
\caption{\it Dynkin indexes $D$, corrections $\delta b$ to the $\beta$-function of $\rm U(1)_Y$, $\rm SU(2)_L$ and $\rm SU(3)_c$ gauge couplings 
for LQs in $\mbf{45}_H$ (upper) and $\mbf{70}_H$ (lower) and the value of the GUT area ratio $A$. 
We define $b_j \equiv b_j^{\SM} + \delta b_j$ and $b^{\SM} = \{ 41/10, -19/6, -7\}$. 
$b_1$ has been normalized to the usual $\sqrt{5/3}$ factor.}
\label{tab:running_couplings}
\end{center}
\end{table}

From this analysis it is clear that the best candidate is $(\mbf{\bar 3}, \mbf{3}, -4/3)$, while $(\mbf{3}, \mbf{3}, 1/3)$ is also 
a good possibility. We stress again that this exercise has to be understood as a very simplistic and qualitative study, 
as many different combinations of fields up to the GUT scale are possible, yielding a large number of possibilities.

\section{Concluding remarks}
\label{sec:discussion}

We have studied different low-energy realizations of $\rm SU(5)$, involving new scalars and leptoquarks at the TeV scale. We have shown that they can be used to address different anomalies. In particular, we have focus on analyzing ways to explain the diphoton excess within an $\rm SU(5)$ framework, each of which leading to a different phenomenology. We have also studied the possibility to have Higgs flavor violation. In addition to studying the different scenarios, we have tried to make definite predictions for the different cases in order to pinpoint the underlying physics beneath the excess, should it be confirmed. We list in the following some concluding remarks and differences of the possible set-ups:
\begin{itemize}
\item The first type of models with the singlet/triplet/degenerate cases have a large phenomenology. 
Pair production of the triplets and octets at the TeV scale are a clear signature to test them.
\item The second type of models with an effective 2HDM can come from either another $\mbf{5}_H$, or a $\mbf{45}_H$ or a $\mbf{70}_H$. 
The $\mbf{45}_H$ is well motivated by down-quark and charged-lepton masses, in the well-known Georgi-Jarlskog model.
The $\mbf{70}_H$ does not couple to fermions at the renormalizable level  and thus naturally evades FCNC, leading to a type-I 2HDM. 
\item Both cases of the singlet/triplet and the 2HDM in the alignment limit (no decays to light Higgses or gauge bosons at tree level) are allowed. 
In the last cases, decays into SM fermions are predicted which, depending on the model and $t_\beta$, are predominantly into tops, b's and/or taus. 
These are absent in the singlet/triplet case, and serve as a clear discriminant of both scenarios.
\item Obtaining a large width is possible both in the case of singlet/triplet/degenerate for some configurations 
(see the left panel of fig.~\ref{fig:region_c70_MLQ_LowMass}), and in the type-III 2HDM, while this is not the case for type-I 2HDM, even if both the 
CP-even and the CP-odd are almost degenerate in mass and contribute significantly to the rate. 
\item For the 2HDM the parameter space able to explain the excess is smaller (also because of the constraint imposed by $\mu^h_{\gamma \gamma}$) 
as the main decays come from low-energy dimension 6 operators, to be compared 
with the dimension 5 ones of the singlet.
\item Regarding fine-tuning, whenever non-supersymmetric $\rm SU(5)$ is present there are un-natural couplings
that must be tuned in order to tackle 
the 
doublet-triplet splitting and/or the hierarchy problem. In the case of the singlet, the new fine-tuning sources are shown in 
fig.~\ref{fig:bar_plot_FT}, and lie below the typical $\rm SU(5)$ ones. For the 2HDM, for any of the $\mbf{5}_H$ or 
$\mbf{45}_H$  representations, the particles that mediate proton decay will need to be at the GUT scale, 
while the doublets need to be at the TeV scale. 
This is another source contributing to the doublet-triplet splitting problem.
Notice that none of the  LQs of the $\mbf{70}_H$ mediate proton decay at tree level but, for some of them like the $(\mbf{\bar 3},\mbf{3},-4/3)$, 
possible loop level contributions may arise, thus their mass can not be too light or the relevant couplings 
must be somehow suppressed.
\item In the  models discussed in sec.~\ref{sec:singlet}, in addition to dijet events, we expect significant decays into gauge bosons, 
all of which cannot be simultaneously reduced. This can help to pinpoint the underlying model. For instance, for singlets, 
we expect decays into 
$WW, ZZ, \gamma Z$, see tab.~\ref{tab:coupling_representation_45_70}. For triplets, decays into $W\gamma$ and $WZ$ are present, 
see tab.~\ref{tab:coupling_representation_45_70_triplet}. And in the case of degenerate spectrum with a light octet, correlated 
decays into photon$+$jet and Z$+$jet can be searched for, see tab.~\ref{tab:coupling_representation_45_70_degenerate_mass_spectrum}.
\item For the 2HDM, the decays into SM fermions and the anti-correlation among the $\gamma\gamma$ and Z$+$photon are the most striking signatures. 
Furthermore, when the 
resolution is larger, both a CP-even and CP-odd almost degenerate in mass could be disentangled.
\item From a measurement of the decay rates, one could disentangle whether the representation behind the resonance is a $\mbf{45}_H$ or a $\mbf{70}_H$. This is due to two reasons: the decays of the resonance have a different dependence on $t_\beta$, and furthermore $t_\beta$ is not the same in the different scenarios, as it depends on the $\rm SU(5)$ representation ($\mbf{45}_H$ or $\mbf{70}_H$) to which the second Higgs doublet belongs. Furthermore, the $\mbf{45}_H$ can have a larger width, while the $\mbf{70}_H$ does not, 
as it can be seen in figs.~\ref{fig:scan_2HDM_45}, \ref{fig:scan_2HDM_70} and \ref{fig:scan_2HDM_AS}.
\item In addition, in the case of the Georgi-Jarlskog model there is flavor violation in Higgs decays. However, we have found that $B_s$ mixing limits imply 
$\Br(h\rightarrow \tau \mu,\,bs)\lesssim 10^{-5}$, beyond any expected sensitivity. Thus a confirmation of a $1\%$ $\Br(h\rightarrow \tau \mu)$ at the LHC would rule-out this model as an explanation, at least in its minimal version.
\end{itemize}

As a last remark, we want to emphasize again that, whether the diphoton excess will be confirmed or not by future LHC data, 
the analysis performed in this paper will remain a useful study of low-energy realizations of $\rm SU(5)$, with many phenomenological implications in different
sectors, ranging from the phenomenology of low-mass colored states to that of Higgs flavour violating interactions.

\subsection*{Acknowledgments}

JHG wants to thank Andreas Crivellin for useful discussions on $h\rightarrow bs$. ADI thanks Gabriele Ria for different numerical checks and useful discussions. We would also like to acknowledge partial support by the DFG Cluster of Excellence `Origin and Structure of the Universe' SEED project ''Neutrino mass generation mechanisms in (grand) unified flavor models and phenomenological imprints".

\appendix
\section{Decay widths}
\label{sec:LoopFunctionsAppendix}
For the decay of $S$ in two photon we use same convention of ref.~\cite{Carena:2012xa} for the loop functions,
that we report here for completeness:
\begin{align}
 \Gamma(S \to \gamma \gamma) = \frac{\alpha_{em}^2G_F m_S^3}{128 \sqrt{2} \pi^3} \bigg| \sum_f N_f Q_f^2\frac{g_{S\overline{f}f}}{ m_f}A_{1/2}(\tau_f)  + \sum_{\LQ}  \frac{g_{S \LQ \LQ}}{M_{\LQ}^2} g_{\gamma\gamma}^{\LQ}  A_0(\tau_{\LQ})\bigg|^2\,,
\end{align}
where the loop functions $A_i$ are discussed below in \ref{sec:LoopFunctions}. The couplings $g_{S\Sigma\Sigma}$ and $g_{S \LQ\LQ}$ 
can be obtained from the potential. Couplings to fermions are zero for the case $S = \Sigma_0, \Sigma_3^0$ but are relevant for 2HDM. 
Here and in the following $\alpha_{em}$ is the fine-structure constant, $\alpha_S$ the strong coupling constant, $G_F$ the Fermi constant
and $\tau_j \equiv 4m_j^2/m_S^2$. In our numerical study, we assume $\alpha_{em}(m_W) = 1/128$ and $\alpha_S(m_Z) = 0.1185$.\\
In the case of $S \to gg$ the relation is quite similar:
\begin{align}
  \Gamma(S \to gg) = \frac{\alpha_S^2G_F m_S^3}{36 \sqrt{2} \pi^3}  \bigg|  \sum_f N_f Q_f^2\frac{g_{S\overline{f}f}}{m_f}A_{1/2}(\tau_f) 
  + \sum_{\LQ}  \frac{g_{S \LQ \LQ}}{M_{\LQ}^2} g_{gg}^{\LQ}  A_0(\tau_{\LQ})\bigg|^2\,.
\end{align}
Another interesting process that occurs at loop level is the decay into $Z$ and $\gamma$, in this case the decay width for $S$ is:
\begin{align}
  \Gamma(S \to Z\gamma) &= \frac{\alpha_{em}^2G_Fm_S^3}{64\sqrt{2}\pi^3}\left[1 - \dfrac{m_Z^2}{m_S^2}\right]^3\bigg| \sum_f N_f 
  Q_f\frac{g_{S\overline{f}f}}{m_f}\frac{2T_3^{(f)}-4s_W^2 Q_f}{s_W c_W}A_{1/2}(\tau_f, \lambda_f)+
  \nextline-2\sum_{\LQ}  \frac{g_{S \LQ \LQ}}{M_{\LQ}^2} g_{Z\gamma}^{\LQ}  A_0(\tau_{\LQ}, \lambda_{\LQ})\bigg|^2\,.
\end{align}
We expect that the decay into $Z\gamma$ is suppressed with respect to the $\gamma\gamma$ process also because of the loop function 
$A_0(\tau, \lambda)$, where $\lambda_j \equiv 4m_j^2/m_Z^2$ since, for  $m_Z \to 0$, we have 
$A_0(\tau) \to 2 A_0(\tau, \lambda)$ (see ref.~\cite{Chen:2013vi} for further details).\\
The decay width into $W^+W^-$ can be obtained using the effective field theory and it reads \cite{Low:2015qho}:
\begin{align}
\label{DecayS_WW}
 \Gamma(S \to W^+ W^-) \simeq \frac{\alpha_{em}^2 G_F m_S^3}{256 \sqrt{2} \pi^3}\left|\sum_{\LQ} 
 \frac{g_{S\LQ \LQ}}{M_{\LQ}^2} g_{WW}^{\LQ}  \right|^2 \times \ord(1)\,,
\end{align}
where the $\ord(1)$ represent the loop contribution. A similar analysis occurs for the decay in $ZZ$. 
If $M_{\LQ} \lesssim m_S/2$ it is possible to have the decay $S$ into a LQ pair. From the effective operator in 
eq.~\eqref{effective_operators_LQs} we get~\cite{Djouadi:2005gj} :
\begin{align}
\label{S_LQLQ_Width}
 \Gamma(S \to  \overline{\LQ} \LQ ) = d_c|c_{\LQ}^{\mbf{r}}|^2\frac{ m_S}{32\pi}\sqrt{1 - \frac{4M_{\LQ}^2}{m_S^2}}\,.
\end{align}

As done in the case of $S$ we can obtain the loop mediated decay widths for $\Sigma_8$. These are:
\begin{align} 
\Gamma(\Sigma_8 \to \gamma g) &\simeq \frac{\alpha_{em} \alpha_SG_F m_8^3}{8\times128 \sqrt{2} \pi^3} \bigg| \sum_{\LQ}  \frac{g_{\Sigma_8 \LQ \LQ}}{M_{\LQ}^2} g_{\gamma g}^{\LQ}  A_0(\overline{\tau}_{\LQ})\bigg|^2\,, \label{eq:S8ggamma} 
\nline
 \Gamma(\Sigma_8  \to Zg) &= \frac{\alpha_{em} \alpha_SG_Fm_8^3}{8\times64\sqrt{2}\pi^3}\left[1 - \dfrac{m_Z^2}{m_8^2}\right]^3\bigg|-2 \sum_{\LQ}  \frac{g_{\Sigma_8 \LQ \LQ}}{M_{\LQ}^2} g_{Zg}^{\LQ}  A_0(\overline{\tau}_{\LQ}, \lambda_{\LQ})\bigg|^2\,, \label{eq:S8gZ}
\end{align}
where $\overline{\tau}_j \equiv 4 m_j^2/m_8^2$. 
The factor of $8$ in the denominator takes into account the average over the initial states. In order to consider the right colour factor for the decay $\Sigma_8 \to gg$ we have to recast the decay width $\Gamma(S \to gg)$. The rescaling factor $\rho_\LQ$ for a single LQ in representation $\mbf{r} \in {\rm SU(3)}$ is given by
\begin{align}
\label{rho_LQ}
 \rho_{\LQ} = \left| \frac{\sum_{\text{colour}}\Tr [\{ T_{\mbf{r}}^a, T_{\mbf{r}}^b\} T_{\mbf{r}}^c ]}{\sum_{\text{colour}}\Tr [ T_{\mbf{r}}^a T_{\mbf{r}}^b ]}\right|^2 = \frac{\kappa^2(\mbf{r}) d^{abc} d_{abc}}{T_R^2(\mbf{r}) \delta^{ab} \delta_{ab}}\,,
\end{align}
where $\kappa(\mbf{r})$ is the coefficient associated to gauge triangle anomaly, $\Tr [\{ T_{\mbf{r}}^a, T_{\mbf{r}}^b\} T_{\mbf{r}}^c ] = \kappa(\mbf{r}) d^{abc}$. It can be easily evaluated with the help of \texttt{Susyno}. The numerator in \eqref{rho_LQ} is given by the two possible diagrams of the decay while the denominator is the usual color structure of a scalar decay into two gluons. We get
\begin{align}
 \rho_\LQ(\mbf{3}) = \frac{7}{12}+ \frac{\sqrt{3}}{8} \simeq 0.80\,, \qquad \rho_\LQ(\mbf{6}) = \frac{49}{600}\left(14 +3\sqrt{3}\right) \simeq 1.57\, .
\end{align}
Hence we have
\begin{align}
 \Gamma(\Sigma_8 \to gg) &\simeq \rho_{\LQ}\frac{\alpha_S^2G_F m_8^3}{8\times36 \sqrt{2} \pi^3}  \bigg| \sum_{\LQ}  \frac{g_{\Sigma_8 \LQ \LQ}}{M_{\LQ}^2} g_{gg}^{\LQ}  A_0(\overline{\tau}_{\LQ})\bigg|^2\,.
 \end{align}
For a LQ in representation $\mbf{r} \in SU(5)$ we have the tree level decay process:
\begin{align}
\label{width_S8_LQLQ}
 \Gamma(\Sigma_8 \to  \overline{\LQ} \LQ ) = \frac{d_c}{8}|\overline{c}_{\LQ}^{\mbf{r}}T_R(\mbf{r})|^2\frac{ m_8}{32\pi}
 \sqrt{1 - \frac{4M_{\LQ}^2}{m_8^2}}\,.
\end{align}

In the context of 2HDM, we need to consider the tree-level decay of $S/A$ 
into a fermion pair. The formulae are \cite{Djouadi:2005gj}:
\begin{align}
\label{SA_decay_fermions}
 \Gamma(S \to \overline{f}f) = d_c|c_{S\overline{f}f}|^2\frac{ m_S }{32\pi}\left[1 - \frac{4m_f^2}{m_S^2}\right]^{3/2}\,,  \qquad \Gamma(A \to \overline{f}f) = d_c|c_{A\overline{f}f}|^2\frac{ m_S }{32\pi}\left[1 - \frac{4m_f^2}{m_S^2}\right]^{1/2}\,,
\end{align}
where $d_c$ is the colour multiplicity of the final state, that is $d_c = 1$ for leptons and $d_c = 3$ for quarks. 

At one loop, there can be decays into gauge bosons. In ref.~\cite{Moretti:2015pbj} 
the authors discussed similar processes and also took into account the interference with the top quark amplitude;
in the  effective theory approach they found:
\begin{align}
\label{sww}
 \Gamma(S \to W^+W^-) \simeq (0.19\ \mathrm{GeV}) \times \frac{|C_{SWW}^{eff}|^2}{m_S^2}\,,
\end{align}
where the dimensionful one-loop induced coupling $C_{SWW}^{eff}$ can be estimated to be proportional to $g_{WW} v^2C_0(m_W^2, m_W^2, m_S^2, M_{\LQ}, M_{\LQ}, M_{\LQ})$,
where $C_0$ is the well-known Passarino-Veltman function. 
Using \texttt{Package-X}~\cite{Patel:2015tea}, we get $C_{SWW}^{eff} \simeq \ord(10^{-6}) \times g_{WW} v^2$ 
for a wide range of values of $M_{\LQ}$ and so  $\Gamma(S \to W^+W^-) = \ord(10^{-12})\ \mathrm{GeV}$ for 
$g_{WW}\sim 10$ while $\Gamma(S \to \gamma \gamma) = \ord(10^{-3})\ \mathrm{GeV}$.

\subsection{Loop functions}
\label{sec:LoopFunctions}
For the decays of $S \to \gamma\gamma/gg$, the scalar, fermion and gauge boson contributions into the loops are: 
\begin{subequations}
\begin{align}
 A_0(\tau) &\equiv -\tau^2 \left[\tau^{-1} - f(\tau^{-1})\right]\,, \nline
 A_{1/2}(\tau) &\equiv 2\tau^2 \left[  \tau^{-1} + (\tau^{-1} -1) f(\tau^{-1})\right]\,, \nline
 A_1(\tau) &\equiv -\tau^2 \left[ 2 \tau^{-2} + 3 \tau^{-1} +3(2 \tau^{-1} -1) f(\tau^{-1})\right] \,,
\end{align}
\end{subequations}
with $\tau \equiv 4 m^2/m_S^2$ ($m$ is the mass particle in the loop),
while in the case of $S \to Z \gamma$ we have:
\begin{subequations}
\begin{align}
 A_0(x, y) &\equiv  I_1(x, y)\,,\nline
 A_{1/2}(x, y) &\equiv I_1(x, y) - I_2(x, y)\,, \nline
 A_1(x, y) &\equiv 4(3 - \tan^2\theta_W) I_2(x, y) +\left[(1 + 2 x^{-1}) \tan^2\theta_W -(5 + 2 x^{-1}) \right]I_1(x, y)\,,
\end{align}
\end{subequations}
where
\begin{subequations}
\begin{align}
 I_1(x, y)&\equiv \frac{xy}{2(x - y)} + \frac{x^2y^2}{2(x - y)^2} \left[f(x^{-1}) - f(y^{-1}) \right] + \frac{x^2y}{(x-y)^2}\left[g(x^{-1}) - g(y^{-1}) \right]\,,\nline
 I_2(x, y)  &\equiv - \frac{xy}{2(x - y)}\left[f(x^{-1}) - f(y^{-1}) \right]\,.
\end{align}
\end{subequations}
In the case of a CP-odd state the spin-1/2 function for the decay $A \to\gamma\gamma/gg$ 
is different: 
 \begin{align}
 \tilde{A}_{1/2}(\tau) &\equiv 2\tau f(\tau^{-1})\,.
 \end{align}
The functions $g$ and $f$  are explicitly given by~\cite{Azatov:2012rd}:
\begin{subequations}
\begin{align}
 f(\tau)&= \left\{ \begin{array}{l l}
                         \arcsin^2\sqrt{\tau} & \tau \geq 1\,, \\
                         -\dfrac{1}{4} \left[\ln \dfrac{1 + \sqrt{1 + \tau^{-1}}}{1 + \sqrt{1 - \tau^{-1}}} - i \pi \right]^2 & \tau < 1 \,,
                        \end{array}
\right.\nline
 g(\tau)&= \left\{ \begin{array}{l l}
                         \sqrt{\tau^{-1} - 1 }\arcsin\sqrt{\tau} & \tau \geq 1\,, \\
                         \dfrac{1}{2}\sqrt{\tau^{-1} - 1 } \left[\ln \dfrac{1 + \sqrt{1 + \tau^{-1}}}{1 + \sqrt{1 - \tau^{-1}}} - i \pi \right] & \tau < 1 \,,
                        \end{array}
\right. .
\end{align}
\end{subequations}
In the limit of large loop masses (compared to the scalar one) we have 
$A_0 \to 1/3$, $A_{1/2} \to 4/3$ and $A_1 \to -7$. For  pseudoscalar particles,
we have $\tilde{A}_{1/2} \to 2$. 
In the case of the decay of the SM Higgs, 
we get $A_1(\tau_W) =  -8.32$ and $N_c Q^2_t A_{1/2}(\tau_t) =  1.83$ hence the dominant contribution comes from the $W$.

\section{LHC production mechanisms}

\subsection{$S/\Sigma_8$ production}
\label{sec:LHC_production}

The total signal for the particle responsible of the diphoton excess is
\begin{align}
\label{cross_section_theo}
 \sigma(pp\to S \to \gamma\gamma) = K_{gg} C_{gg}(\mu_F = m_S) \frac{\Gamma(S \to gg)}{m_S s} \Br(S \to \gamma\gamma) + 
 \sigma_{\gamma\gamma}\,,
\end{align}
where the photoproduction $\sigma_{\gamma \gamma}$ can be expressed as \cite{Csaki:2015vek}
\begin{align}
\label{cross_section_photon}
 \sigma_{\gamma\gamma} = 10.8 \ \mathrm{pb} \ \left( \frac{\Gamma_S}{45\ \mathrm{GeV}}\right)\Br^2(S \to \gamma\gamma)\,.
\end{align}
In eq.~\eqref{cross_section_theo} we introduce the parameter $C_{gg}$, that is given by the gluon PDF $f_g(x; \mu_F^2)$ of the proton at the factorization scale $\mu_F$
\begin{align}
\label{Cgg_gluon_PDF}
 C_{gg} = \frac{\pi^2}{8}\int_{m_S^2/s}^{1}\frac{\de x}{x}f_g(x; \mu_F^2)f_g\left(\frac{m_S^2}{xs}; \mu_F^2\right)\,.
\end{align}
The values used in our paper and obtained with the PDF set \texttt{mstw2008nlo}, see ref.~\cite{Martin:2009iq}, are reported in tab.~\ref{tab:cgg}.
\begin{table}[h!]
\begin{center}
\begin{tabular}{c c c}
\toprule
\toprule
			  & $C_{gg} @\ \sqrt{s} = 8\ \mathrm{TeV}$ & $C_{gg} @\ \sqrt{s} = 13\ \mathrm{TeV}$\\
\midrule
$\mu_F = 730\ \mathrm{GeV}$   & 202	 &  2445 \\ 
\midrule
$\mu_F = 750\ \mathrm{GeV}$   & 174 	 & 2137\\   
\bottomrule
\bottomrule
\end{tabular}
\caption{\it $C_{gg}$ coefficients at the two scales $\mu_F = 730\ \mathrm{GeV}$ and $\mu_F = 750\ \mathrm{GeV}$.}
\label{tab:cgg}
\end{center}
\end{table}
The factor $K_{gg}$ is introduced to taking into account QCD corrections, the typical value is $K_{gg} \simeq 1.48$.\\
A relation similar to eq.~\eqref{cross_section_theo} also holds for $\Sigma_8$; for any possible final state $XY$ we have
\begin{align}
\label{cross_section_theo_S8}
 \sigma(pp \to \Sigma_8 \to XY) = K_{gg}C_{gg}(\mu_F = m_8) \frac{\Gamma(\Sigma_8 \to gg)}{m_8 s}\Br(\Sigma_8 \to XY)\,.
\end{align}

\subsection{Scalar pair production}
\label{sec:LQAppendix}
Let us consider the pair production of 
LQs or $\Sigma_8$. In the following $P = \{\LQ, \Sigma_8\}$ and $M_{P}$ is the related mass. 
The partonic cross sections are \cite{Eichten:1984eu, Manohar:2006ga}:
\begin{subequations}
\begin{align}
 \hat{\sigma}(\overline{q} q \to \overline{P} P) &= \alpha_S^2\xi C(\mbf{r}) d \frac{\pi}{54 \hat{s}} \beta_{P}^3\,,\label{partonic_cross_section_pair} \nline
\hat{\sigma}(gg \to \overline{P} P) &=  \alpha_S^2\xi \frac{C(\mbf{r})^2}{d}\frac{\pi}{6 \hat{s}} \left[27 \beta_{P} -17 \beta_{P}^3 + 3(\beta_{P}^4+2\beta_{P}^2-3)\ln \frac{1+ \beta_{P}}{1-\beta_{P}} \right] \label{partonic_cross_section_pairb}\,,
\end{align}
\end{subequations}
where $\beta_{P} \equiv \sqrt{1 + 4M_{P}^2/\hat{s}}$ is the velocity of $P$ in the center of mass frame and $\hat{s}$ is the partonic energy, $d = d_c \times d_L$ is the number of states in representation $\rm (\mbf{r}, \mbf{r'}) \in SU(3)_c \otimes \rm SU(2)_L$ and $C(\mbf{r})$ is the Casimir invariant, we have $C(\mbf{3}) = 4/3$, $C(\mbf{6}) = 10/3$, $C(\mbf{8}) = 3$ and $C(\mbf{15}) = 16/3$. The factor $\xi$ is the multiplicity, which is equal to $1/2$ for real representations of all quantum numbers or one otherwise. The total cross section is given by the sum of the partial cross sections
\begin{align}
 \sigma(pp \to \overline{P} P) = K_{\overline{q}q}\sigma(\overline{q} q \to \overline{P} P) + K_{gg}\sigma(gg \to \overline{P} P)\,,
\end{align}
where typical value of the $K$-factor for quarks is $K_{\overline{q}q} \sim 1.2$ ($K_{gg} \simeq 1.48$ for gluons). 
The integrated partonic cross sections are \cite{Quigg:2009gg}
\begin{subequations}
\label{integrated_partonic_cross_section}
\begin{align}
\sigma(gg \to \overline{P} P) &= \int_{4M_{P}^2/s}^{1}\frac{\de x}{x} \left[\frac{\tau}{\hat{s}}
\frac{\de \mathcal{L}_{gg}}{\de \tau}\right]\left[\hat{s} \hat{\sigma}(gg \to \overline{P} P)\right]\,, \nline
 \sigma(\overline{q} q \to\overline{P} P) &= \sum_{ \overline{q}q}\int_{4M_{P}^2/s}^{1}\frac{\de x}{x}
 \left[\frac{\tau}{\hat{s}}\frac{\de \mathcal{L}_{\overline{q}q}}{\de \tau}\right]\left[\hat{s} 
 \hat{\sigma}(\overline{q} q \to \overline{P} P)\right]\,,
\end{align}
\end{subequations}
where the parton luminosity for partons $i$ and $j$ is defined as:
\begin{align}
\label{partonic_luminosity}
 \frac{\tau}{\hat{s}}\frac{\de \mathcal{L}_{ij}}{\de \tau} = \frac{\tau/\hat{s}}{1 + \delta_{ij}} \int_{\tau}^1 \frac{\de x}{x} 
 \left[f_i(x; \mu_F^2) f_j(\tau/x; \mu_F^2) + f_i(\tau/x; \mu_F^2) f_j(x; \mu_F^2)\right]\,, \qquad \tau \equiv \frac{\hat{s}}{s}.
\end{align}



\bibliographystyle{elsarticle-num} 
\bibliography{manuscript}

\begin{thebibliography}{10}
\expandafter\ifx\csname url\endcsname\relax
  \def\url#1{\texttt{#1}}\fi
\expandafter\ifx\csname urlprefix\endcsname\relax\def\urlprefix{URL }\fi
\expandafter\ifx\csname href\endcsname\relax
  \def\href#1#2{#2} \def\path#1{#1}\fi

\bibitem{Aad:2015mna}
G.~Aad, et~al., {Search for high-mass diphoton resonances in $pp$ collisions at
  $\sqrt{s}=8$ TeV with the ATLAS detector}, Phys. Rev. D92~(3) (2015) 032004.
\newblock \href {http://arxiv.org/abs/1504.05511} {\path{arXiv:1504.05511}},
  \href {http://dx.doi.org/10.1103/PhysRevD.92.032004}
  {\path{doi:10.1103/PhysRevD.92.032004}}.

\bibitem{ATLASconf2}
{The ATLAS Collaboration}, Atlas-conf-2016-018.

\bibitem{ATLASconf}
{The ATLAS Collaboration}, Atlas-conf-2015-081.

\bibitem{Aaboud:2016tru}
M.~Aaboud, et~al., {Search for resonances in diphoton events at $\sqrt{s}$=13
  TeV with the ATLAS detector}\href {http://arxiv.org/abs/1606.03833}
  {\path{arXiv:1606.03833}}.

\bibitem{CMSconf}
{The CMS Collaboration}, Cms-pas-exo-15-004.

\bibitem{CMSconf2}
{The CMS Collaboration}, Cms-pas-exo-16-018.

\bibitem{Aad:2016blu}
G.~Aad, et~al., {Search for lepton-flavour-violating decays of the Higgs and
  $Z$ bosons with the ATLAS detector}\href {http://arxiv.org/abs/1604.07730}
  {\path{arXiv:1604.07730}}.

\bibitem{Khachatryan:2015kon}
V.~Khachatryan, et~al., {Search for Lepton-Flavour-Violating Decays of the
  Higgs Boson}, Phys. Lett. B749 (2015) 337--362.
\newblock \href {http://arxiv.org/abs/1502.07400} {\path{arXiv:1502.07400}},
  \href {http://dx.doi.org/10.1016/j.physletb.2015.07.053}
  {\path{doi:10.1016/j.physletb.2015.07.053}}.

\bibitem{Franceschini:2015kwy}
R.~Franceschini, G.~F. Giudice, J.~F. Kamenik, M.~McCullough, A.~Pomarol,
  R.~Rattazzi, M.~Redi, F.~Riva, A.~Strumia, R.~Torre, {What is the $\gamma
  \gamma$ resonance at 750 GeV?}, JHEP 03 (2016) 144.
\newblock \href {http://arxiv.org/abs/1512.04933} {\path{arXiv:1512.04933}},
  \href {http://dx.doi.org/10.1007/JHEP03(2016)144}
  {\path{doi:10.1007/JHEP03(2016)144}}.

\bibitem{Kamenik:2016tuv}
J.~F. Kamenik, B.~R. Safdi, Y.~Soreq, J.~Zupan, {Comments on the diphoton
  excess: critical reappraisal of effective field theory interpretations}\href
  {http://arxiv.org/abs/1603.06566} {\path{arXiv:1603.06566}}.

\bibitem{Ellis:2015oso}
J.~Ellis, S.~A.~R. Ellis, J.~Quevillon, V.~Sanz, T.~You, {On the Interpretation
  of a Possible $\sim 750$ GeV Particle Decaying into $\gamma \gamma$}, JHEP 03
  (2016) 176.
\newblock \href {http://arxiv.org/abs/1512.05327} {\path{arXiv:1512.05327}},
  \href {http://dx.doi.org/10.1007/JHEP03(2016)176}
  {\path{doi:10.1007/JHEP03(2016)176}}.

\bibitem{Low:2015qho}
I.~Low, J.~Lykken, {Implications of Gauge Invariance on a Heavy Diphoton
  Resonance}\href {http://arxiv.org/abs/1512.09089} {\path{arXiv:1512.09089}}.

\bibitem{Buttazzo:2015txu}
D.~Buttazzo, A.~Greljo, D.~Marzocca, {Knocking on new physics' door with a
  scalar resonance}, Eur. Phys. J. C76~(3) (2016) 116.
\newblock \href {http://arxiv.org/abs/1512.04929} {\path{arXiv:1512.04929}},
  \href {http://dx.doi.org/10.1140/epjc/s10052-016-3970-7}
  {\path{doi:10.1140/epjc/s10052-016-3970-7}}.

\bibitem{Gupta:2015zzs}
R.~S. Gupta, S.~J{\"a}ger, Y.~Kats, G.~Perez, E.~Stamou, {Interpreting a 750
  GeV Diphoton Resonance}\href {http://arxiv.org/abs/1512.05332}
  {\path{arXiv:1512.05332}}.

\bibitem{Dev:2015vjd}
P.~S.~B. Dev, R.~N. Mohapatra, Y.~Zhang, {Quark Seesaw, Vectorlike Fermions and
  Diphoton Excess}, JHEP 02 (2016) 186.
\newblock \href {http://arxiv.org/abs/1512.08507} {\path{arXiv:1512.08507}},
  \href {http://dx.doi.org/10.1007/JHEP02(2016)186}
  {\path{doi:10.1007/JHEP02(2016)186}}.

\bibitem{Angelescu:2015uiz}
A.~Angelescu, A.~Djouadi, G.~Moreau, {Scenarii for interpretations of the LHC
  diphoton excess: two Higgs doublets and vector-like quarks and leptons},
  Phys. Lett. B756 (2016) 126--132.
\newblock \href {http://arxiv.org/abs/1512.04921} {\path{arXiv:1512.04921}},
  \href {http://dx.doi.org/10.1016/j.physletb.2016.02.064}
  {\path{doi:10.1016/j.physletb.2016.02.064}}.

\bibitem{Djouadi:2016eyy}
A.~Djouadi, J.~Ellis, R.~Godbole, J.~Quevillon, {Future Collider Signatures of
  the Possible 750 GeV State}, JHEP 03 (2016) 205.
\newblock \href {http://arxiv.org/abs/1601.03696} {\path{arXiv:1601.03696}},
  \href {http://dx.doi.org/10.1007/JHEP03(2016)205}
  {\path{doi:10.1007/JHEP03(2016)205}}.

\bibitem{Bauer:2015boy}
M.~Bauer, M.~Neubert, {Flavor Anomalies, the Diphoton Excess and a Dark Matter
  Candidate}\href {http://arxiv.org/abs/1512.06828} {\path{arXiv:1512.06828}}.

\bibitem{Murphy:2015kag}
C.~W. Murphy, {Vector Leptoquarks and the 750 GeV Diphoton Resonance at the
  LHC}, Phys. Lett. B757 (2016) 192--198.
\newblock \href {http://arxiv.org/abs/1512.06976} {\path{arXiv:1512.06976}},
  \href {http://dx.doi.org/10.1016/j.physletb.2016.03.076}
  {\path{doi:10.1016/j.physletb.2016.03.076}}.

\bibitem{Hati:2016thk}
C.~Hati, {Explaining the diphoton excess in Alternative Left-Right Symmetric
  Model}, Phys. Rev. D93~(7) (2016) 075002.
\newblock \href {http://arxiv.org/abs/1601.02457} {\path{arXiv:1601.02457}},
  \href {http://dx.doi.org/10.1103/PhysRevD.93.075002}
  {\path{doi:10.1103/PhysRevD.93.075002}}.

\bibitem{Dey:2016sht}
U.~K. Dey, S.~Mohanty, G.~Tomar, {Leptoquarks: 750 GeV Diphoton Resonance and
  IceCube Events}\href {http://arxiv.org/abs/1606.07903}
  {\path{arXiv:1606.07903}}.

\bibitem{Strumia:2016wys}
A.~Strumia,
  \href{http://inspirehep.net/record/1466435/files/arXiv:1605.09401.pdf}{{Interpreting
  the 750 GeV digamma excess: a review}}, 2016.
\newblock \href {http://arxiv.org/abs/1605.09401} {\path{arXiv:1605.09401}}.
\newline\urlprefix\url{http://inspirehep.net/record/1466435/files/arXiv:1605.09401.pdf}

\bibitem{Georgi:1974sy}
H.~Georgi, S.~L. Glashow, {Unity of All Elementary Particle Forces}, Phys. Rev.
  Lett. 32 (1974) 438--441.
\newblock \href {http://dx.doi.org/10.1103/PhysRevLett.32.438}
  {\path{doi:10.1103/PhysRevLett.32.438}}.

\bibitem{Dorsner:2016ypw}
I.~Dorsner, S.~Fajfer, N.~Kosnik, {Is symmetry breaking of SU(5) theory
  responsible for the diphoton excess?}\href {http://arxiv.org/abs/1601.03267}
  {\path{arXiv:1601.03267}}.

\bibitem{Giveon:1991zm}
A.~Giveon, L.~J. Hall, U.~Sarid, {SU(5) unification revisited}, Phys. Lett.
  B271 (1991) 138--144.
\newblock \href {http://dx.doi.org/10.1016/0370-2693(91)91289-8}
  {\path{doi:10.1016/0370-2693(91)91289-8}}.

\bibitem{Patel:2015ulo}
K.~M. Patel, P.~Sharma, {Interpreting 750 GeV diphoton excess in SU(5) grand
  unified theory}, Phys. Lett. B757 (2016) 282--288.
\newblock \href {http://arxiv.org/abs/1512.07468} {\path{arXiv:1512.07468}},
  \href {http://dx.doi.org/10.1016/j.physletb.2016.04.006}
  {\path{doi:10.1016/j.physletb.2016.04.006}}.

\bibitem{Dutta:2016jqn}
B.~Dutta, Y.~Gao, T.~Ghosh, I.~Gogoladze, T.~Li, Q.~Shafi, J.~W. Walker,
  {Diphoton Excess in Consistent Supersymmetric SU(5) Models with Vector-like
  Particles}\href {http://arxiv.org/abs/1601.00866} {\path{arXiv:1601.00866}}.

\bibitem{Dorsner:2012pp}
I.~Dorsner, S.~Fajfer, A.~Greljo, J.~F. Kamenik, {Higgs Uncovering Light Scalar
  Remnants of High Scale Matter Unification}, JHEP 11 (2012) 130.
\newblock \href {http://arxiv.org/abs/1208.1266} {\path{arXiv:1208.1266}},
  \href {http://dx.doi.org/10.1007/JHEP11(2012)130}
  {\path{doi:10.1007/JHEP11(2012)130}}.

\bibitem{Bai:2010dj}
Y.~Bai, B.~A. Dobrescu, {Heavy octets and Tevatron signals with three or four b
  jets}, JHEP 1107 (2011) 100.
\newblock \href {http://arxiv.org/abs/1012.5814} {\path{arXiv:1012.5814}},
  \href {http://dx.doi.org/10.1007/JHEP07(2011)100}
  {\path{doi:10.1007/JHEP07(2011)100}}.

\bibitem{Badziak:2015zez}
M.~Badziak, {Interpreting the 750 GeV diphoton excess in minimal extensions of
  Two-Higgs-Doublet models}\href {http://arxiv.org/abs/1512.07497}
  {\path{arXiv:1512.07497}}.

\bibitem{Georgi:1979df}
H.~Georgi, C.~Jarlskog, {A New Lepton - Quark Mass Relation in a Unified
  Theory}, Phys. Lett. B86 (1979) 297--300.
\newblock \href {http://dx.doi.org/10.1016/0370-2693(79)90842-6}
  {\path{doi:10.1016/0370-2693(79)90842-6}}.

\bibitem{Guth:1981uk}
A.~H. Guth, E.~J. Weinberg, {Cosmological Consequences of a First Order Phase
  Transition in the SU(5) Grand Unified Model}, Phys. Rev. D23 (1981) 876.
\newblock \href {http://dx.doi.org/10.1103/PhysRevD.23.876}
  {\path{doi:10.1103/PhysRevD.23.876}}.

\bibitem{Guth:1979bh}
A.~H. Guth, S.~H.~H. Tye, {Phase Transitions and Magnetic Monopole Production
  in the Very Early Universe}, Phys. Rev. Lett. 44 (1980) 631, [Erratum: Phys.
  Rev. Lett.44,963(1980)].
\newblock \href {http://dx.doi.org/10.1103/PhysRevLett.44.631}
  {\path{doi:10.1103/PhysRevLett.44.631}}.

\bibitem{Buras:1977yy}
A.~J. Buras, J.~R. Ellis, M.~K. Gaillard, D.~V. Nanopoulos, {Aspects of the
  Grand Unification of Strong, Weak and Electromagnetic Interactions}, Nucl.
  Phys. B135 (1978) 66--92.
\newblock \href {http://dx.doi.org/10.1016/0550-3213(78)90214-6}
  {\path{doi:10.1016/0550-3213(78)90214-6}}.

\bibitem{Casas:1996de}
J.~A. Casas, S.~Dimopoulos, {Stability bounds on flavor violating trilinear
  soft terms in the MSSM}, Phys. Lett. B387 (1996) 107--112.
\newblock \href {http://arxiv.org/abs/hep-ph/9606237}
  {\path{arXiv:hep-ph/9606237}}, \href
  {http://dx.doi.org/10.1016/0370-2693(96)01000-3}
  {\path{doi:10.1016/0370-2693(96)01000-3}}.

\bibitem{Sierra:2015zma}
D.~Aristizabal~Sierra, J.~Herrero-Garcia, D.~Restrepo, A.~Vicente, {Diboson
  anomaly: Heavy Higgs resonance and QCD vectorlike exotics}, Phys. Rev. D93
  (2016) 015012.
\newblock \href {http://arxiv.org/abs/1510.03437} {\path{arXiv:1510.03437}},
  \href {http://dx.doi.org/10.1103/PhysRevD.93.015012}
  {\path{doi:10.1103/PhysRevD.93.015012}}.

\bibitem{Carena:2012xa}
M.~Carena, I.~Low, C.~E.~M. Wagner, {Implications of a Modified Higgs to
  Diphoton Decay Width}, JHEP 1208 (2012) 060.
\newblock \href {http://arxiv.org/abs/1206.1082} {\path{arXiv:1206.1082}},
  \href {http://dx.doi.org/10.1007/JHEP08(2012)060}
  {\path{doi:10.1007/JHEP08(2012)060}}.

\bibitem{Csaki:2015vek}
C.~Csaki, J.~Hubisz, J.~Terning, {The Minimal Model of a Diphoton Resonance:
  Production without Gluon Couplings}, Phys. Rev. D93 (2016) 035002.
\newblock \href {http://arxiv.org/abs/1512.05776} {\path{arXiv:1512.05776}},
  \href {http://dx.doi.org/10.1103/PhysRevD.93.035002}
  {\path{doi:10.1103/PhysRevD.93.035002}}.

\bibitem{Dorsner:2012nq}
I.~Dorsner, S.~Fajfer, N.~Kosnik, {Heavy and light scalar leptoquarks in proton
  decay}, Phys. Rev. D86 (2012) 015013.
\newblock \href {http://arxiv.org/abs/1204.0674} {\path{arXiv:1204.0674}},
  \href {http://dx.doi.org/10.1103/PhysRevD.86.015013}
  {\path{doi:10.1103/PhysRevD.86.015013}}.

\bibitem{Djouadi:2005gi}
A.~Djouadi, {The Anatomy of electro-weak symmetry breaking. I: The Higgs boson
  in the standard model}, Phys. Rept. 457 (2008) 1--216.
\newblock \href {http://arxiv.org/abs/hep-ph/0503172}
  {\path{arXiv:hep-ph/0503172}}, \href
  {http://dx.doi.org/10.1016/j.physrep.2007.10.004}
  {\path{doi:10.1016/j.physrep.2007.10.004}}.

\bibitem{Fonseca:2011sy}
R.~M. Fonseca, {Calculating the renormalisation group equations of a SUSY model
  with Susyno}, Comput. Phys. Commun. 183 (2012) 2298--2306.
\newblock \href {http://arxiv.org/abs/1106.5016} {\path{arXiv:1106.5016}},
  \href {http://dx.doi.org/10.1016/j.cpc.2012.05.017}
  {\path{doi:10.1016/j.cpc.2012.05.017}}.

\bibitem{Asakawa:2006gm}
E.~Asakawa, S.~Kanemura, {The $H^{\pm}W^{\mp} Z^0$ vertex and single charged
  Higgs boson production via W Z fusion at the large hadron collider}, Phys.
  Lett. B626 (2005) 111--119.
\newblock \href {http://arxiv.org/abs/hep-ph/0506310}
  {\path{arXiv:hep-ph/0506310}}, \href
  {http://dx.doi.org/10.1016/j.physletb.2005.08.091}
  {\path{doi:10.1016/j.physletb.2005.08.091}}.

\bibitem{Emmanuel-Costa:2013gia}
D.~Emmanuel-Costa, C.~Simoes, M.~Tortola, {The minimal adjoint-SU(5) x $Z_{4}$
  GUT model}, JHEP 1310 (2013) 054.
\newblock \href {http://arxiv.org/abs/1303.5699} {\path{arXiv:1303.5699}},
  \href {http://dx.doi.org/10.1007/JHEP10(2013)054}
  {\path{doi:10.1007/JHEP10(2013)054}}.

\bibitem{Becirevic:2015fmu}
D.~Becirevic, E.~Bertuzzo, O.~Sumensari, R.~Zukanovich~Funchal, {Can the new
  resonance at LHC be a CP-Odd Higgs boson?}, Phys. Lett. B757 (2016) 261--267.
\newblock \href {http://arxiv.org/abs/1512.05623} {\path{arXiv:1512.05623}},
  \href {http://dx.doi.org/10.1016/j.physletb.2016.03.073}
  {\path{doi:10.1016/j.physletb.2016.03.073}}.

\bibitem{Khalil:2013ixa}
S.~Khalil, S.~Salem, {Enhancement of $H \rightarrow \gamma \gamma$ in SU(5)
  model with $45_H$ plet}, Nucl. Phys. B876 (2013) 473--492.
\newblock \href {http://arxiv.org/abs/1304.3689} {\path{arXiv:1304.3689}},
  \href {http://dx.doi.org/10.1016/j.nuclphysb.2013.08.016}
  {\path{doi:10.1016/j.nuclphysb.2013.08.016}}.

\bibitem{Branco:2011iw}
G.~C. Branco, P.~M. Ferreira, L.~Lavoura, M.~N. Rebelo, M.~Sher, J.~P. Silva,
  {Theory and phenomenology of two-Higgs-doublet models}, Phys. Rept. 516
  (2012) 1--102.
\newblock \href {http://arxiv.org/abs/1106.0034} {\path{arXiv:1106.0034}},
  \href {http://dx.doi.org/10.1016/j.physrep.2012.02.002}
  {\path{doi:10.1016/j.physrep.2012.02.002}}.

\bibitem{ATLAS_CMS_HIGGS_CONFERENCE_2015}
{The ATLAS and CMS Collaborations}, Atlas-conf-2015-044.

\bibitem{Gunion:2002zf}
J.~F. Gunion, H.~E. Haber, {The CP conserving two Higgs doublet model: The
  Approach to the decoupling limit}, Phys. Rev. D67 (2003) 075019.
\newblock \href {http://arxiv.org/abs/hep-ph/0207010}
  {\path{arXiv:hep-ph/0207010}}, \href
  {http://dx.doi.org/10.1103/PhysRevD.67.075019}
  {\path{doi:10.1103/PhysRevD.67.075019}}.

\bibitem{Aad:2014vgg}
G.~Aad, et~al., {Search for neutral Higgs bosons of the minimal supersymmetric
  standard model in pp collisions at $\sqrt{s}$ = 8 TeV with the ATLAS
  detector}, JHEP 11 (2014) 056.
\newblock \href {http://arxiv.org/abs/1409.6064} {\path{arXiv:1409.6064}},
  \href {http://dx.doi.org/10.1007/JHEP11(2014)056}
  {\path{doi:10.1007/JHEP11(2014)056}}.

\bibitem{Khachatryan:2014wca}
V.~Khachatryan, et~al., {Search for neutral MSSM Higgs bosons decaying to a
  pair of tau leptons in pp collisions}, JHEP 10 (2014) 160.
\newblock \href {http://arxiv.org/abs/1408.3316} {\path{arXiv:1408.3316}},
  \href {http://dx.doi.org/10.1007/JHEP10(2014)160}
  {\path{doi:10.1007/JHEP10(2014)160}}.

\bibitem{Aad:2015fna}
G.~Aad, et~al., {A search for $ t\overline{t} $ resonances using
  lepton-plus-jets events in proton-proton collisions at $ \sqrt{s}=8 $ TeV
  with the ATLAS detector}, JHEP 08 (2015) 148.
\newblock \href {http://arxiv.org/abs/1505.07018} {\path{arXiv:1505.07018}},
  \href {http://dx.doi.org/10.1007/JHEP08(2015)148}
  {\path{doi:10.1007/JHEP08(2015)148}}.

\bibitem{Khachatryan:2015sma}
V.~Khachatryan, et~al., {Search for resonant $t \bar t$ production in
  proton-proton collisions at $\sqrt s=$ 8  TeV}, Phys. Rev. D93~(1) (2016)
  012001.
\newblock \href {http://arxiv.org/abs/1506.03062} {\path{arXiv:1506.03062}},
  \href {http://dx.doi.org/10.1103/PhysRevD.93.012001}
  {\path{doi:10.1103/PhysRevD.93.012001}}.

\bibitem{Blankenburg:2012ex}
G.~Blankenburg, J.~Ellis, G.~Isidori, {Flavour-Changing Decays of a 125 GeV
  Higgs-like Particle}, Phys. Lett. B712 (2012) 386--390.
\newblock \href {http://arxiv.org/abs/1202.5704} {\path{arXiv:1202.5704}},
  \href {http://dx.doi.org/10.1016/j.physletb.2012.05.007}
  {\path{doi:10.1016/j.physletb.2012.05.007}}.

\bibitem{Harnik:2012pb}
R.~Harnik, J.~Kopp, J.~Zupan, {Flavor Violating Higgs Decays}, JHEP 03 (2013)
  026.
\newblock \href {http://arxiv.org/abs/1209.1397} {\path{arXiv:1209.1397}},
  \href {http://dx.doi.org/10.1007/JHEP03(2013)026}
  {\path{doi:10.1007/JHEP03(2013)026}}.

\bibitem{Herrero-Garcia:2016uab}
J.~Herrero-Garcia, N.~Rius, A.~Santamaria, {Higgs lepton flavour violation: UV
  completions and connection to neutrino masses}\href
  {http://arxiv.org/abs/1605.06091} {\path{arXiv:1605.06091}}.

\bibitem{Davidson:2010xv}
S.~Davidson, G.~J. Grenier, {Lepton flavour violating Higgs and tau to mu
  gamma}, Phys. Rev. D81 (2010) 095016.
\newblock \href {http://arxiv.org/abs/1001.0434} {\path{arXiv:1001.0434}},
  \href {http://dx.doi.org/10.1103/PhysRevD.81.095016}
  {\path{doi:10.1103/PhysRevD.81.095016}}.

\bibitem{Sierra:2014nqa}
D.~Aristizabal~Sierra, A.~Vicente, {Explaining the CMS Higgs flavor violating
  decay excess}, Phys. Rev. D90~(11) (2014) 115004.
\newblock \href {http://arxiv.org/abs/1409.7690} {\path{arXiv:1409.7690}},
  \href {http://dx.doi.org/10.1103/PhysRevD.90.115004}
  {\path{doi:10.1103/PhysRevD.90.115004}}.

\bibitem{Dorsner:2015mja}
I.~Dor{\v s}ner, S.~Fajfer, A.~Greljo, J.~F. Kamenik, N.~Ko{\v s}nik, I.~Ni{\v
  s}and{\v z}ic, {New Physics Models Facing Lepton Flavor Violating Higgs
  Decays at the Percent Level}, JHEP 06 (2015) 108.
\newblock \href {http://arxiv.org/abs/1502.07784} {\path{arXiv:1502.07784}},
  \href {http://dx.doi.org/10.1007/JHEP06(2015)108}
  {\path{doi:10.1007/JHEP06(2015)108}}.

\bibitem{Omura:2015nja}
Y.~Omura, E.~Senaha, K.~Tobe, {Lepton-flavor-violating Higgs decay $h \to
  \mu\tau$ and muon anomalous magnetic moment in a general two Higgs doublet
  model}, JHEP 05 (2015) 028.
\newblock \href {http://arxiv.org/abs/1502.07824} {\path{arXiv:1502.07824}},
  \href {http://dx.doi.org/10.1007/JHEP05(2015)028}
  {\path{doi:10.1007/JHEP05(2015)028}}.

\bibitem{Botella:2015hoa}
F.~J. Botella, G.~C. Branco, M.~Nebot, M.~N. Rebelo, {Flavour Changing Higgs
  Couplings in a Class of Two Higgs Doublet Models}, Eur. Phys. J. C76~(3)
  (2016) 161.
\newblock \href {http://arxiv.org/abs/1508.05101} {\path{arXiv:1508.05101}},
  \href {http://dx.doi.org/10.1140/epjc/s10052-016-3993-0}
  {\path{doi:10.1140/epjc/s10052-016-3993-0}}.

\bibitem{Bizot:2015qqo}
N.~Bizot, S.~Davidson, M.~Frigerio, J.~L. Kneur, {Two Higgs doublets to explain
  the excesses $pp\rightarrow \gamma\gamma(750\ {\rm GeV})$ and $h \to \tau^\pm
  \mu^\mp$}, JHEP 03 (2016) 073.
\newblock \href {http://arxiv.org/abs/1512.08508} {\path{arXiv:1512.08508}},
  \href {http://dx.doi.org/10.1007/JHEP03(2016)073}
  {\path{doi:10.1007/JHEP03(2016)073}}.

\bibitem{Arganda:2015naa}
E.~Arganda, M.~J. Herrero, X.~Marcano, C.~Weiland, {Enhancement of the lepton
  flavor violating Higgs boson decay rates from SUSY loops in the inverse
  seesaw model}, Phys. Rev. D93~(5) (2016) 055010.
\newblock \href {http://arxiv.org/abs/1508.04623} {\path{arXiv:1508.04623}},
  \href {http://dx.doi.org/10.1103/PhysRevD.93.055010}
  {\path{doi:10.1103/PhysRevD.93.055010}}.

\bibitem{Efrati:2016uuy}
A.~Efrati, J.~F. Kamenik, Y.~Nir, {The phenomenology of the di-photon excess
  and $h\to\tau\mu$ within 2HDM}\href {http://arxiv.org/abs/1606.07082}
  {\path{arXiv:1606.07082}}.

\bibitem{Han:2015qqj}
X.-F. Han, L.~Wang, {Implication of the 750 GeV diphoton resonance on
  two-Higgs-doublet model and its extensions with Higgs field}, Phys. Rev.
  D93~(5) (2016) 055027.
\newblock \href {http://arxiv.org/abs/1512.06587} {\path{arXiv:1512.06587}},
  \href {http://dx.doi.org/10.1103/PhysRevD.93.055027}
  {\path{doi:10.1103/PhysRevD.93.055027}}.

\bibitem{Han:2016bus}
X.-F. Han, L.~Wang, L.~Wu, J.~M. Yang, M.~Zhang, {Explaining 750 GeV diphoton
  excess from top/bottom partner cascade decay in two-Higgs-doublet model
  extension}, Phys. Lett. B756 (2016) 309--316.
\newblock \href {http://arxiv.org/abs/1601.00534} {\path{arXiv:1601.00534}},
  \href {http://dx.doi.org/10.1016/j.physletb.2016.03.035}
  {\path{doi:10.1016/j.physletb.2016.03.035}}.

\bibitem{Han:2016bvl}
X.-F. Han, L.~Wang, J.~M. Yang, {An extension of two-Higgs-doublet model and
  the excesses of 750 GeV diphoton, muon g-2 and $h\to\mu\tau$}, Phys. Lett.
  B757 (2016) 537--547.
\newblock \href {http://arxiv.org/abs/1601.04954} {\path{arXiv:1601.04954}},
  \href {http://dx.doi.org/10.1016/j.physletb.2016.04.036}
  {\path{doi:10.1016/j.physletb.2016.04.036}}.

\bibitem{Chang:1993kw}
D.~Chang, W.~S. Hou, W.-Y. Keung, {Two loop contributions of flavor changing
  neutral Higgs bosons to mu $\to$ e gamma}, Phys. Rev. D48 (1993) 217--224.
\newblock \href {http://arxiv.org/abs/hep-ph/9302267}
  {\path{arXiv:hep-ph/9302267}}, \href
  {http://dx.doi.org/10.1103/PhysRevD.48.217}
  {\path{doi:10.1103/PhysRevD.48.217}}.

\bibitem{Goudelis:2011un}
A.~Goudelis, O.~Lebedev, J.-h. Park, {Higgs-induced lepton flavor violation},
  Phys. Lett. B707 (2012) 369--374.
\newblock \href {http://arxiv.org/abs/1111.1715} {\path{arXiv:1111.1715}},
  \href {http://dx.doi.org/10.1016/j.physletb.2011.12.059}
  {\path{doi:10.1016/j.physletb.2011.12.059}}.

\bibitem{Gupta:2009wn}
R.~S. Gupta, J.~D. Wells, {Next Generation Higgs Bosons: Theory, Constraints
  and Discovery Prospects at the Large Hadron Collider}, Phys. Rev. D81 (2010)
  055012.
\newblock \href {http://arxiv.org/abs/0912.0267} {\path{arXiv:0912.0267}},
  \href {http://dx.doi.org/10.1103/PhysRevD.81.055012}
  {\path{doi:10.1103/PhysRevD.81.055012}}.

\bibitem{Bazavov:2016nty}
A.~Bazavov, et~al., {$B^0_{(s)}$-mixing matrix elements from lattice QCD for
  the Standard Model and beyond}\href {http://arxiv.org/abs/1602.03560}
  {\path{arXiv:1602.03560}}.

\bibitem{Agashe:2014kda}
K.~A. Olive, et~al., {Review of Particle Physics}, Chin. Phys. C38 (2014)
  090001.
\newblock \href {http://dx.doi.org/10.1088/1674-1137/38/9/090001}
  {\path{doi:10.1088/1674-1137/38/9/090001}}.

\bibitem{Aubert:2009ag}
B.~Aubert, et~al., {Searches for Lepton Flavor Violation in the Decays
  $\tau^\pm \to e^\pm \gamma$ and $\tau^\pm \to \mu^\pm \gamma$}, Phys. Rev.
  Lett. 104 (2010) 021802.
\newblock \href {http://arxiv.org/abs/0908.2381} {\path{arXiv:0908.2381}},
  \href {http://dx.doi.org/10.1103/PhysRevLett.104.021802}
  {\path{doi:10.1103/PhysRevLett.104.021802}}.

\bibitem{Aushev:2010bq}
T.~Aushev, et~al., {Physics at Super B Factory}\href
  {http://arxiv.org/abs/1002.5012} {\path{arXiv:1002.5012}}.

\bibitem{Dorsner:2016wpm}
I.~Dor{\v s}ner, S.~Fajfer, A.~Greljo, J.~F. Kamenik, N.~Ko{\v s}nik, {Physics
  of leptoquarks in precision experiments and at particle colliders}\href
  {http://arxiv.org/abs/1603.04993} {\path{arXiv:1603.04993}}.

\bibitem{Chivukula:1991zk}
R.~S. Chivukula, M.~Golden, E.~H. Simmons, {Multi - jet physics at hadron
  colliders}, Nucl. Phys. B363 (1991) 83--96.
\newblock \href {http://dx.doi.org/10.1016/0550-3213(91)90235-P}
  {\path{doi:10.1016/0550-3213(91)90235-P}}.

\bibitem{Chivukula:2013hga}
R.~S. Chivukula, E.~H. Simmons, N.~Vignaroli, {Same-Sign Dileptons from Colored
  Scalars in the Flavorful Top-Coloron Model}, Phys. Rev. D88 (2013) 034006.
\newblock \href {http://arxiv.org/abs/1306.2248} {\path{arXiv:1306.2248}},
  \href {http://dx.doi.org/10.1103/PhysRevD.88.034006}
  {\path{doi:10.1103/PhysRevD.88.034006}}.

\bibitem{Koh:1983ir}
I.~G. Koh, S.~Rajpoot, {FINITE N=2 EXTENDED SUPERSYMMETRIC FIELD THEORIES},
  Phys. Lett. B135 (1984) 397--401.
\newblock \href {http://dx.doi.org/10.1016/0370-2693(84)90302-2}
  {\path{doi:10.1016/0370-2693(84)90302-2}}.

\bibitem{Jones:1981we}
D.~R.~T. Jones, {The Two Loop beta Function for a G(1) $\times$ G(2) Gauge
  Theory}, Phys. Rev. D25 (1982) 581.
\newblock \href {http://dx.doi.org/10.1103/PhysRevD.25.581}
  {\path{doi:10.1103/PhysRevD.25.581}}.

\bibitem{Chen:2013vi}
C.-S. Chen, C.-Q. Geng, D.~Huang, L.-H. Tsai, {New Scalar Contributions to
  $h\to Z\gamma$}, Phys. Rev. D87 (2013) 075019.
\newblock \href {http://arxiv.org/abs/1301.4694} {\path{arXiv:1301.4694}},
  \href {http://dx.doi.org/10.1103/PhysRevD.87.075019}
  {\path{doi:10.1103/PhysRevD.87.075019}}.

\bibitem{Djouadi:2005gj}
A.~Djouadi, {The Anatomy of electro-weak symmetry breaking. II. The Higgs
  bosons in the minimal supersymmetric model}, Phys. Rept. 459 (2008) 1--241.
\newblock \href {http://arxiv.org/abs/hep-ph/0503173}
  {\path{arXiv:hep-ph/0503173}}, \href
  {http://dx.doi.org/10.1016/j.physrep.2007.10.005}
  {\path{doi:10.1016/j.physrep.2007.10.005}}.

\bibitem{Moretti:2015pbj}
S.~Moretti, K.~Yagyu, {750 GeV diphoton excess and its explanation in
  two-Higgs-doublet models with a real inert scalar multiplet}, Phys. Rev.
  D93~(5) (2016) 055043.
\newblock \href {http://arxiv.org/abs/1512.07462} {\path{arXiv:1512.07462}},
  \href {http://dx.doi.org/10.1103/PhysRevD.93.055043}
  {\path{doi:10.1103/PhysRevD.93.055043}}.

\bibitem{Patel:2015tea}
H.~H. Patel, {Package-X: A Mathematica package for the analytic calculation of
  one-loop integrals}, Comput. Phys. Commun. 197 (2015) 276--290.
\newblock \href {http://arxiv.org/abs/1503.01469} {\path{arXiv:1503.01469}},
  \href {http://dx.doi.org/10.1016/j.cpc.2015.08.017}
  {\path{doi:10.1016/j.cpc.2015.08.017}}.

\bibitem{Azatov:2012rd}
A.~Azatov, R.~Contino, D.~Del~Re, J.~Galloway, M.~Grassi, S.~Rahatlou,
  {Determining Higgs couplings with a model-independent analysis of $h \to
  \gamma \gamma$}, JHEP 1206 (2012) 134.
\newblock \href {http://arxiv.org/abs/1204.4817} {\path{arXiv:1204.4817}},
  \href {http://dx.doi.org/10.1007/JHEP06(2012)134}
  {\path{doi:10.1007/JHEP06(2012)134}}.

\bibitem{Martin:2009iq}
A.~D. Martin, W.~J. Stirling, R.~S. Thorne, G.~Watt, {Parton distributions for
  the LHC}, Eur. Phys. J. C63 (2009) 189--285.
\newblock \href {http://arxiv.org/abs/0901.0002} {\path{arXiv:0901.0002}},
  \href {http://dx.doi.org/10.1140/epjc/s10052-009-1072-5}
  {\path{doi:10.1140/epjc/s10052-009-1072-5}}.

\bibitem{Eichten:1984eu}
E.~Eichten, I.~Hinchliffe, K.~D. Lane, C.~Quigg, {Super Collider Physics}, Rev.
  Mod. Phys. 56 (1984) 579--707, [Addendum: Rev. Mod. Phys.58,1065(1986)].
\newblock \href {http://dx.doi.org/10.1103/RevModPhys.56.579,
  10.1103/RevModPhys.58.1065} {\path{doi:10.1103/RevModPhys.56.579,
  10.1103/RevModPhys.58.1065}}.

\bibitem{Manohar:2006ga}
A.~V. Manohar, M.~B. Wise, {Flavor changing neutral currents, an extended
  scalar sector, and the Higgs production rate at the CERN LHC}, Phys. Rev. D74
  (2006) 035009.
\newblock \href {http://arxiv.org/abs/hep-ph/0606172}
  {\path{arXiv:hep-ph/0606172}}, \href
  {http://dx.doi.org/10.1103/PhysRevD.74.035009}
  {\path{doi:10.1103/PhysRevD.74.035009}}.

\bibitem{Quigg:2009gg}
C.~Quigg, {LHC Physics Potential versus Energy}\href
  {http://arxiv.org/abs/0908.3660} {\path{arXiv:0908.3660}}.

\end{thebibliography}
\end{document}